%% file: B2Dprd.tex
\documentclass[prd,preprint,tightenlines,preprintnumbers,aps,eqsecnum,nofootinbib,superscriptaddress,
showpacs,floatfix]{revtex4-1}

\usepackage{amsmath}    
\usepackage{bm}         
\usepackage{graphicx}   
\usepackage{verbatim}   
\usepackage{color}      
\usepackage{subfigure}  
\usepackage[colorlinks=true,citecolor=blue,linkcolor=blue,urlcolor=blue]{hyperref} 

\definecolor{red}    {rgb}{0.8,0.0,0.0}
\definecolor{green}  {rgb}{0.0,0.6,0.0}
\definecolor{blue}   {rgb}{0.0,0.2,0.6}
\definecolor{yellow} {rgb}{1.0,0.8,0.0}
\definecolor{orange} {rgb}{1.0,0.4,0.2}
\definecolor{violet} {rgb}{0.2,0.0,0.2}

\newcommand{\bea}{\begin{eqnarray}}
\newcommand{\eea}{\end{eqnarray}}

\newcommand{\asqtad}{asqtad}
\newcommand{\fm}{\text{fm}}

\newcommand{\logs}{\mathop{\text{logs}}\nolimits}

\renewcommand{\case}[2]{\ensuremath{{\textstyle\frac{#1}{#2}}}}
\newcommand{\half}{\case{1}{2}}
\newcommand{\third}{\case{1}{3}}

\newcommand{\rhoA}[1]{\ensuremath{\rho_{A^{#1}}}}
\newcommand{\rhoV}[1]{\ensuremath{\rho_{V^{#1}}}}

\newcommand{\ZVbb}{\ensuremath{Z_{V^4_{bb}}}}
\newcommand{\ZVcc}{\ensuremath{Z_{V^4_{cc}}}}
\newcommand{\ZAbc}{\ensuremath{Z_{A^i_{bc}}}}
\newcommand{\ZAcb}{\ensuremath{Z_{A^i_{cb}}}}

\newcommand{\gDDp}{\ensuremath{g_{D^*D\pi}}}

\newcommand{\hs}{\phantom{5}}

\newcommand{\sealight}{\hat{m}'} 
\newcommand{\seaheavy}{m'_s}
\newcommand{\ratlight}{\hat{x}} 
\newcommand{\ratheavy}{x_s}

\newcommand{\ie}{\emph{i.e.}}
\newcommand{\eg}{\emph{e.g.}}

\newcommand{\Dslash}{\ensuremath{D\kern-0.65em/\kern0.15em}}
\newcommand{\Eslash}{\ensuremath{E\kern-0.65em/\kern0.15em}}
\newcommand{\dslash}{\ensuremath{\partial\kern-0.65em/\kern0.15em}}
\newcommand{\vslash}{\ensuremath{v\kern-0.65em/\kern0.15em}}
\newcommand{\eslash}{\ensuremath{\kern0.1em\epsilon\kern-0.4em/\kern-0.1em}}

\begin{document}
\bibliographystyle{apsrev4-1}

\pacs{12.38.Gc, 
      13.20.He, 
	  12.15.Hh  
}

\title{\boldmath Update of $|V_{cb}|$ from the $\bar{B}\to D^*\ell\bar{\nu}$ 
form factor at zero recoil with three-flavor lattice QCD}

\author{Jon A.~Bailey}
\affiliation{Department of Physics and Astronomy, Seoul National University, Seoul, South Korea}
\author{A.~Bazavov}
\affiliation{Physics Department, Brookhaven National Laboratory, Upton, New York, USA}
\author{C.~Bernard} 
\affiliation{Department of Physics, Washington University, St.~Louis, Missouri, USA}
\author{C.~M.~Bouchard}
\affiliation{Department of Physics, The Ohio State University, Columbus, Ohio, USA}
\author{C.~DeTar} 
\affiliation{Department Physics and Astronomy, University of Utah, Salt Lake City, Utah, USA}
\author{Daping~Du}
\affiliation{Physics Department, University of Illinois, Urbana, Illinois, USA}
\affiliation{Department of Physics, Syracuse University, Syracuse, New York, USA}
\author{A.~X.~El-Khadra}
\affiliation{Physics Department, University of Illinois, Urbana, Illinois, USA}
\affiliation{Fermi National Accelerator Laboratory, Batavia, Illinois, USA}
\author{J.~Foley}
\affiliation{Department Physics and Astronomy, University of Utah, Salt Lake City, Utah, USA}
\author{E.~D.~Freeland}
\affiliation{Department of Physics, Benedictine University, Lisle, Illinois, USA}
\affiliation{Liberal Arts Department, School of the Art Institute of Chicago, Chicago, Illinois, USA}
\author{E.~G\'amiz}
\affiliation{CAFPE and Departamento de F\'{\i}sica Te\'orica y del Cosmos, Universidad de Granada, Granada, Spain}
\author{Steven~Gottlieb} 
\affiliation{Department of Physics, Indiana University, Bloomington, Indiana, USA}
\author{U.~M.~Heller}
\affiliation{American Physical Society, Ridge, New York, USA}
\author{A.~S.~Kronfeld}\email{ask@fnal.gov}
\affiliation{Fermi National Accelerator Laboratory, Batavia, Illinois, USA}
\author{J.~Laiho}\email{jlaiho@fnal.gov}
\affiliation{SUPA, Department of Physics and Astronomy, University of Glasgow, Glasgow, United Kingdom}
\affiliation{Department of Physics, Syracuse University, Syracuse, New York, USA}
\author{L.~Levkova}
\affiliation{Department Physics and Astronomy, University of Utah, Salt Lake City, Utah, USA}
\author{P.~B.~Mackenzie} 
\affiliation{Fermi National Accelerator Laboratory, Batavia, Illinois, USA}
\author{E.~T.~Neil}
\affiliation{Fermi National Accelerator Laboratory, Batavia, Illinois, USA}
\affiliation{Department of Physics, University of Colorado, Boulder, Colorado, USA}
\affiliation{RIKEN-BNL Research Center, Brookhaven National Laboratory, Upton, NY 11973, USA}
\author{Si-Wei Qiu}
\affiliation{Department Physics and Astronomy, University of Utah, Salt Lake City, Utah, USA}
\author{J.~Simone}
\affiliation{Fermi National Accelerator Laboratory, Batavia, Illinois, USA}
\author{R.~Sugar}
\affiliation{Department of Physics, University of California, Santa Barbara, California, USA}
\author{D.~Toussaint}
\affiliation{Department of Physics, University of Arizona, Tucson, Arizona, USA}
\author{R.~S.~Van~de~Water}
\affiliation{Fermi National Accelerator Laboratory, Batavia, Illinois, USA}
\author{Ran Zhou}
\affiliation{Department of Physics, Indiana University, Bloomington, Indiana, USA}
\affiliation{Fermi National Accelerator Laboratory, Batavia, Illinois, USA}
\collaboration{Fermilab Lattice and MILC Collaborations}
\noaffiliation

\begin{abstract}
We compute the zero-recoil form factor for the semileptonic decay $\bar{B}^0\to D^{*+}\ell^-\bar{\nu}$ (and
modes related by isospin and charge conjugation) using lattice QCD with three flavors of sea quarks.
We use an improved staggered action for the light valence and sea quarks (the MILC \asqtad\ configurations),
and the Fermilab action for the heavy quarks.
Our calculations incorporate higher statistics, finer lattice spacings, and lighter quark masses than our
2008 work.
As a byproduct of tuning the new data set, we obtain the $D_s$ and $B_s$ hyperfine splittings with few-MeV 
accuracy.
For the zero-recoil form factor, we obtain $\mathcal{F}(1)=0.906(4)(12)$, where the first error is
statistical and the second is the sum in quadrature of all systematic errors.
With the latest HFAG average of experimental results and a cautious treatment of QED effects, we find
$|V_{cb}| = (39.04 \pm 0.49_\text{expt} \pm 0.53_\text{QCD} \pm 0.19_\text{QED})\times10^{-3}$.
The QCD error is now commensurate with the experimental error.
\end{abstract}

\date{\today} 

\maketitle

\section{Introduction}

The Cabibbo-Kobayashi-Maskawa (CKM) matrix element $|V_{cb}|$ is one of the fundamental parameters of the
Standard Model (SM).
Together with $|V_{us}|$, $|V_{ub}|$, and $\arg V^*_{ub}$, it allows for a full SM determination of flavor
and $CP$ violation via processes that proceed at the tree level of the electroweak interaction.
In the case of $|V_{cb}|$, one requires a measurement of the differential rate of $B$ mesons decaying
semileptonically to a charmed final state.
The hadronic part of the final state can be exclusive---\eg, a $D^*$ or $D$ meson---or inclusive.

The 2012 edition of the Review of Particle Physics by the Particle Data Group (PDG)~\cite{Beringer:1900zz}
notes that the exclusive and inclusive values of $|V_{cb}|$ are marginally consistent with each other.
Furthermore, global fits to a comprehensive range of flavor- and $CP$-violating observables tend to prefer
the inclusive value~\cite{Laiho:2009eu,Lunghi:2009ke,Lenz:2010gu}: when direct information on $|V_{cb}|$ is
omitted from the fit, one of the outputs of the fit is a value of $|V_{cb}|$ that agrees better with the 
inclusive than the exclusive value.
One should bear in mind that some tension in the global fits to the whole CKM
paradigm has been seen~\cite{Lunghi:2010gv}.
A full discussion of the possible resolutions of the discrepancy lies beyond the scope of this article.
We conclude merely that it is important and timely to revisit the theoretical and experimental ingredients of both
determinations.

\begin{figure}[b]
    \vspace*{-6pt}
    \includegraphics[width=0.5\textwidth]{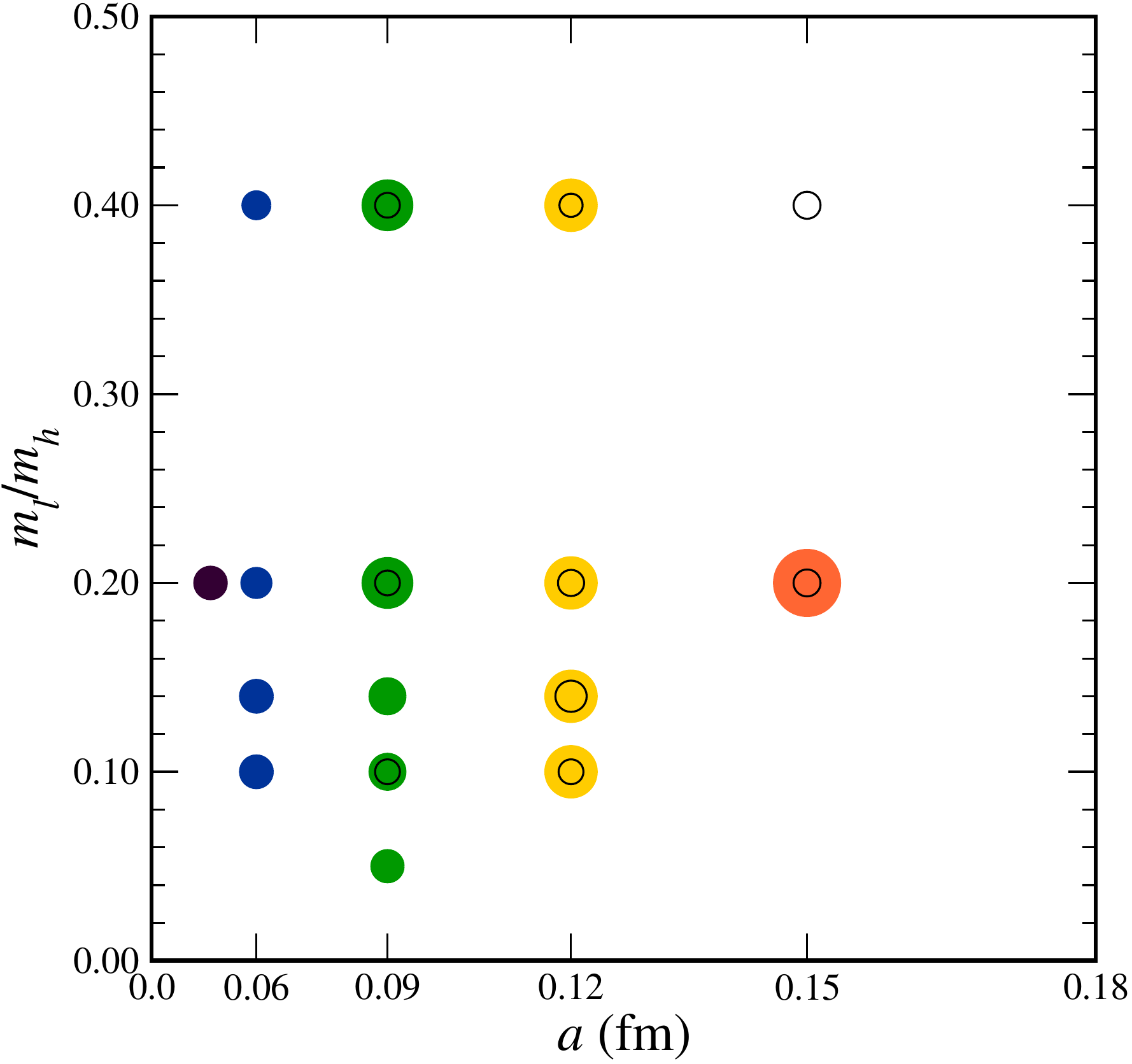}
    \vspace*{-6pt}
    \caption[fig:a-vs-m]{(color online) Range of lattice spacings and light-quark masses used here 
        (colored or gray discs) and in Ref.~\cite{Bernard:2008dn} (black circles).
        The area is proportional to the size of the ensemble.
        The lattice spacings are $a\approx0.15$, 0.12, 0.09, 0.06, and 0.045~fm.
        Reference~\cite{Bailey:2010gb} did not yet include the ensembles with 
        $(a,\sealight/m_s)=(0.045~\text{fm},0.20)$, $(0.06~\text{fm},0.14)$, $(0.06~\text{fm},0.10)$, and 
        $(0.09~\text{fm},0.05)$.} 
    \label{fig:a-vs-m}
\end{figure}

In this paper, we improve the lattice-QCD calculation~\cite{Hashimoto:2001nb,Bernard:2008dn,Bailey:2010gb}
of the zero-recoil form factor for the exclusive decay $\bar{B}\to D^*\ell\bar{\nu}$ (and isopin-partner 
and charge-conjugate modes).
Our analysis strategy is very similar to our previous work~\cite{Bernard:2008dn}, but the lattice-QCD
data set is much more extensive, with higher statistics on all ensembles, smaller lattice spacings (as small
as $a\approx0.045$~fm) and light-quark masses as small as $\sealight=m_s/20$ (at lattice spacing
$a\approx0.09$~fm).
Figure~\ref{fig:a-vs-m} provides a simple overview of the new and old data sets;
further details are given in Sec.~\ref{sec:lattice}.
Our preliminary status report~\cite{Bailey:2010gb} encompassed the higher statistics but not yet four of the
ensembles in the lower left-hand corner of Fig.~\ref{fig:a-vs-m}.

With this work, we improve the precision of $|V_{cb}|$ as determined from exclusive decays to that claimed 
for the determination from inclusive decays: 2\%.
Moreover, we reduce the QCD uncertainty on $|V_{cb}|$ to the same level as the experimental uncertainty.
Because $|V_{cb}|$ normalizes the unitarity triangle, it appears throughout flavor physics.
For example, the SM expressions for $\varepsilon_K$ and for the branching ratios of the golden modes 
$K^+\to\pi^+\nu\bar{\nu}$ and $K_L\to\pi^0\nu\bar{\nu}$ all contain $|V_{cb}|^4$.
Therefore, further improvements---beyond what is achieved here---are warranted, particularly during the 
course of the Belle~II experiment~\cite{Aushev:2010bq}.

The amplitude for $B\to D^*$ semileptonic decay is expressed in terms of form factors,
\begin{eqnarray}
    \frac{\langle D^*(p_{D^*},\epsilon^{(\alpha)})|\mathcal{A}^\mu|B(p_B)\rangle}%
    {\quad\sqrt{2M_{D^*}}\hfill\sqrt{2M_B}\;} & = & \frac{i}{2} {\epsilon^{(\alpha)}_\nu}^*
        \left[g^{\mu\nu}(1+w)h_{A_1}(w) - 
        v^\nu_B\left(v^\mu_Bh_{A_2}(w) + v^\mu_{D^*}h_{A_3}^{}(w)\right)\right], \hspace*{24pt}
    \label{eq:hAi} \\
    \frac{\langle D^*(p_{D^*},\epsilon^{(\alpha)})|\mathcal{V}^\mu|B(p_B)\rangle}%
    {\quad\sqrt{2M_{D^*}}\hfill\sqrt{2M_B}\;} & = & \frac{1}{2} 
        \varepsilon^{\mu\nu}_{\hphantom{\mu\nu}\rho\sigma} {\epsilon^{(\alpha)}_\nu}^*
        v^\rho_B v^\sigma_{D^*} h_V(w),
    \label{eq:hV}
\end{eqnarray}
where $\mathcal{A}^\mu$ and $\mathcal{V}^\mu$ are the (continuum QCD) $b\to c$ electroweak currents,
$v_B^\mu=p_B^\mu/M_B$, $v_{D^*}^\mu=p_{D^*}^\mu/M_{D^*}$, the velocity transfer $w=v_B\cdot v_{D^*}$, and
$\epsilon^{(\alpha)}$ is the polarization vector of the $D^*$ meson.
In the SM, the differential rate for $B^-\to D^{0*}\ell^-\bar{\nu}$ (and the charge-conjugate mode) is given
by
\begin{equation}
    \frac{d\Gamma}{dw} = \frac{G_F^2 M_{D^*}^3}{4\pi^3}(M_B-M_{D^*})^2(w^2-1)^{1/2} 
        |\eta_{\text{EW}}|^2|V_{cb}|^2 \chi(w)|\mathcal{F}(w)|^2,
    \label{eq:dGdw}
\end{equation}
where $\eta_{\text{EW}}$ provides a structure-independent electroweak correction from
next-to-leading-order box diagrams, in which a photon or $Z$ boson is exchanged along with the $W$ 
boson~\cite{Sirlin:1981ie}.
(See Sec.~\ref{sec:ew} for details.)
The rate for $\bar{B}^0\to D^{+*}\ell^-\bar{\nu}$ (and charge conjugate) is the same as Eq.~(\ref{eq:dGdw})
but with an additional factor on the right-hand side
$(1+\pi\alpha)$~\cite{Ginsberg:1968pz,Atwood:1989em}, which accounts for the Coulomb attraction of the
final-state charged particles.

The notation $\chi(w)|\mathcal{F}(w)|^2$ is conventional, motivated by the heavy-quark limit.
In the zero-recoil limit, $w\to1$, one has $\chi(w)\to1$, and only one form factor survives:
\begin{equation}
    \mathcal{F}(1) = h_{A_1}(1).
    \label{eq:F=hA1}
\end{equation}
From Eq.~(\ref{eq:hAi}), one sees that the needed matrix element is
$\langle D^*|\epsilon^{(\alpha)}\cdot\mathcal{A}|B\rangle$ with initial and final states both at rest.

For nonvanishing lepton mass~$m_\ell$, the rate is multiplied by $(1-m_\ell^2/q^2)^2$, and the expressions
for $\chi(w)$ and $|\mathcal{F}(w)|^2$ receive corrections proportional to $m_\ell^2/q^2$
\cite{Korner:1989qb}.
At zero recoil, these corrections reduce to an additional factor $(1+m_\ell^2/q^2_{\text{max}})$ on the
right-hand side of Eq.~(\ref{eq:F=hA1}).
Except for $\ell=\tau$, lepton mass effects are not important even at the current level of accuracy.

Because precision is so crucial, the lattice-QCD calculation must be set up in a way that ensures 
considerable cancellation of all sources of uncertainty.
The pioneering work of Hashimoto \emph{et al.}~\cite{Hashimoto:1999yp,Hashimoto:2001nb} introduced several
double ratios to this end.
Here, we follow Ref.~\cite{Bernard:2008dn} and use a single, direct double ratio
\begin{equation}
    \mathcal{R}_{A_1} = \frac{\langle D^*|\bar{c}\gamma^j\gamma^5b|\bar{B}\rangle
        \langle\bar{B}|\bar{b}\gamma^j\gamma^5c|D^*\rangle}%
        {\langle D^*|\bar{c}\gamma^4c|D^*\rangle \langle\bar{B}|\bar{b}\gamma^4b|\bar{B}\rangle} = 
        \left|h_{A_1}(1)\right|^2
    \label{eq:contRA1}
\end{equation}
with all states at rest and the polarization of the $D^*$ aligned with~$j$.
In the continuum, the denominator of Eq.~(\ref{eq:contRA1}) is unity, by the definition of the flavor 
quantum numbers.
On the lattice, however, it normalizes the flavor numbers \emph{and} cancels statistical fluctuations.
The main uncertainties stem, then, from the chiral extrapolation (the light-quark masses in our data exceed 
the up and down masses) and discretization and matching errors.
In particular, we show how the discretization errors of the analogous ratio of lattice-QCD correlation 
functions are reduced by use of the ratio.

The rest of this paper is organized as follows.
Section~\ref{sec:lattice} describes the details of the lattice-QCD calculation.
We discuss the lattice implementation of Eq.~(\ref{eq:contRA1}), the details of the numerical data, and the
general structure of the computed correlation functions.
Section~\ref{sec:correlator} describes our fits to a ratio of correlation functions.
Section~\ref{sec:PT} discusses perturbative matching.
Section~\ref{sec:kappa} summarizes the tuning of the bottom- and charm-quark masses and presents results for 
the $D_s$ and $B_s$ hyperfine splittings.
Our extrapolation to the continuum limit and physical light-quark mass is described in Sec.~\ref{sec:extrap}.
Section~\ref{sec:systematics} gives full details of our systematic error analysis.
Section~\ref{sec:ew} provides a discussion of electroweak and electromagnetic effects, which, though separate
from our QCD calculation, are needed to obtain~$|V_{cb}|$.
Section~\ref{sec:end} concludes with final results for $h_{A_1}(1)$ and~$|V_{cb}|$.
The appendices contain additional material, including the formulas used for the chiral extrapolation
(Appendix~\ref{app:chpt}), an estimate of heavy-quark discretization errors (Appendix~\ref{app:hqetdisc}),
and a thorough discussion of our procedure for tuning the bottom- and charm-quark masses
(Appendix~\ref{app:kappa-tuning}), which also yields the hyperfine splitting.

\section{Lattice setup}
\label{sec:lattice}

In this section we discuss the ingredients of our lattice-QCD calculation.
We outline first the generation of ensembles of lattice gauge fields, and then the procedures for computing 
the three-point correlation functions needed to obtain the double ratio~$R_{A_1}$, which is the lattice 
correlation-function analog of~$\mathcal{R}_{A_1}$.

\subsection{Simulation parameters}

We use the MILC ensembles~\cite{Bazavov:2009bb} of lattice gauge fields listed in Table~\ref{tab:params}.
The ensembles were generated with a Symanzik-improved gauge action~\cite{Weisz:1982zw,Curci:1983an,%
Weisz:1983bn,Luscher:1984xn} and 2+1 flavors of sea quarks.
The couplings in the gauge action include the one-loop effects of gluons~\cite{Luscher:1985zq} but not of sea
quarks~\cite{Hao:2007iz}; the latter were not yet available when the gauge-field generation
began~\cite{Aubin:2004fs}.
\begin{table}
\centering
    \caption{Parameters of the lattice gauge fields.
        The columns from left to right are the approximate lattice spacing in fm, the sea-quark masses
        $a\sealight/a\seaheavy$, the linear spatial dimension of the lattice ensemble in fm, the dimensionless 
        factor $m_\pi L$ (with $m_\pi$ from the Goldstone pion), 
        the gauge coupling, the dimensions of the lattice in lattice units, the number of sources and
        configurations in each ensemble, and the tadpole improvement factor~$u_0$ 
        (obtained from the average plaquette).\vspace*{4pt}}
    \label{tab:params}
    \begin{tabular}{c@{\quad}r@{/}l@{\quad}c@{\quad}c@{\quad}c@{\quad}c@{\quad}r@{$\times$}l@{\quad}l}
      \hline \hline
       $a$ (fm) & $a\sealight$&$a\seaheavy$ & $L$~(fm) & $m_\pi L$ & $10/g^2$ & Volume & Sources&Configs & ~~$u_0$ \\
      \hline
      $0.15\hs$ & $0.0097$&$0.0484$   & 2.4 & 3.9 & 6.572 & \ $16^3\times 48$ & 24&628 & 0.8604 \\
      \hline
      $0.12\hs$ & $0.02$&$0.05$       & 2.4 & 6.2 & 6.79  & \ $20^3\times 64$ & 4&2052 & 0.8688 \\
      $0.12\hs$ & $0.01$&$0.05$       & 2.4 & 4.5 & 6.76  & \ $20^3\times 64$ & 4&2256 & 0.8677 \\
      $0.12\hs$ & $0.007$&$0.05$      & 2.4 & 3.8 & 6.76  & \ $20^3\times 64$ & 4&2108 & 0.8678 \\
      $0.12\hs$ & $0.005$&$0.05$      & 2.9 & 3.8 & 6.76  & \ $24^3\times 64$ & 4&2096 & 0.8678 \\
      \hline
      $0.09\hs$ & $0.0124$&$0.031$    & 2.4 & 5.8 & 7.11  & \ $28^3\times 96$ & 4&1992 & 0.8788 \\
      $0.09\hs$ & $0.0062$&$0.031$    & 2.4 & 4.1 & 7.09  & \ $28^3\times 96$ & 4&1928 & 0.8782 \\
      $0.09\hs$ & $0.00465$&$0.031$   & 2.7 & 4.1 & 7.085 & \ $32^3\times 96$ & 4&984  & 0.8781 \\
      $0.09\hs$ & $0.0031$&$0.031$    & 3.4 & 4.2 & 7.08  & \ $40^3\times 96$ & 4&1012 & 0.8779 \\
      $0.09\hs$ & $0.00155$&$0.031$   & 5.5 & 4.8 & 7.075 & \ $64^3\times 96$ & 4&788  & 0.877805 \\
      \hline
      $0.06\hs$ & $0.0072$&$0.018$    & 2.9 & 6.3 & 7.48  & \ $48^3\times144$ & 4&576  & 0.8881  \\
      $0.06\hs$ & $0.0036$&$0.018$    & 2.9 & 4.5 & 7.47  & \ $48^3\times144$ & 4&672  & 0.88788 \\
      $0.06\hs$ & $0.0025$&$0.018$    & 3.4 & 4.4 & 7.465 & \ $56^3\times144$ & 4&800  & 0.88776 \\
      $0.06\hs$ & $0.0018$&$0.018$    & 3.8 & 4.3 & 7.46  & \ $64^3\times144$ & 4&824  & 0.88764 \\
      \hline
      $0.045$ & $0.0028$&$0.014$      & 2.9 & 4.6 & 7.81  & \ $64^3\times192$ & 4&800 & 0.89511 \\
      \hline \hline
    \end{tabular}
\end{table}
The sea-quark action is the order~$a^2$, tadpole-improved (\asqtad)
action~\cite{Blum:1996uf,Orginos:1998ue,Lagae:1998pe,Lepage:1998vj,Orginos:1999cr} for staggered
quarks~\cite{Susskind:1976jm,Sharatchandra:1981si}.
To reduce the species content from the four that come with staggered fermions, the light quarks (strange
quark) are simulated with the square root (fourth root) of the determinant~\cite{Hamber:1983kx}.
At nonzero lattice spacing this procedure introduces small violations of unitarity~\cite{Prelovsek:2005rf,%
Bernard:2006zw,Bernard:2007qf,Aubin:2008wk} and locality~\cite{Bernard:2006ee}.
Considerable numerical and theoretical evidence suggests that these effects go away in the continuum limit,
so that the procedure yields~QCD~\cite{Adams:2004mf,Shamir:2004zc,Durr:2005ax,Bernard:2006zw,Shamir:2006nj,%
Sharpe:2006re, Bernard:2007eh,Kronfeld:2007ek,Golterman:2008gt,Donald:2011if}.

As one can see from Table~\ref{tab:params}, some ensembles contain $\sim2000$ independent gauge fields,
others $\sim600$--1000.
To increase statistics, we re-use each field four times (for $a\approx0.15$~fm, 24 times) by computing 
quark propagators that are evenly spaced in the time direction with a spatial source origin that is chosen at random from one configuration to the next.

We also use the \asqtad\ action for the light valence (spectator) quark.
In this paper, we denote the physical quark masses by $m_u$, $m_d$, $\hat{m}=\half(m_u+m_d)$, and $m_s$; the
variable spectator mass by~$m_x$; and the sea-quark masses $\sealight$ and $\seaheavy$, which are fixed within
each ensemble.
The bare spectator masses $am_x$ are listed in Table~\ref{tab:params2}.
\begin{table}
\centering
    \caption{Valence-quark parameters used in the simulations.
        The (approximate) lattice spacings~$a$ and the sea-quark masses $a\sealight/a\seaheavy$ (first two 
        columns) identify the ensemble.
        Here, $am_x$ denotes the bare masses for the light spectator quarks, 
        $c_\text{SW}$ and $\kappa$ denote the parameters in the SW action,
        and $d_1$ the rotation parameter in the current.
        The primes on $\kappa$ and $d_1$ distinguishes the simulation from the physical values.
        \vspace*{4pt}}
    \label{tab:params2}
    \begin{tabular}{c@{\quad}r@{/}l@{\quad}l@{ }l@{\quad}l@{\quad}l@{\quad}l@{\quad}l@{\quad}l}
      \hline \hline
       $a$ (fm) & $a\sealight$&$a\seaheavy$ & \multicolumn{2}{c}{$am_x$} & \multicolumn{1}{c}{$c_\text{SW}$~~} 
           & \multicolumn{1}{c}{$\kappa'_b$~~~} & \multicolumn{1}{c}{$d'_{1b}$~~~}    
           & \multicolumn{1}{c}{$\kappa'_c$~~~} & \multicolumn{1}{c}{$d'_{1c}$} \\
      \hline
      $0.15\hs$ & $0.0097$&$0.0484$ & 0.0097,&0.0194 & 1.567  & 0.0781 & 0.08354 & 0.1218 & 0.08825 \\
      \hline
      $0.12\hs$ & $0.02$&$0.05$     & 0.02   &       & 1.525  & 0.0918 & 0.09439 & 0.1259 & 0.07539 \\
      $0.12\hs$ & $0.01$&$0.05$     & 0.01,  & 0.02  & 1.531  & 0.0901 & 0.09334 & 0.1254 & 0.07724 \\
      $0.12\hs$ & $0.007$&$0.05$    & 0.007, & 0.02  & 1.530  & 0.0901 & 0.09332 & 0.1254 & 0.07731 \\
      $0.12\hs$ & $0.005$&$0.05$    & 0.005, & 0.02  & 1.530  & 0.0901 & 0.09332 & 0.1254 & 0.07733 \\
      \hline
      $0.09\hs$ & $0.0124$&$0.031$  & 0.0124 &       & 1.473  & 0.0982 & 0.09681 & 0.1277 & 0.06420 \\
      $0.09\hs$ & $0.0062$&$0.031$  & 0.0062,&0.0124 & 1.476  & 0.0979 & 0.09677 & 0.1276 & 0.06482 \\
      $0.09\hs$ & $0.00465$&$0.031$ & 0.00465 &      & 1.477  & 0.0977 & 0.09671 & 0.1275 & 0.06523 \\
      $0.09\hs$ & $0.0031$&$0.031$  & 0.0031,&0.0124 & 1.478  & 0.0976 & 0.09669 & 0.1275 & 0.06537 \\
      $0.09\hs$ & $0.00155$&$0.031$ & 0.00155 &      & 1.4784 & 0.0976 & 0.09669 & 0.1275 & 0.06543 \\
      \hline
      $0.06\hs$ & $0.0072$&$0.018$  & 0.0072 &       & 1.4276 & 0.1048 & 0.09636 & 0.1295 & 0.05078 \\
      $0.06\hs$ & $0.0036$&$0.018$  & 0.0036,&0.0072 & 1.4287 & 0.1052 & 0.09631 & 0.1296 & 0.05055 \\
      $0.06\hs$ & $0.0025$&$0.018$  & 0.0025  &      & 1.4293 & 0.1052 & 0.09633 & 0.1296 & 0.05070 \\
      $0.06\hs$ & $0.0018$&$0.018$  & 0.0018  &      & 1.4298 & 0.1052 & 0.09635 & 0.1296 & 0.05076 \\
      \hline
      $0.045$   & $0.0028$&$0.014$  & 0.0028  &      & 1.3943 & 0.1143 & 0.08864 & 0.1310 & 0.03842 \\
      \hline \hline
    \end{tabular}
\end{table}
In every case, we compute light-quark propagators with the valence mass equal to the light mass,
$am_x=a\sealight$, and, in several cases, we also compute a partially quenched propagator with 
$am_x=0.4a\seaheavy$.

For the heavy $b$ and $c$ quarks we use Wilson fermions~\cite{Wilson:1977nj} with the Sheikholeslami-Wohlert
(SW) action~\cite{Sheikholeslami:1985ij}, adjusting the parameters in the action according to the Fermilab
method~\cite{ElKhadra:1996mp}.
Table~\ref{tab:params2} also lists the parameters of the heavy-quark action: the hopping parameter $\kappa$
(for each quark) and the clover coefficient of the SW action.
We use $\kappa'_b$ and $\kappa'_c$ to denote the values used in the computations, reserving $\kappa_b$ and
$\kappa_c$ for those that reproduce the $B_s$ and $D_s$ meson masses most accurately.
We set $c_{\text{SW}}$ to the value from tree-level tadpole-improved perturbation theory,
$c_{\text{SW}}=u_0^{-3}$, with $u_0$ from Table~\ref{tab:params}.
Table~\ref{tab:params1.5} gives the values of $\kappa_{\rm crit}$ where the quark mass vanishes for the SW
action on each of our ensembles.
\begin{table}
\centering
    \caption{Derived parameters that enter the simulations.
        The (approximate) lattice spacings~$a$ and the sea-quark masses $a\sealight/a\seaheavy$ (first two 
        columns) identify the ensemble.
        Values for $r_1/a$ are given in column three, and $\kappa_{\rm crit}$ values for the SW action 
        evaluated on our ensembles are given in column four.
        For $r_1/a$, statistical errors are 0.1 to $0.3\%$, and the systematic errors are comparable.
        For $\kappa_{\rm crit}$ the errors are a few in the last quoted digit. 
        \vspace*{4pt}}
    \label{tab:params1.5}
    \begin{tabular}{c@{\quad}r@{/}l@{\quad}l@{ }l@{\quad}l@{\quad}l}
      \hline \hline
       $a$ (fm) & $a\sealight$&$a\seaheavy$ & $r_1/a$ & $\kappa_{\rm crit}$ \\
      \hline
      $0.15\hs$ & $0.0097$&$0.0484$ & 2.2215 \ \ & 0.142432  \\
      \hline
      $0.12\hs$ & $0.02$&$0.05$     & 2.8211   & 0.14073       \\
      $0.12\hs$ & $0.01$&$0.05$     & 2.7386  & 0.14091   \\
      $0.12\hs$ & $0.007$&$0.05$    & 2.7386 & 0.14095  \\
      $0.12\hs$ & $0.005$&$0.05$    & 2.7386 & 0.14096   \\
      \hline
      $0.09\hs$ & $0.0124$&$0.031$  & 3.8577 & 0.139052       \\
      $0.09\hs$ & $0.0062$&$0.031$  & 3.7887 & 0.139119  \\
      $0.09\hs$ & $0.00465$&$0.031$ & 3.7716 & 0.139134     \\
      $0.09\hs$ & $0.0031$&$0.031$  & 3.7546 & 0.139173    \\
      $0.09\hs$ & $0.00155$&$0.031$ & 3.7376 & 0.13919   \\
      \hline
      $0.06\hs$ & $0.0072$&$0.018$  & 5.3991 & 0.137582     \\
      $0.06\hs$ & $0.0036$&$0.018$  & 5.3531 & 0.137632    \\
      $0.06\hs$ & $0.0025$&$0.018$  & 5.3302  & 0.137667      \\
      $0.06\hs$ & $0.0018$&$0.018$  & 5.3073  & 0.137678      \\
      \hline
      $0.045$   & $0.0028$&$0.014$  & 7.2082  &  0.13664    \\
      \hline \hline
    \end{tabular}
\end{table}
These values were determined using the methods discussed in Ref.~\cite{Bernard:2010fr}; note that
$\kappa_{\rm crit}$ is only needed in the present work to fix the improvement coefficients that correct the
lattice currents described below.

The relative lattice spacing is determined by calculating $r_1/a$ on each ensemble, where $r_1$ is related to
the heavy-quark potential and is defined such that the force between static quarks, $r_1^2F(r_1)=1.0$
\cite{Sommer:1993ce, Bernard:2000gd}.
A mass-independent procedure is used to set $r_1/a$.
This procedure takes the measured values $r_1(\hat{m}',m'_s,\beta)/a$ and constructs a smooth 
interpolation/extrapolation, which we use to replace the measured values with
$r_1(\hat{m},m_s,\beta)/a$, evaluated now at the physical masses $\hat{m}$, $m_s$.
Table~\ref{tab:params1.5} lists $r_1/a$ values for each of the ensembles that results from fitting the
calculated $r_1/a$ to the smooth function and extrapolating/interpolating to physical masses.
The absolute lattice spacing requires a physical quantity to set the scale.
We take the absolute lattice spacing to be $r_1=0.3117(22)$ fm from the MILC determination of $f_\pi$.
The value used is explained and justified in Ref.~\cite{Bazavov:2011aa}.

We have to adjust the light-quark bare masses and the heavy-quark hopping parameters to their physical values
\emph{a~posteriori}.
The adjustment of the light-quark masses is carried out in the chiral extrapolation, discussed in
Sec.~\ref{sec:extrap}.
For the heavy quarks, we have chosen $\kappa'_b$ and $\kappa'_c$ in Table~\ref{tab:params2} close to the
physical value based on an initial set of runs that studied a range of $\kappa$ but computed only the
two-point functions for heavy-strange meson masses.
After the full runs, including three-point functions, we re-analyzed the two-point functions to determine
more precise $\kappa$ values, as discussed in detail in Appendix~\ref{app:kappa-tuning}.
Using information on the $\kappa$ dependence, we can then fine-tune our result.

\subsection{\boldmath $B\to D^*$ correlation functions}
\label{sec:corrs}

To obtain the matrix elements in Eq.~(\ref{eq:contRA1}), we compute the correlation functions
\begin{eqnarray}
    C^{B\to D^*}(t_s,t_f) & = & \sum_{\bm{x},\bm{y}} \langle\mathcal{O}_{D^*_j}(\bm{x},t_f) 
        A^j_{cb}(\bm{y},t_s) \mathcal{O}_B^\dagger(\bm{0},0) \rangle,
    \label{eq:BAD} \\
    C^{B\to B}(t_s,t_f)   & = & \sum_{\bm{x},\bm{y}} \langle\mathcal{O}_B(\bm{x},t_f) 
        V^4_{bb}(\bm{y},t_s) \mathcal{O}_B^\dagger(\bm{0},0) \rangle,
    \label{eq:BVB} 
\end{eqnarray}
and similarly $C^{D^*\to B}$ and $C^{D^*\to D^*}$.
Here, $\mathcal{O}_B$ and $\mathcal{O}_{D^*_j}$ are lattice operators with quantum numbers needed to
annihilate $B$ and $D^*$ mesons, in the case of $D^*$ with polarization in the $j$~direction; $V^\mu_{cb}$
and $A^\mu_{cb}$ are lattice currents for $b\to c$ transitions.  The lattices are gauge-fixed before evaluating the correlation functions so that we can use a smearing function that is extended over a spatial slice.

We form the interpolating operators from a staggered fermion field~$\chi$ and heavy-quark field~$\psi$ in the SW action:
\begin{eqnarray}
    \mathcal{O}_{D^*_j}(\bm{x},t)   & = & \sum_{\bm{w}} \bar{\chi}(\bm{x},t)\Omega^\dagger(\bm{x},t)i\gamma_j
        S(\bm{x},\bm{w})\psi_c(\bm{w},t), \\
    \mathcal{O}_B^\dagger(\bm{x},t) & = & \sum_{\bm{w}} \bar{\psi}_b(\bm{w},t)S(\bm{w},\bm{x})
        \gamma_5\Omega(\bm{x},t)\chi(\bm{x},t), \\
    \Omega(x) & = & \gamma_1^{x_1/a}\gamma_2^{x_2/a}\gamma_3^{x_3/a}\gamma_4^{x_4/a}, \quad
        [x=(\bm{x},t)],
\end{eqnarray}
where $S(\bm{x},\bm{y})$ is a spatial smearing function.  .
The free Dirac index on $\Omega$ can be interpreted as a taste index (in which case we average over
taste)~\cite{Kronfeld:2007ek}, or one can promote $\chi$ to a four-component field~\cite{Wingate:2003nn},
which leads to the same results for the correlation functions of bilinear operators.

We employ two smearing functions.
One is the local $S(\bm{x},\bm{y})=\delta(\bm{x}-\bm{y})$.
The other is the ground-state 1S wavefunction of the Richardson potential.
See Ref.~\cite{Bazavov:2011aa} for details.

We define the lattice vector and axial-vector currents to be
\begin{eqnarray}
    V^\mu_{hh} & = & \bar{\Psi}_h\gamma^\mu\Psi_h,
    \label{eq:Vlat} \\
    A^\mu_{cb} & = & \bar{\Psi}_c\gamma^\mu\gamma^5\Psi_b,
    \label{eq:Alat}
\end{eqnarray}
where $h=b$, $c$ are flavor indices.
The fermion field $\Psi$ includes a correction factor to reduce discretization
effects~\cite{ElKhadra:1996mp},
\begin{equation}
    \Psi_h = \left(1 + d_1\bm{\gamma}\cdot\bm{D}_{\text{lat}}\right)\psi_h,
    \label{eq:Psi=psi}
\end{equation}
where $D^\mu_{\text{lat}}$ is a nearest-neighbor covariant difference operator.  Its coefficient $d_1$ is 
set to its value in tree-level tadpole-improved perturbation theory, where it does not depend on the other 
quark in the current.
The matrix elements of the lattice currents satisfy ($\doteq$ means ``has the same matrix elements as'')~%
\cite{Kronfeld:2000ck,Harada:2001fj}
\begin{equation}
    Z_{J^\mu_{cb}} J^\mu \doteq \mathcal{J}^\mu + \text{O}(\alpha_s^{1+\ell_Z}, \alpha_s^{1+\ell_d} a, a^2),
    \label{eq:ZJ}
\end{equation}
where $\mathcal{J}^\mu$ is the continuum current corresponding to the lattice current $J^\mu$ and the matching factors $Z_{J^\mu_{cb}}$ are defined such that Eq.~(\ref{eq:ZJ}) holds.
In practice, $Z_{J^\mu_{cb}}$ can be determined only approximately, via either perturbative or 
nonperturbative methods.
Thus, $\ell_{Z\,\text{or}\,d}=0$ for tree-level matching of $Z$ or $d_1$, $\ell_{Z\,\text{or}\,d}=1$ for
one-loop matching, etc.
Nonperturbative matching schemes could be set up, which would remove all powers of $\alpha_s$.
Here we implicitly use nonperturbative matching for flavor-diagonal $Z_{V^4_{hh}}$,
one-loop matching for suitable ratios of $Z_J$ factors (see below), and tree-level matching for $d_1$.
Higher-loop and nonperturbative calculations, except for $Z_{V^4_{hh}}$, are not available.

In the double ratio like Eq.~(\ref{eq:contRA1}) but with matrix elements of lattice currents, the following
ratio of matching factors remains:
\begin{equation}
    \rhoA{i}^2 = \frac{\ZAcb\ZAbc}{\ZVcc\ZVbb}.
    \label{eq:rho_ratio}
\end{equation}
In this ratio, all corrections associated with wave-function renormalization cancel out, leaving only vertex
diagrams.
Each $Z$ contains the difference between continuum and lattice vertex diagrams, and the ratio introduces
further cancellations.
It is not surprising, then, that one-loop calculations of $\rhoA{j}$ yield very small coefficients
of~$\alpha_s$~\cite{Harada:2001fj}.

With the Fermilab method applied to the SW action, the Lagrangian also leads to discretization effects of
order $\text{O}(\alpha_s^{1+\ell_c} a, a^2)$, where $\ell_c$ counts, as above, the matching of the SW 
(clover)~term.
Again, one-loop matching is not completely available (see Ref.~\cite{Nobes:2005dz}), so we use tree-level
matching.
Table~\ref{tab:params2} lists the values of $c_\text{SW}$ used in this work.
Appendix~\ref{app:hqetdisc} discusses the discretization effects in $h_{A_1}(1)$ (as extracted here) in
detail.

For large enough time separations $t_s$ and $t_f-t_s$, the correlation function
\begin{equation}
    C^{B\to D^*}(t_s,t_f) = \mathcal{Z}^{1/2}_{D^*}\mathcal{Z}^{1/2}_{\bar{B}}
        \frac{\langle D^*|A^j_{cb}|\bar{B}\rangle}{\sqrt{2M_{D^*}}\sqrt{2M_B}} \;
        e^{-M_B t_s} e^{-M_{D^*}(t_f-t_s)} + \cdots, 
    \label{eq:grndstate}
\end{equation}
where $M_B$ and $M_{D^*}$ are the masses of the $B$ and $D^*$ mesons and
$\mathcal{Z}_H=|\langle0|\mathcal{O}_H|H\rangle|^2/2M_H$.
The omitted terms from higher-mass states are discussed in Sec.~\ref{sec:correlator}.
The other correlation functions $C^{D^*\to B}$, $C^{B\to B}$, $C^{D^*\to D^*}$ have analogous large-time 
behavior.
Therefore, the ratio of correlation functions
\begin{equation}
     R(t_s,t_f) \equiv \frac{C^{B\to D^*}(t_s,t_f)C^{D^*\to B}(t_s,t_f)}%
        {C^{D^*\to D^*}(t_s,t_f)C^{B\to B}(t_s,t_f)}
        \to R_{A_1},
    \label{eq:Rlatdef} 
\end{equation}
where
\begin{equation}
    R_{A_1} = \frac{\langle D^*|A^j_{cb}|\bar{B}\rangle\langle\bar{B}|A^j_{bc}|D^*\rangle}%
        {\langle D^*|V^4_{cc}|D^*\rangle \langle\bar{B}|V^4_{bb}|\bar{B}\rangle}
        = \left|\frac{h_{A_1}(1)}{\rhoA{j}}\right|^2 + \cdots,
    \label{eq:RlathA1}
\end{equation}
is a lattice version of $\mathcal{R}_{A_1}$, up to the matching factor~$\rhoA{j}$ and discretization errors.
The analysis of $R(t_s,t_f)$ to extract $R_{A_1}$ is discussed in Sec.~\ref{sec:correlator}, the calculation
of $\rhoA{j}$ is discussed in Sec.~\ref{sec:PT}, the light-quark discretization errors are analyzed in
Sec.~\ref{sec:extrap}, and the heavy-quark discretization errors are derived in Appendix~\ref{app:hqetdisc}.

Above we mentioned that we increase statistics by choosing four (24 at $a\approx0.15$~fm) sources.
This means we choose four (24) origins $(\bm{0},0)$ in Eqs.~(\ref{eq:BAD}) and~(\ref{eq:BVB}).
We do so by picking at random four (24) equally separated timeslices for $t=0$.
On each timeslice, we choose a completely random point for $\bm{x}=\bm{0}$.

Starting at each origin [$(\bm{0},0)$ in Eqs.~(\ref{eq:BAD}) and~(\ref{eq:BVB})], we construct the
three-point correlation functions as follows.
We compute the parent heavy-quark propagator from smeared $(\bm{0},0)$ to all points, in particular
$(\bm{y},t_s)$.
We also compute the spectator staggered-quark propagator from $(\bm{0},0)$ to all points.
At time $t_f$, we convolve this propagator with the Dirac matrix and smearing function of the sink,
projecting onto a fixed momentum (here, $\bm{p}=\bm{0}$).
This combination is used for a further inversion for the daughter heavy quark; this inversion yields a
sequential propagator encoding the propagation of the spectator quark, a flavor change at the sink, and
(reverse) propagation of the daughter quark back to the decay.
This sequential propagator and the parent propagator are then inserted into the appropriate trace over color
and Dirac indices.

\section{Analysis of correlation functions}
\label{sec:correlator}

To obtain $R_{A_1}$ from $R(t_s,t_f)$ with sufficient accuracy, we have to treat the 
excited states [denoted by $\cdots$ in Eq.~(\ref{eq:grndstate})] carefully.
From the transfer-matrix formalism, one finds
\begin{equation}
    C^{X\to Y}(t_s,t_f) = \sum_{r=0}^\infty \sum_{s=0}^\infty (-1)^{rt_s/a}(-1)^{s(t_f-t_s)/a}\,
        A_{sr}\, e^{-M^{(r)}_{X}t_s}e^{-M^{(s)}_{Y}(t_f-t_s)}
    \label{eq:oscil}
\end{equation}
where even $r$ and $s$ label excitations of desired parity, and odd $r$ and $s$ label excitations of 
opposite parity.
The appearance of the opposite-parity states and their oscillating time dependence are consequences of 
using staggered fermions for the spectator quark.
The $A_{rs}$ are transition matrix elements, multiplied by uninteresting factors.
For the desired $A_{00}$, these factors cancel in $R(t_s,t_f)$.

In practice, we can choose the time separations such that only the lowest-lying states of each parity make a 
significant contribution.
As discussed in detail in Ref.~\cite{Bernard:2008dn}, it is advantageous to smear over time in a way that 
suppresses the opposite-parity state, and define
\begin{equation}
    \bar{R}(t_s,t_f) \equiv \frac{1}{2}R(t_s,t_f) +
        \frac{1}{4}R(t_s,t_f+1) + \frac{1}{4}R(t_s+1,t_f+1),
    \label{eq:avg} 
\end{equation}
which is very close to $R_{A_1}$, with small time-dependent effects that one can disentangle via a fit to the
$t_s$ dependence.

The average in Eq.~(\ref{eq:avg}) is designed to suppress the contribution from oscillating states that
changes sign only when the total source-sink separation is varied (the ``same-sign'' oscillating-state
contributions).
The double ratio, including the leading effects of the wrong-parity states, is
\bea\label{eq:Ra1avg} 
    \overline{R}_{A_1}(t_s,t_f) &=& 
        \frac{A^{B\to D^*}_{00}A^{D^*\to B}_{00}}{A^{D^*\to D^*}_{00}A^{B\to B}_{00}}
        \left[1+\overline{c}^{B\to D^*}(t_s,t_f)+\overline{c}^{D^*\to B}(t_s,t_f)
    \right. \nonumber \\ & & \hspace*{7.5em} - \left. 
        \overline{c}^{D^*\to D^*}(t_s,t_f) - \overline{c}^{B\to B}(t_s,t_f)+...\right], 
\eea
where the function $\overline{c}^{X\to Y}$ contains the oscillating-state contributions, and is
given by
\bea\label{eq:average} \overline{c}^{X\to Y}(t_s,t_f) &\equiv &
\frac{A^{X\to Y}_{01}}{A^{X\to Y}_{00}}(-1)^{t_f-t_s}e^{-\Delta
m_Y(t_f-t_s)}\left[\frac{1}{2}+\frac{1}{4}(1-e^{-\Delta m_Y})\right]
\nonumber
\\ && + \frac{A^{X\to Y}_{10}}{A^{X\to Y}_{00}}(-1)^{t_s} e^{-\Delta m_X
t_s}\left[\frac{1}{2}+\frac{1}{4}(1-e^{-\Delta m_X})\right]
\nonumber
\\ && + \frac{A^{X\to Y}_{11}}{A^{X\to Y}_{00}}(-1)^{t_f} e^{-\Delta
m_X t_s -\Delta m_Y(t_f-t_s)}\left[\frac{1}{2}-\frac{1}{4}(e^{-\Delta
m_Y}+e^{-\Delta m_X})\right]. \eea
The terms in square brackets in Eq.~(\ref{eq:average}) are the suppression factors for the oscillating state
contributions.
The $\Delta m_{X,Y}$  are the splittings between the ground-state masses and the opposite-parity masses, and their values can be computed precisely from fits to two-point correlators.  We find values
for these splittings in the range between about 0.1 and 0.4 in lattice units.
With these values of the parameters the ``same-sign'' contributions [the third term in
Eq.~(\ref{eq:average})] are suppressed by a factor of $\sim 6$--20 by Eq.~(\ref{eq:avg}), where the suppression is greater at finer lattice spacings.
The other oscillating-state contributions change sign as a function of $t_s$ and are given by the first two
terms in Eq.~(\ref{eq:average}).
These contributions are very small for our double ratio, and they are further suppressed by a factor of
$\sim$2 by the average in Eq.~(\ref{eq:avg}).

\begin{figure}[b]
\centering
    \includegraphics[width=0.48\textwidth]{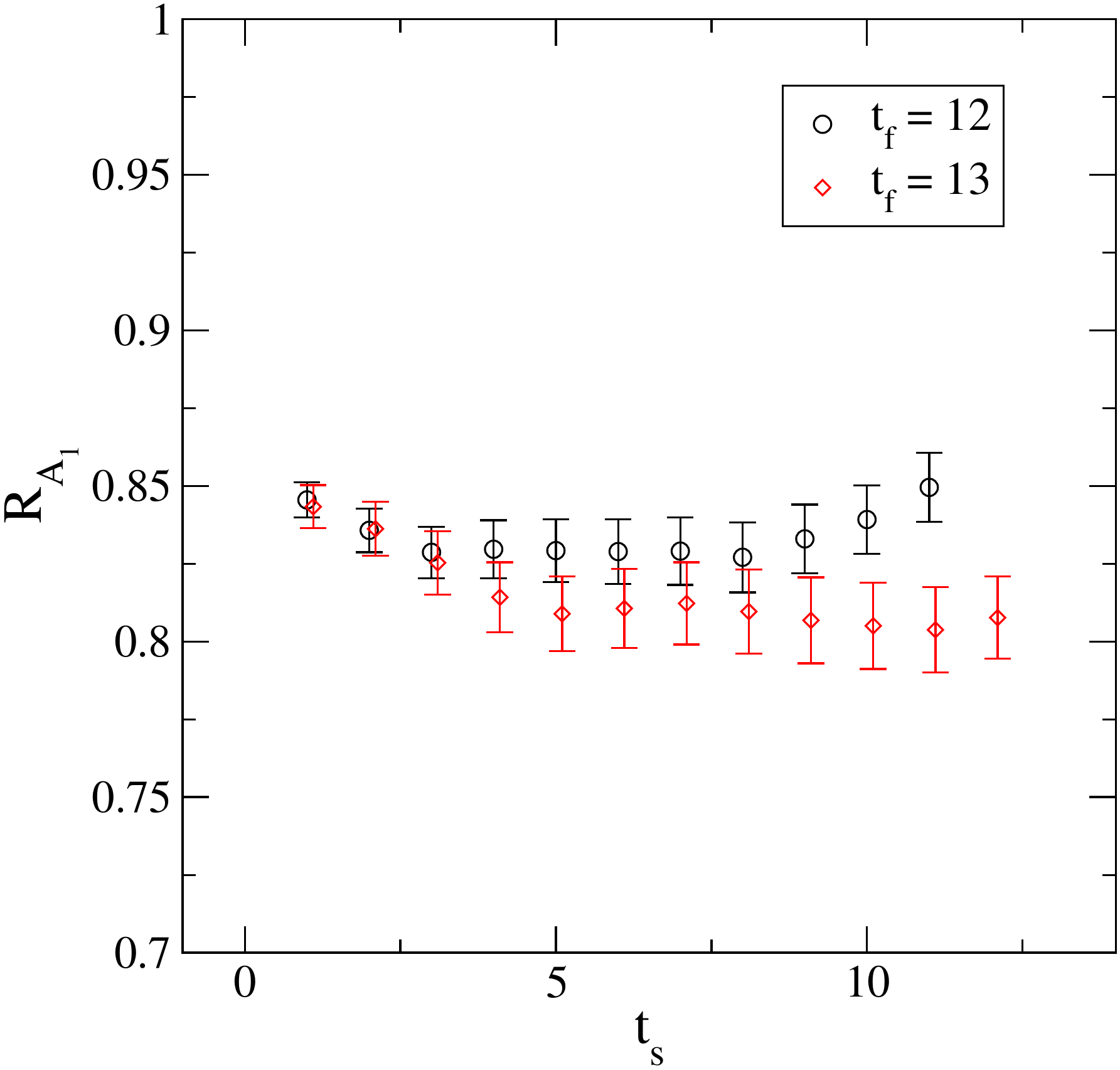} \hfill
    \includegraphics[width=0.48\textwidth]{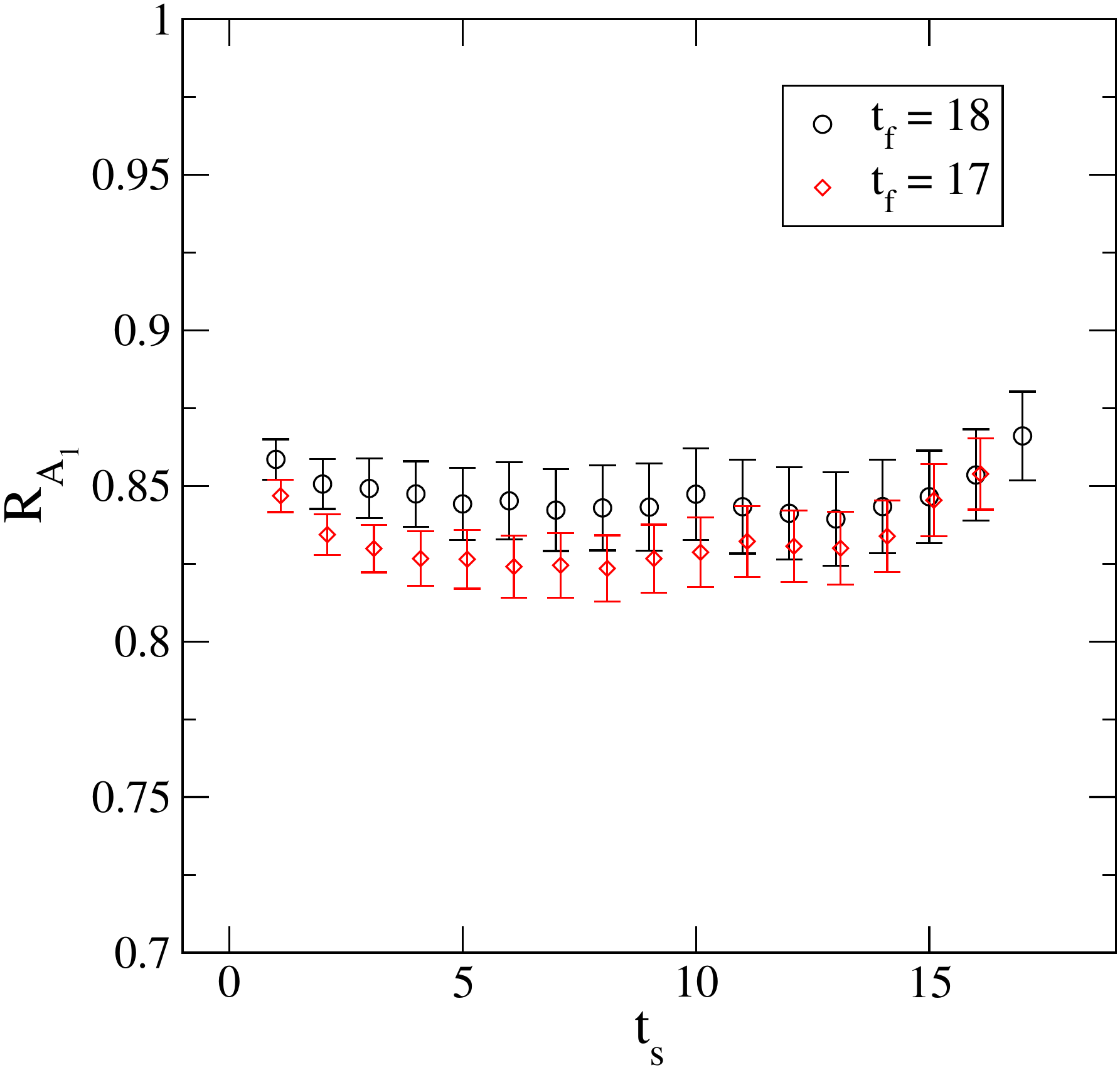}
    \caption{$R_{A_1}$ at $m_x=0.2\seaheavy$ on 0.12~fm (left) and on 0.09~fm (right) lattice 
        spacings.}
    \label{fig:oscillate}
\end{figure}

\begin{figure}[b]
\centering
    \includegraphics[width=0.48\textwidth]{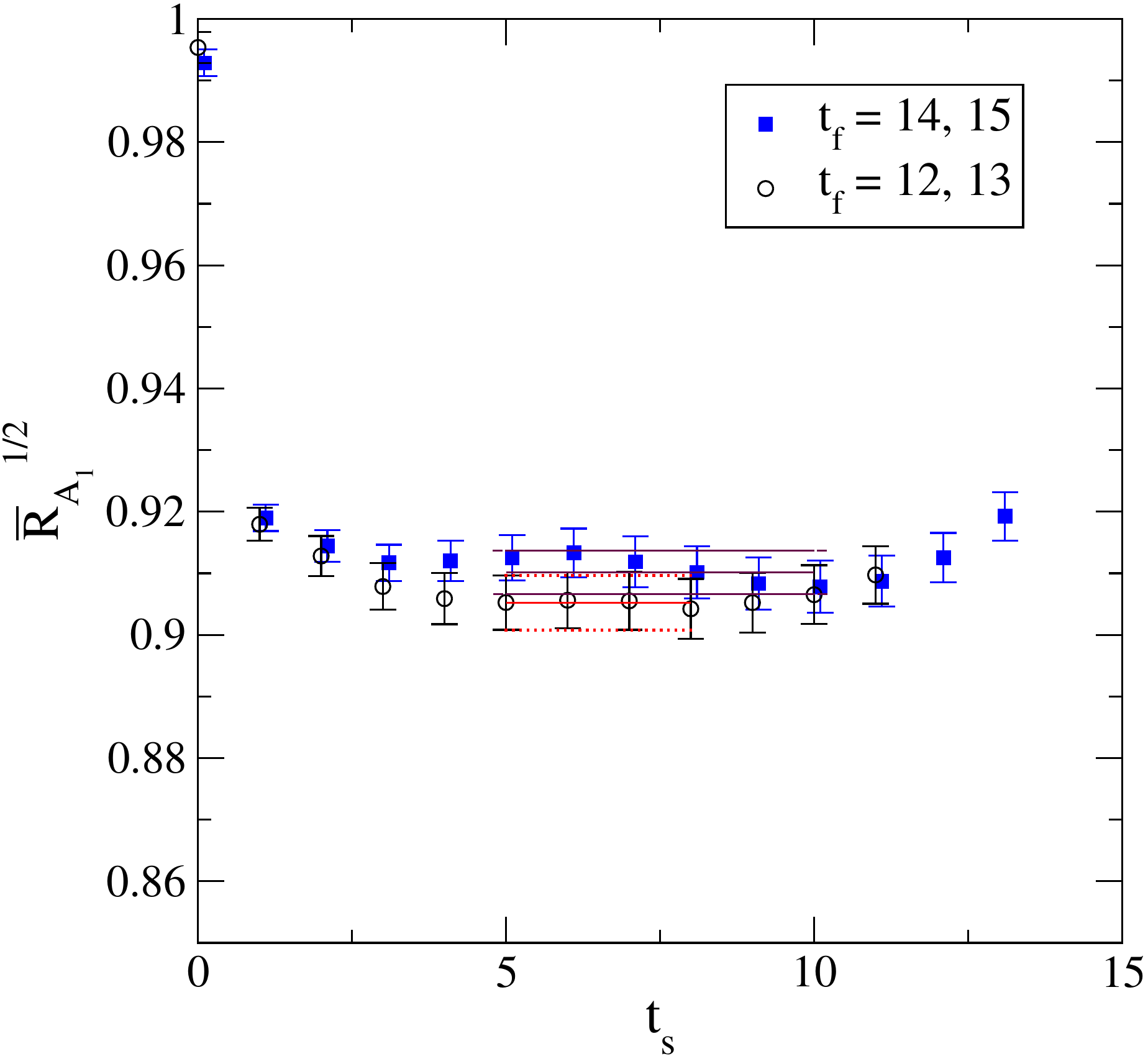} \hfill
    \caption{$R_{A_1}^{1/2}$ at $m_x=0.2\seaheavy$ on 0.12~fm for two different combinations of source-sink separations.}
    \label{fig:compare}
\end{figure}

Figure~\ref{fig:oscillate} shows $R_{A_1}(t_s, t_f)$ and $R_{A_1}(t_s, t_f+1)$ for two different, representative
ensembles.
One can see that the plateau is lower for odd total source-sink separation than for even total
source-sink separation, whether the odd source-sink separation is larger or smaller than the even
source-sink separation.
This feature holds for all ensembles.
It suggests that the ``same-sign'' oscillating states are visible in our data, and are comparable to, but
somewhat larger than, our current statistical errors.
The average of Eq.~(\ref{eq:avg}) suppresses this effect to around $0.1\%$ on our coarser lattices and around $0.03\%$ on our finest lattices.  This effect is negligible compared to other errors.
The fact that this effect is visible independently of whether the odd source-sink separation is larger or
smaller than the even source-sink separation indicates that this effect is larger than other excited-state
contributions and that within the current statistical precision of our data, these can also be neglected.  That this is the case is verified by a calculation at a larger source-sink separation on the 0.12~fm $0.2m_s$ ensemble.  Figure~\ref{fig:compare} shows a comparison between the square-root of the average Eq.~(\ref{eq:avg}) for two different combinations of source-sink separations.  The larger source-sink separation is computed with 16 time sources on 2256 configurations, compared with four time sources on the same configurations for the smaller separation.  The source was moved around the lattice randomly with a different seed for the two calculations, so we expect the ratios to be less correlated than is typical for quantities computed on the same configurations. The agreement between the best fits to the different source-sink separations is good to the 1 $\sigma$ level, as expected if residual excited state contamination is small.  Since ordinary excited state contamination would tend to cause the plateau fit to be too high, as can be seen by the higher values of $t_s$ near the source and sink, this contamination must be negligible within our current statistics because the fit with the larger separation and smaller contamination gives a slightly higher plateau value.

\begin{figure}[b]
\centering
    \includegraphics[width=0.48\textwidth]{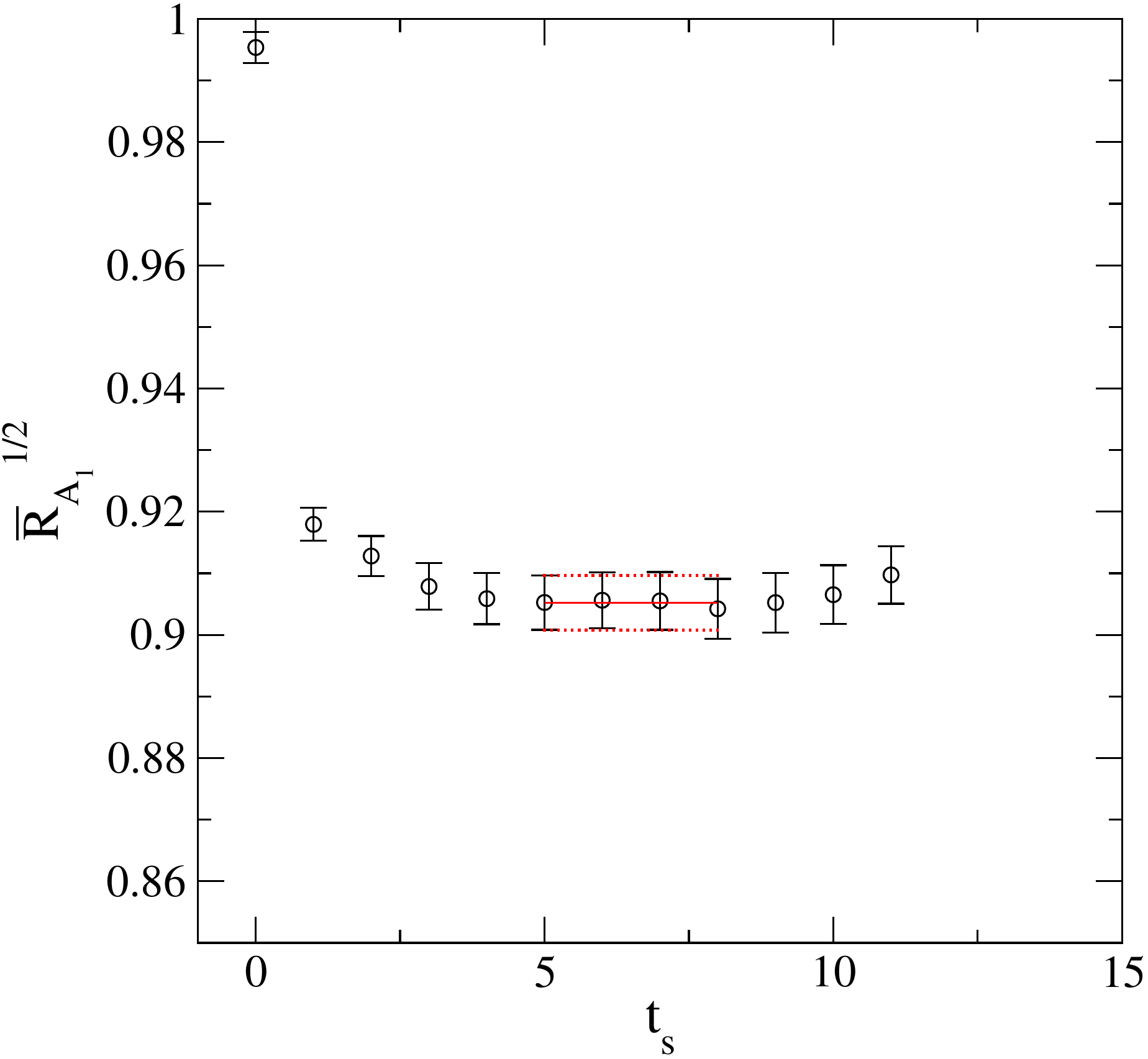} \hfill
    \includegraphics[width=0.48\textwidth]{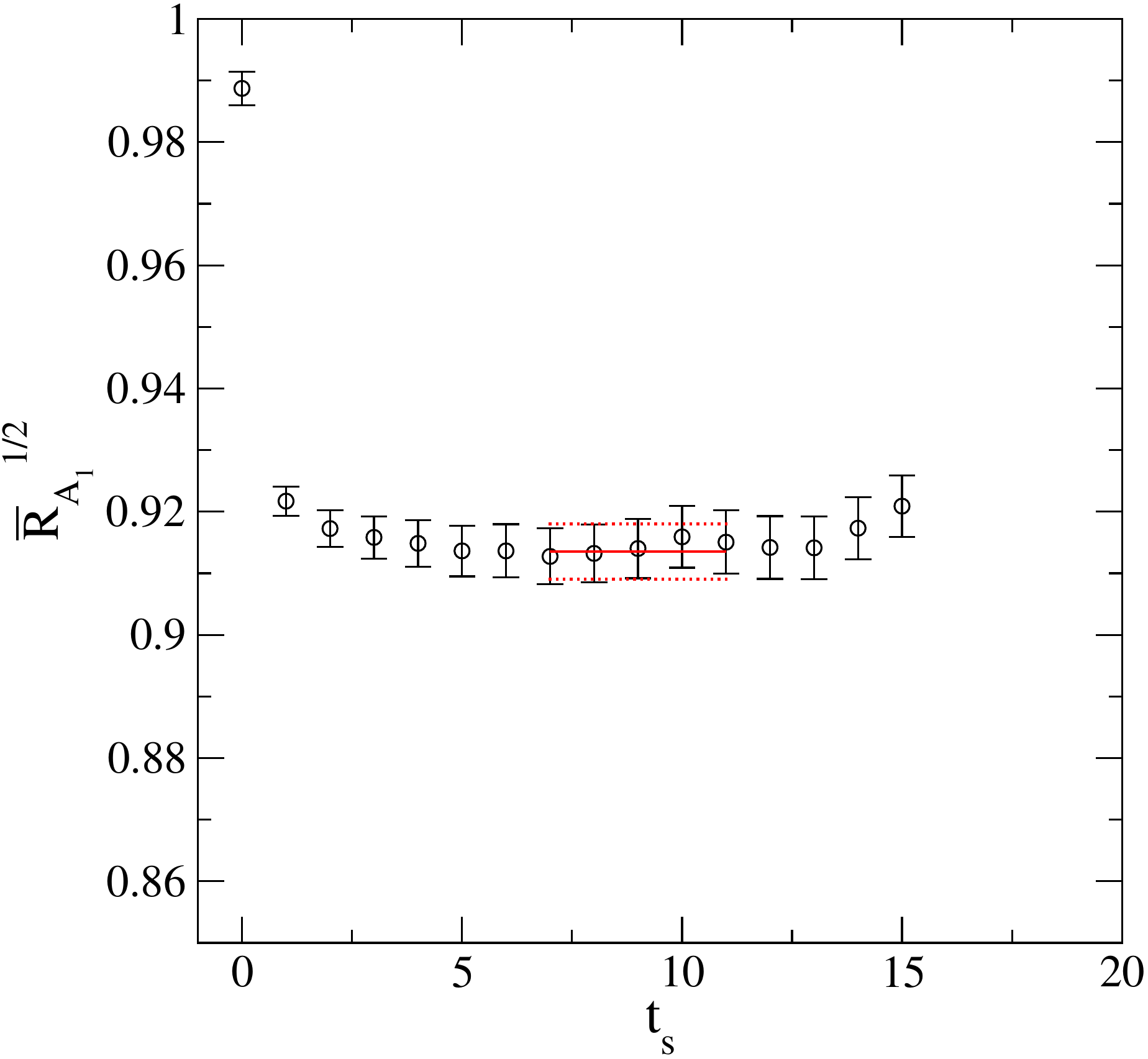} \\
    \includegraphics[width=0.48\textwidth]{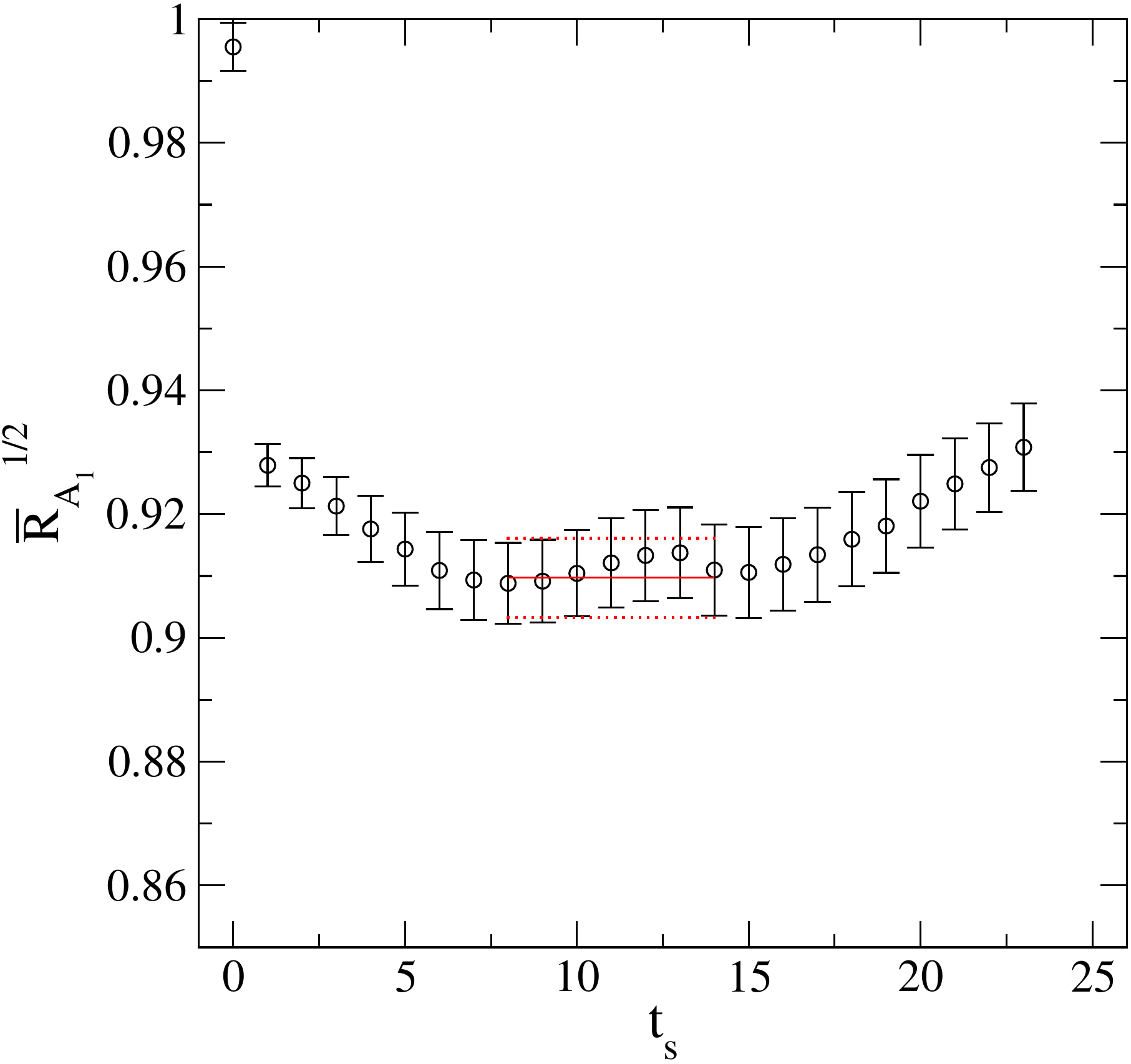}
    \caption{$\overline{R}_{A_1}^{1/2}$ at $m_x=0.2\seaheavy$ on 0.12~fm (left), 
        0.09~fm (right), and 0.06~fm (bottom).
       The plateau fits are shown with 1$\sigma$ error bands.}
    \label{fig:plateau}
\end{figure}

The square-root of the average Eq.~(\ref{eq:avg}) is shown in Fig.~\ref{fig:plateau} for 0.12~fm,
0.09~fm, and 0.06~fm lattice spacings.
These plots show data at unitary (full QCD) points, with valence spectator- and light sea-quark masses equal
to $0.2\seaheavy$.
The square-root of $\overline{R}_{A_1}(t_s,t_f)$ is fit to a constant in the identified plateau region, including
the full covariance matrix to determine the correlated $\chi^2$ and to ensure that the fits yielded
acceptable $p$~values.
The fits are shown in Fig.~\ref{fig:plateau} superimposed over the data with $1\sigma$ error bands.
Source-sink separations and plateau ranges are approximately the same in physical units for all lattice
spacings.
Time ranges for fits, their $p$~values, and the raw values for $h_{A_1}(1)$ are given in
Table~\ref{tab:fit_results}.

\begin{table}
\centering
    \caption{Fit results for double ratios at the full QCD points.
        The (approximate) lattice spacings~$a$ and the sea-quark masses $a\sealight/a\seaheavy$ (first two columns) identify the ensemble.
        The third column is the pair of spectator quark source-sink separations, the fourth is the time-slice fit range, the fifth is the $p$~value of the fit, and the sixth is the
        value of $h_{A_1}(1)/\rho_{A^j}$ determined from the fit.}
    \label{tab:fit_results}
    \vspace*{4pt}
    \begin{tabular}{c@{\quad}r@{/}l@{\quad}l@{\quad}ccc}
      \hline \hline
       $a$ (fm) & $a\sealight$&$a\seaheavy$ & \ \ $t_f$ & fit range & ~$p$~value & $h_{A_1}(1)/\rho_{A^j}$   \\
      \hline
      $0.15\hs$ & $0.0097$&$0.0484$  & 10, 11 & \ 5-7  &  0.85     &  0.9141(51)    \\
      \hline
      $0.12\hs$ & $0.02$&$0.05$   &  12, 13  &  \ 5-8     &  0.80   & 0.9035(28) \\
      $0.12\hs$ & $0.01$&$0.05$    &  12, 13 &  \ 5-8  &   0.97  &  0.9052(44) \\
      $0.12\hs$ & $0.007$&$0.05$   &  12, 13  &  \ 5-8    &  0.63   & 0.9160(53)   \\
      $0.12\hs$ & $0.005$&$0.05$   & 12, 13  & \ 5-8    &   0.68   & 0.9143(55)    \\
      \hline
      $0.09\hs$ & $0.0124$&$0.031$  & 17, 18  & \ 7-11         &   0.63   & 0.9162(31)     \\
      $0.09\hs$ & $0.0062$&$0.031$  & 17, 18 & \ 7-11       &   0.54   & 0.9135(45)  \\
      $0.09\hs$ & $0.00465$&$0.031$ & 17, 18 & \ 7-11        &   0.78   & 0.9212(73)    \\
      $0.09\hs$ & $0.0031$&$0.031$  & 17, 18 &  \ 7-11 &   0.95  & 0.9092(68) \\
      $0.09\hs$ & $0.00155$&$0.031$ & 17, 18 & \ 7-11        &   0.79   & 0.9208(90)   \\
      \hline
      $0.06\hs$ & $0.0072$&$0.018$  & 24, 25  & \ 8-14        &   0.84   & 0.9126(50)  \\
      $0.06\hs$ & $0.0036$&$0.018$  & 24, 25 & \ 8-14 &   0.93   & 0.9097(64)  \\
      $0.06\hs$ & $0.0025$&$0.018$  & 24, 25 & \ 8-14    &   0.13   & 0.9073(67)   \\
      $0.06\hs$ & $0.0018$&$0.018$  & 24, 25 & \ 8-14     &   0.55   & 0.9147(64)  \\
      \hline
      $0.045$   & $0.0028$&$0.014$  & 32, 33 & \ 7-14 & 0.87 & 0.9029(45)   \\
      \hline \hline
    \end{tabular}
\end{table}

\section{Perturbation theory for \boldmath$\rhoA{}$}
\label{sec:PT}

As discussed in Sec.~\ref{sec:corrs}, we need the ratio of matching factors, $\rhoA{j}$, defined in
Eq.~(\ref{eq:rho_ratio}).
This ratio has been calculated in one-loop perturbation theory, which will be discussed in detail in another
publication.
The perturbative expansion for $\rhoA{j}$ is
\begin{equation}
    \rhoA{j} = 1 +  \sum_{\ell} \rhoA{j}^{[\ell]} \alpha_V^{\ell}(q^*),
\end{equation}
where we make explicit a choice of scheme and scale for the perturbative series.
The calculation of $\rhoA{j}^{[1]}$ is a straightforward extension of the work in  
Ref.~\cite{Harada:2001fj}, modified to use the improved gluon propagator.

For the expansion parameter $\alpha_V(q^*)$, we would like to make a choice that prevents large logarithms
associated with the $\beta$ function from making the neglected terms unnecessarily large.
Brodsky, Lepage, and Mackenzie~\cite{Brodsky:1982gc} discussed how to do so by exploiting the $n_f$
dependence of the second order in $\alpha_V$, and Lepage and Mackenzie~\cite{Lepage:1992xa} explained how to
define an equivalent scale choice when the second order is not yet available.
The Lepage-Mackenzie version requires a coefficient ${}^*\rhoA{j}^{[1]}$ defined by weighting the Feynman
integral for $\rhoA{j}^{[1]}$ with an additional factor of~$\ln(q^2a^2)$, where $q$ is the gluon momentum in
the one-loop diagram(s).
Then the recommended (and empirically successful~\cite{Lepage:1992xa,Harada:2002jh}) scale~$q^*$ is given
through
\begin{equation}
    \ln(q^*a) = \frac{{}^*\rhoA{j}^{[1]}}{2\rhoA{j}^{[1]}},
    \label{eq:BLM}
\end{equation}
when the scheme is the $V$~scheme, such that the interquark potential in momentum space is 
$C_F\alpha_V(q^2)/q^2$.

Unfortunately, as the heavy-quark masses vary over the range of interest, nearby zeroes of the numerator
and denominator in Eq.~(\ref{eq:BLM}) lead to physically unreasonable values for~$q^*$.
Fortunately, the way to deal with such cases has been spelled out by Hornbostel, Lepage, and Morningstar 
(HLM)~\cite{Hornbostel:2002af}.
The HLM method requires integrals weighted by higher powers of~$\ln(q^2a^2)$.
This prescription results in values for $q_{\text{HLM}}^*$ that are close to $2/a$.
We therefore use $q^*=2/a$ to obtain the $\rhoA{j}$ listed in Table~\ref{tab:rho}.
\begin{table}
\centering
    \caption{One-loop estimate of $\rhoA{j}$.
        The first two columns label each ensemble with the approximate lattice spacing in fm and the sea
        simulation light- and strange-quark masses.
        The third column is $\alpha_V(q^*)$ with $q^*=2/a$.
        The fourth column is $\rhoA{j}$ on that ensemble with statistical errors from 
        the VEGAS evaluation of the one-loop coefficients.}
    \label{tab:rho}
    \begin{tabular}{l@{\quad}r@{/}l@{\quad}l@{\quad}l}
    \hline \hline
        $a$ (fm) & $a\sealight$ & $a\seaheavy$ & ~~$\alpha_V(q^*)$ &  ~~~$\rhoA{j}$ \\
    \hline
        $0.15$ & 0.0097  & 0.0484 & 0.3589 & 0.99422(4) \\
    \hline
        $0.12$ & 0.02    & 0.05   & 0.3047 & 0.99650(5) \\
        $0.12$ & 0.01    & 0.05   & 0.3108 & 0.99623(5) \\
        $0.12$ & 0.007   & 0.05   & 0.3102 & 0.99618(5) \\
        $0.12$ & 0.005   & 0.05   & 0.3102 & 0.99617(5) \\
    \hline
        $0.09$ & 0.0124  & 0.031  & 0.2582 & 0.99978(4) \\
        $0.09$ & 0.0062  & 0.031  & 0.2607 & 0.99963(4) \\
        $0.09$ & 0.00465 & 0.031  & 0.2611 & 0.99957(4) \\
        $0.09$ & 0.0031  & 0.031  & 0.2619 & 0.99950(4) \\
        $0.09$ & 0.00155 & 0.031  & 0.2623 & 0.99946(4) \\
    \hline
        $0.06$ & 0.0072  & 0.018  & 0.2238 & 1.00334(3) \\
        $0.06$ & 0.0036  & 0.018  & 0.2245 & 1.00323(3) \\
        $0.06$ & 0.0025  & 0.018  & 0.2249 & 1.00317(3) \\
        $0.06$ & 0.0018  & 0.018  & 0.2253 & 1.00312(3) \\
    \hline
        $0.045$ & 0.0028 & 0.014  & 0.2013 & 1.00608(2) \\
    \hline\hline
    \end{tabular}
\end{table}
As expected, $\rhoA{j}$ varies somewhat as a function of lattice spacing.
It is even slightly different from ensemble to ensemble at the same nominal lattice spacing, because these
ensembles have slightly different lattice spacings.

\section{Heavy-quark mass tuning}
\label{sec:kappa}

Our approach to tuning $\kappa_{b,c}$ is similar to that described in Ref.~\cite{Bernard:2010fr}, and a
detailed description of the current approach is given in Appendix~\ref{app:kappa-tuning}.
We start with the lattice dispersion relation
\begin{equation}
    E^2(\bm{p}) = M_1^2 + \frac{M_1}{M_2}\bm{p}^2 + \frac{1}{4}A_4 (a\bm{p}^2)^2 + 
        \frac{1}{3}A_{4'}a^2\sum_{j=1}^{3}|p_j|^4 + \cdots,
    \label{eq:dis}
\end{equation}
where $M_1\equiv E(\bm{0})$ defines the meson rest mass and the kinetic mass is given by
\begin{equation}
    M_2^{-1} \equiv 2\left.\frac{\partial E(\bm{p})}{\partial p_j^2}\right|_{\bm{p}=\bm{0}}.
\end{equation}
The meson masses differ from the corresponding quark masses, $m_1$ and $m_2$, by binding-energy effects.
In the Fermilab method, the lattice pole energy is fit to the dispersion relation Eq.~(\ref{eq:dis}), and
$\kappa$ is adjusted so that the kinetic mass agrees with experiment.
We tune to the experimental $D_s$ and $B_s$ meson masses to obtain $\kappa_c$ and $\kappa_b$, respectively.

The simulation values $\kappa'_{b,c}$ differ from our current best estimates of these parameters because of
improvements in statistics and methodology since the initial tuning runs.
Table~\ref{tab:kappas} shows our best estimates of $\kappa_{b,c}$, along with errors.
The first error is a combination of statistical and fitting systematics, and the second error is that due to
fixing the lattice scale.
For comparison, Table~\ref{tab:kappas} also shows the $\kappa'_{b,c}$ values used in the runs.
\begin{table}
    \centering
    \caption{Errors in the tuned $\kappa_{b,c}$ parameters.
        The (approximate) lattice spacings~$a$ and the sea-quark masses $a\sealight/a\seaheavy$ (first two columns) identify the 
        ensemble.
        The third and fourth columns are the tuned $\kappa$ values for the $b$ and $c$ quarks, respectively.
        The first error is the statistics plus fitting error, and the second is an error due to the 
        uncertainty in the lattice scale.
        The fifth and six columns are the $\kappa$ values used in the simulations.}
    \label{tab:kappas}
    \vspace*{4pt}
    \begin{tabular}{c@{\quad}r@{/}l@{\quad}lllll}
      \hline \hline
       $a$ (fm) & $a\sealight$&$a\seaheavy$ & \multicolumn{1}{c}{$\kappa_b$} & 
           \multicolumn{1}{c}{$\kappa_c$} & \multicolumn{1}{c}{$\kappa'_b$} & 
           \multicolumn{1}{c}{$\kappa'_c$} \\
      \hline
      $0.15\hs$ & $0.0097$&$0.0484$  & 0.0775(16)(3)  &  0.12237(26)(20)  & 0.0781 & 0.1218  \\
      \hline
      $0.12\hs$ & $0.02$&$0.05$      & 0.0879(9)(3)   &  0.12452(15)(16)  & 0.0918 & 0.1259  \\
      $0.12\hs$ & $0.01$&$0.05$      & 0.0868(9)(3)   &  0.12423(15)(16)  & 0.0901 & 0.1254  \\
      $0.12\hs$ & $0.007$&$0.05$     & 0.0868(9)(3)   &  0.12423(15)(16)  & 0.0901 & 0.1254  \\
      $0.12\hs$ & $0.005$&$0.05$     & 0.0868(9)(3)   &  0.12423(15)(16)  & 0.0901 & 0.1254  \\
      \hline
      $0.09\hs$ & $0.0124$&$0.031$   & 0.0972(7)(3)   &  0.12737(9)(14)   & 0.0982 & 0.1277  \\
      $0.09\hs$ & $0.0062$&$0.031$   & 0.0967(7)(3)   &  0.12722(9)(14)   & 0.0979 & 0.1276  \\
      $0.09\hs$ & $0.00465$&$0.031$  & 0.0966(7)(3)   &  0.12718(9)(14)   & 0.0977 & 0.1275  \\
      $0.09\hs$ & $0.0031$&$0.031$   & 0.0965(7)(3)   &  0.12714(9)(14)   & 0.0976 & 0.1275  \\
      $0.09\hs$ & $0.00155$&$0.031$  & 0.0964(7)(3)   &  0.12710(9)(14)   & 0.0976 & 0.1275  \\
      \hline
      $0.06\hs$ & $0.0072$&$0.018$   & 0.1054(5)(2)   &  0.12964(4)(11)   & 0.1048 & 0.1295  \\
      $0.06\hs$ & $0.0036$&$0.018$   & 0.1052(5)(2)   &  0.12960(4)(11)   & 0.1052 & 0.1296  \\
      $0.06\hs$ & $0.0025$&$0.018$   & 0.1051(5)(2)   &  0.12957(4)(11)   & 0.1052 & 0.1296  \\
      $0.06\hs$ & $0.0018$&$0.018$   & 0.1050(5)(2)   &  0.12955(4)(11)   & 0.1052 & 0.1296  \\
      \hline
      $0.045$   & $0.0028$&$0.014$   & 0.1116(3)(2)   &  0.130921(16)(7)  & 0.1143 & 0.1310 \\
      \hline \hline
    \end{tabular}
\end{table}
A detailed discussion of how the tuned values of $\kappa_{b,c}$ are obtained is given in
Appendix~\ref{app:kappa-tuning}.
As a cross-check of our tuning procedure, we calculate the hyperfine splittings
$\Delta M(D_s)= M(D_s^*)-M(D_s)$ and $\Delta M(B_s)=M(B_s^*)-M(B_s)$.
In Appendix~\ref{app:hfs} we find 
\begin{equation}
    \Delta M(D_s) = 146 \pm 4~\text{MeV},\quad
    \Delta M(B_s) =  44 \pm 3~\text{MeV}, 
    \label{eq:HFS}
\end{equation}
where the error includes statistics and the sum of all systematic errors in quadrature.
These are in good agreement with the experimental values $\Delta M(D_s)=143.8\pm0.4$~MeV and 
$\Delta M(B_s)=48.7^{+2.3}_{-2.1}$~MeV.

\begin{figure}[b]
    \centering
    \includegraphics[width=0.48\textwidth]{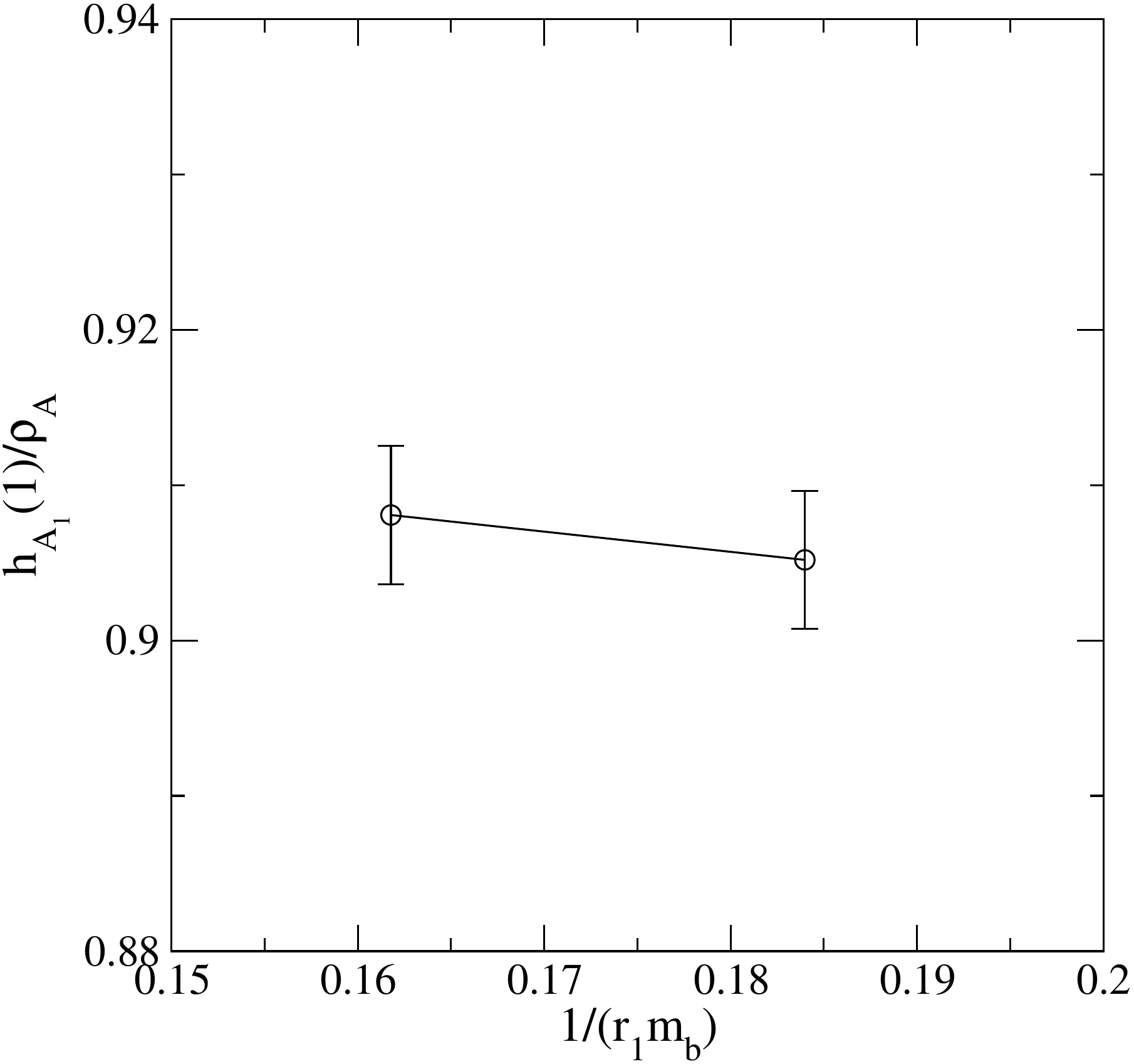} \hfill
    \includegraphics[width=0.48\textwidth]{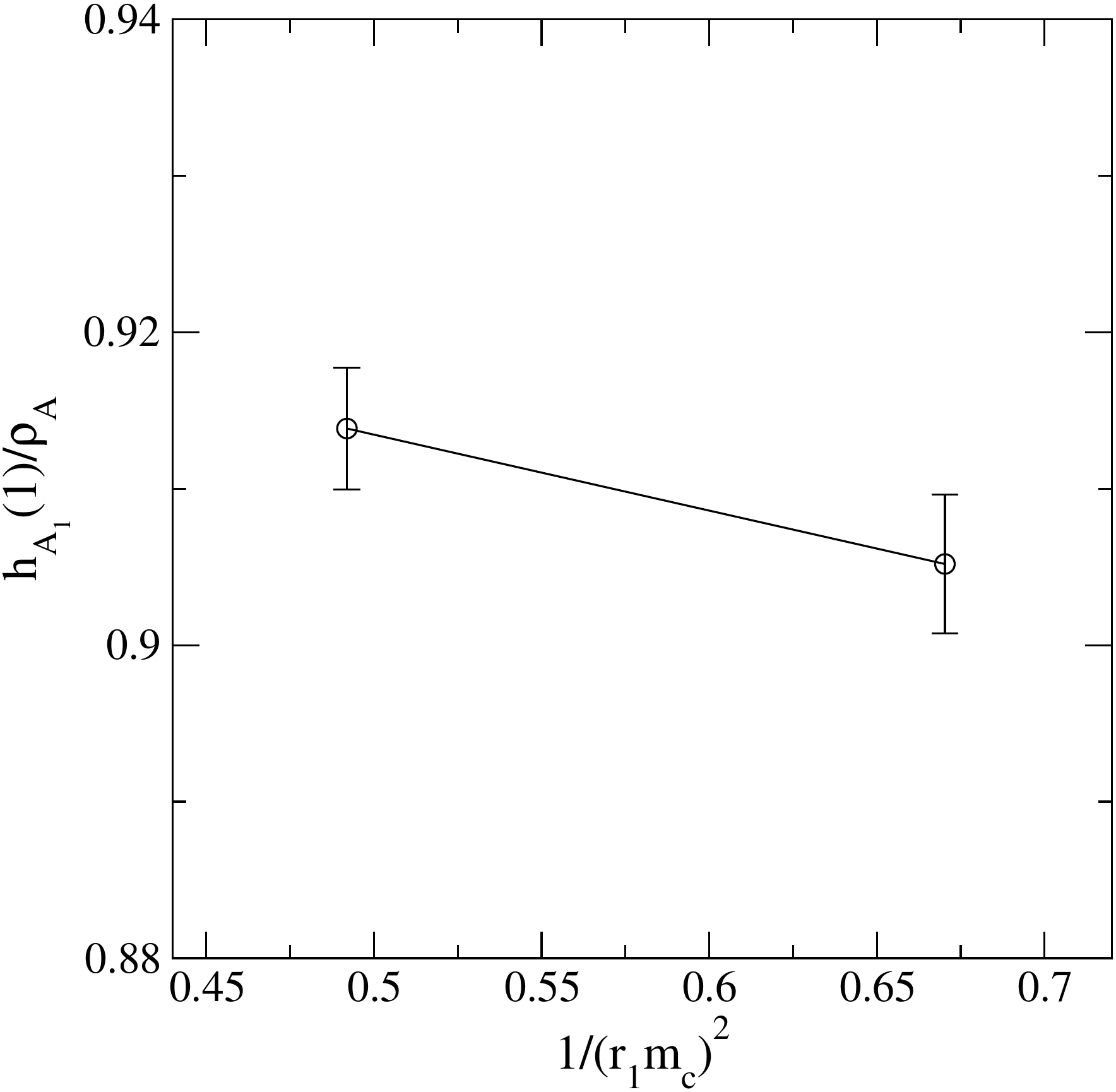}
    \caption[huh]{$h_{A_1}(1)$ at different values close to the tuned $b$ and $c$ quark masses.
    Each is plotted as a function of the leading (assuming $m_b$ is sufficiently heavier than $m_c$) 
    heavy-quark mass dependence in Eq.~(\ref{eq:HQET}), $1/(r_1m_b)$ and $1/(r_1m_c)^2$ for $b$ and $c$, 
    respectively.}
    \label{fig:kappa}
\end{figure}

We correct our values of $h_{A_1}(1)$ for the mistuning of $\kappa$ using information on the heavy-quark 
mass dependence from an additional run with $\kappa'_{b,c}$ nearer their physical values on the coarse ensemble 
with $a\sealight/a\seaheavy=0.01/0.05$.
To apply the correction we exploit information from heavy-quark effective theory (HQET); the form factor at
zero-recoil has the heavy-quark expansion \cite{Falk:1992wt, Mannel:1994kv}
\bea\label{eq:HQET}  
h_{A_1}(1) = \eta_A\left[1-\frac{\ell_V}{(2m_c)^2}+\frac{2\ell_A}{2m_c2m_b}-\frac{\ell_P}{(2m_b)^2}\right],
\eea
up to order $1/m_Q^2$, where $\eta_A$ is a factor that matches HQET to QCD and the $\ell$'s are long-distance
matrix elements of the HQET.
Heavy-quark symmetry forbids terms of order $1/m_Q$ at zero-recoil \cite{Luke:1990eg}.
The form factor depends on both the bottom quark mass and the charm quark mass; we correct for this
dependence and propagate the uncertainty due to the error in $\kappa_{b,c}$ to the form factor before
performing the chiral/continuum extrapolation.
The leading $m_b$ dependence is given by the term that is inversely proportional to $m_cm_b$ in brackets in
Eq.~(\ref{eq:HQET}), and this dependence, inversely proportional to $m_b$ for fixed charm-quark mass, is the
one used to correct the form factor for the mistuning in $m_b$.
The leading charm-quark mass dependence is, however, given by the term that is inversely proportional to the
charm quark mass squared.
Thus, we determine the adjustment that must be made from the simulated form factor $h_\text{sim}$ to the tuned
value $h_\text{tuned}$ using
\begin{eqnarray}
    h_\text{tuned} &=& h_\text{sim} + 
        \frac{\partial h}{\partial[1/(r_1 m_b)]}
            \left[\frac{1}{r_1 m_{b,\text{tuned}}} - \frac{1}{r_1 m_{b,\text{sim}}}\right] \nonumber \\
    & & {} + \frac{\partial h}{\partial[1/(r_1 m_c)^2]}
        \left[\frac{1}{(r_1 m_{c,\text{tuned}})^2} - \frac{1}{(r_1 m_{c,\text{sim}})^2}\right],
    \label{eq:shift}
\end{eqnarray}
where $m_{b,c}$ is the kinetic $b$ or $c$ quark mass, and $r_1$ sets the relative lattice spacing on
different ensembles.
The slope parameters are determined by a linear interpolation between the two sets of points shown in
Fig.~\ref{fig:kappa}.
One of these points in each of these plots is from our original production run, while the other points are
from runs where $\kappa_{b,c}$ were separately varied and chosen to be closer to their tuned values.

The slopes are also used to propagate the errors in the tuned kappa values due to ``statistics and fitting"
to the errors in each individual $h_{A_1}(1)$ data point before performing the chiral/continuum extrapolation.
This is done by inflating the jackknife error of $h_{A_1}(1)$ on each data point by adding to it in quadrature the parametric error in $h_{A_1}(1)$ due to the ``statistics and fitting" part of the
$\kappa$ tuning error.
We make the assumption that the statistics and fitting errors in the tuned $\kappa$ values on different
ensembles are independent of one another, though we also test the size of the additional error induced if
this assumption is not true and find that it is small.
The $\kappa$ tuning ``statistics and fitting" error is thus directly incorporated into the statistical error
of $h_{A_1}$.
The scale error in the tuned $\kappa$ values, however, is $100\%$ correlated across ensembles, and is therefore
treated as a separate systematic error.

\section{Chiral-continuum extrapolation}
\label{sec:extrap}

Because the light $u$ and $d$-quark masses used in the calculation are heavier than the physical ones, an
extrapolation in quark mass is necessary.
This extrapolation can be controlled using an appropriate chiral effective theory, where one can also
incorporate discretization effects particular to staggered quarks.
The chiral effective theory that incorporates these effects is rooted staggered chiral perturbation theory
(rS$\chi$PT), which was extended to include heavy-light quantities in Ref.~\cite{Aubin:2005aq}.

There are discretization effects that are particular to staggered quark actions.
The staggered quark discretization only partially solves the fermion doubling problem, reducing the number of
species from 16 to 4.
There remain unphysical species of quarks, commonly referred to as tastes.
Quarks of different tastes can exchange high momentum gluons with momenta of order the lattice cutoff, and
this exchange breaks the degeneracy in the pion spectrum for pions made of quarks of different tastes.
This taste-symmetry breaking leads to the staggered theory having 16 light pseudoscalar mesons instead of~1.

The tree-level relation in the chiral theory between the pseudoscalar meson masses and the quark masses is
given by
\begin{equation}
    M_{xy,\xi}^2=B_0(m_x+m_y)+a^2\Delta_\xi,
    \label{eq:treelevel}
\end{equation}
where $\xi$ labels the meson taste, $m_x$ and $m_y$ are the staggered quark masses, $B_0$ is the continuum
low-energy constant, and $a^2\Delta_\xi$ are the splittings of the 16~tastes.
An additional SO(4) taste-symmetry, which is broken only at $\text{O}(a^4)$, leads to some degeneracy among
the 16 pions, such that the taste index $\xi$ runs over the multiplets $P$, $A$, $T$, $V$, $I$ with
degeneracies 1, 4, 6, 4,~1, respectively.
The splitting $a^2\Delta_P$ vanishes because of an exact nonsinglet lattice axial symmetry.

Eq.~(34) of Ref.~\cite{Laiho:2005ue} gives the result for $h_{A_1}(1)$ in partially-quenched $\chi$PT with
degenerate up and down quark masses (the 2+1 case) in the rooted staggered theory.
The result is
\begin{equation}
    \frac{h_{A_1}^{(B_x)PQ,2+1}(1)}{\eta_A} = 1 + \frac{X_{A}(\Lambda_\chi)}{m_c^2} + 
        \frac{\gDDp^2}{48\pi^2f^2}\times\logs_\text{1-loop}(\Lambda_\chi),
    \label{eq:schpt} 
\end{equation}
where the term $\logs_\text{1-loop}(\Lambda_\chi)$ stands for the one-loop staggered chiral logarithms, the
detailed expression for which is given in Appendix~\ref{app:chpt}.
$X_A(\Lambda_\chi)$ is a low-energy constant of the chiral effective theory, independent of the light-quark
mass, and its dependence on the chiral scale $\Lambda_\chi$ cancels that of the chiral logarithms.
The $X_A(\Lambda_\chi)$ term is suppressed by a factor of $1/m_c^2$ in the heavy-quark power counting.
The term $\eta_A$ is a factor that matches HQET to QCD, and contains perturbative-QCD logarithmic dependence
on the heavy-quark masses.
It is independent of the light-quark mass.
The coefficient of the chiral logarithm term contains $f$, the pion decay constant and $\gDDp$, the 
$D^*D\pi$ coupling in the chiral effective theory.
\begin{table}
    \centering
    \caption{Parameters used in the chiral extrapolation, including the staggered taste-splittings for the different taste mesons.
        The first column is the approximate lattice spacing, and the second through fifth columns are the
        taste-splittings for the taste scalar, axial-vector, tensor, and vector mesons, respectively.  The sixth column is the tree-level low energy constant appearing in Eq.~(\ref{eq:treelevel}).}
    \label{tab:stag}
    \begin{tabular}{llrrlrr|r}
    \hline\hline
    $a$ (fm) &  $r_1^2 a^2\Delta_I$  & $r_1^2 a^2\Delta_V$ &  $r_1^2 a^2\Delta_T$  & 
        $r_1^2 a^2\Delta_A$ & \ \ $r_1B_0$ \\
    \hline
       0.15 & 0.9851 & \ \ 0.7962   &  \ \ 0.6178  & \ \ 0.3915 & 6.761 \\
       0.12 & 0.6008 & 0.4803   & 0.3662 & \ \ 0.2270 & 6.832\\  
       0.09 & 0.2207 & 0.1593   & 0.1238 & \ \ 0.0747 & 6.639 \\  
       0.06 & 0.0704 & 0.0574   &  0.0430 & \ \ 0.0263 & 6.487\\  
       0.045 & 0.0278 &  0.0227   &  0.0170 & \ \ 0.0104 & 6.417 \\  
      \hline\hline
  \end{tabular}
\end{table}

\begin{table}
\centering
\caption{Values of physical quark masses and $r_1B_0$ with discretization errors removed in a mass independent scheme.  The masses are in units of the 0.09 fm lattice spacing with the 0.09 fm lattice value of the mass renormalization. The first column is the physical $s$ quark mass, the second is the average of the $u$ and $d$ quark masses, the third is the $u$ quark mass, and the fourth is the $d$ quark mass.  The fifth column is the value of the low energy constant $r_1B_0$ evaluated at the same scale within the same scheme and with discretization errors removed.}
\label{tab:masses}
\begin{tabular}{ccccc}
\hline \hline
$am_s \times 10^2$ & $a\hat{m}\times 10^3$ & $am_u\times 10^3$ & $am_d\times 10^3$ & $r_1B_0$ \\
\hline
2.65(8) & 0.965(33)  & 0.610(26) & 1.32(5) & 6.736 \\
\hline\hline
\end{tabular}
\end{table}
The one-loop logarithm term depends on the light valence- and sea-quark masses, including the taste-breaking
discretization effects from the light-quark sector.
The expression contains explicit dependence on the lattice spacing $a$, and requires as inputs the parameters
of the staggered chiral Lagrangian $\delta'_V$ and $\delta'_A$, which are determined from chiral fits to pion
masses and decay constants on the same ensembles.
The chiral formula for $h_{A_1}(1)$ also requires as input the taste-splittings $\Delta_\xi$, which are
obtained from separate spectrum calculations of the various taste mesons.
The values of the staggered taste-splittings are given in Table~\ref{tab:stag}.
We take the values of the hairpin parameters $\delta'_V$ and $\delta'_A$ on the $a\approx 0.12$ fm lattices
to be $r_1^2a^2\delta'_V=0.00$ and $r_1^2a^2\delta'_A=-0.28$.
Their values at other lattice spacings are determined by scaling these numbers by the ratio of the root-mean-square splitting at the target lattice spacing and at $a\approx 0.12$ fm.  We find that varying the staggered parameters within their uncertainties produces a negligible error in $h_{A_1}$, as further discussed in Section~\ref{sec:chpt}.
The continuum low-energy constant $\gDDp$ is taken as an input in our fits.
We take a value with an error that encompasses recent lattice-QCD calculations and the latest measurements of
the $D^*$ decay width (See Sec.~\ref{sec:chpt} for details).
The $D^*$-$D$ mass splitting $\Delta^{(c)}$ is well determined from experiment.
In summary, the only free parameter in the next-to-leading order (NLO) chiral formula is the constant
$X_A(\Lambda)$, which is determined by fits to our lattice data for the form factor $h_{A_1}(1)$.  

The errors in the light quark masses lead to negligible uncertainty in $h_{A_1}$; these masses are presented in Table~\ref{tab:masses} in the ``continuum," where the values have been extrapolated to the continuum, i.e. discretization errors have been removed.  The masses are in units of the 0.09 fm lattice spacing with the 0.09 fm lattice value of the mass renormalization in a mass independent scheme.  The value of $r_1B_0$ evaluated at the same scale within the same scheme and with discretization errors removed is also given in Table~\ref{tab:masses}.

Table~\ref{tab:ChPT} shows our results for the lattice form factor $h_{A_1}(1)$ for various light-quark masses
on the different ensembles.
We computed the form factor at the full QCD points on all of the ensembles, and on some of the ensembles we
included a partially quenched point with the spectator light-quark mass equal to $0.4 \seaheavy$ in order to
help constrain the fits.
Because these points have small statistical errors due to the heavier spectator-quark mass, they are
especially useful in constraining the lattice-spacing dependence.
Table~\ref{tab:ChPT} also presents the values of the pion mass corresponding to the light spectator-quark
mass for the full QCD points.
Both the pseudoscalar-taste pion mass and the root-mean-square pion mass are given.
Note that the RMS and Goldstone pion masses presented in Table~\ref{tab:ChPT} use the mass-independent
determination of $r_1/a$ to fix the relative lattice scale, and thus differ somewhat from an earlier set of
masses on the same ensembles appearing in supporting material of the Flavor Lattice Averaging Group
\cite{Colangelo:2010et}.
This earlier set of masses used mass-dependent $r_1/a$ values to set the relative scale.
As Table~\ref{tab:ChPT} shows, our lightest taste-Goldstone pion mass is 180~MeV, while the lightest
root-mean-squared (RMS) pion mass is 260~MeV.
Previous work on MILC ensembles \cite{Aubin:2004fs, Bazavov:2010hj} suggests that when masses in these ranges
are combined with staggered $\chi$PT then the systematic error from the resulting chiral/continuum
extrapolation can be estimated reliably.
Although staggered $\chi$PT allows us to remove the leading discretization effects from the light quarks, the
heavy-quark discretization effects are more complicated; see Appendix~\ref{app:hqetdisc} for details.

\begin{table}
\centering
    \caption{Results for $h_{A_1}(1)$ at various light-quark masses, including partially-quenched points.
        The (approximate) lattice spacings~$a$ and the sea-quark masses $a\sealight/a\seaheavy$ identify the 
        ensemble (first two columns).
        The third column labels the valence spectator-quark mass.
        The fourth and fifth columns are the approximate taste-Goldstone and root-mean-square pion masses 
        associated with the valence spectator mass (values are only given for the unitary points). 
        The sixth column is the value of $h_{A_1}(1)$ at that valence mass (corrected for $\kappa$ mistuning 
        and including the perturbative matching factor).
        The error on $h_{A_1}(1)$ is statistical only.}
    \label{tab:ChPT}
    \vspace*{4pt}
    \begin{tabular}{c@{\quad}r@{/}l@{\quad}lcccc}
      \hline \hline
       $a$ (fm) & $a\sealight$&$a\seaheavy$ & ~~$am_x$ & ~$M_{\pi,P}$(MeV) & ~$M_{\pi,RMS}$(MeV) & ~~~$h_{A_1}(1)$  \\
      \hline
      $0.15\hs$ & $0.0097$&$0.0484$   & 0.0097  &  340     &  590  &  \ 0.9077(52)    \\
       $0.15\hs$ & $0.0097$&$0.0484$   & 0.0194  &  -     &  -  &  \ 0.9085(35)    \\
      \hline
      $0.12\hs$ & $0.02$&$0.05$       &  0.02    &  560   & 670 & 0.9068(29)   \\
      $0.12\hs$ & $0.01$&$0.05$       & 0.01 &   390  &  540 & 0.9068(45)  \\
      $0.12\hs$ & $0.01$&$0.05$       & 0.02 &   -  &  - & 0.9068(30)  \\
      $0.12\hs$ & $0.007$&$0.05$      & 0.007   &  320   & 500 &  0.9175(53)  \\
      $0.12\hs$ & $0.007$&$0.05$      & 0.02   &  -   & - &  0.9131(28) \\
      $0.12\hs$ & $0.005$&$0.05$      &  0.005   &   270   & 470 & 0.9158(56)   \\
      $0.12\hs$ & $0.005$&$0.05$      & 0.02     &   -   & - & 0.9108(28)  \\
      \hline
      $0.09\hs$ & $0.0124$&$0.031$    &  0.0124    &   500   & 550  &  0.9180(32)    \\
      $0.09\hs$ & $0.0062$&$0.031$    & 0.0062  &   350   & 420  &  0.9155(46)    \\
       $0.09\hs$ & $0.0062$&$0.031$    & 0.0124  &   -   &  -  &  0.9147(31)  \\
      $0.09\hs$ & $0.00465$&$0.031$   &   0.00465     &   310   & 380  &  0.9227(73)    \\
      $0.09\hs$ & $0.0031$&$0.031$    & 0.0031   &   250  & 330 &  0.9108(69)  \\
      $0.09\hs$ & $0.0031$&$0.031$    & 0.0124   &   -   &  - &  0.9125(37)  \\
      $0.09\hs$ & $0.00155$&$0.031$   &  0.00155  &   180   & 280  &  0.9227(90)   \\
      \hline
      $0.06\hs$ & $0.0072$&$0.018$    &  0.0072      &   450   & 470  &  0.9142(51) \\
      $0.06\hs$ & $0.0036$&$0.018$    &  0.0036   &   320   & 340  & 0.9127(65)  \\
       $0.06\hs$ & $0.0036$&$0.018$    &  0.0072   &   -   &  -  & 0.9130(45)  \\
      $0.06\hs$ & $0.0025$&$0.018$    &  0.0025   &   260  & 290  &  0.9105(88) \\
      $0.06\hs$ & $0.0018$&$0.018$    &  0.0018    &   220   & 260  & 0.9182(65) \\
      \hline
      $0.045$   & $0.0028$&$0.014$    & 0.0028   & 320 & 330  & 0.9121(46) \\
      \hline \hline
    \end{tabular}
\end{table}

If we restrict ourselves to a strictly NLO $\chi$PT (one-parameter) fit we find a not-so-good $p$~value of
0.05, but if we modify our fit so that it includes the NLO terms and a free parameter proportional to $a^2$
[a next-to-next-to-leading order (NNLO) analytic term] then we find a reasonably good $p$~value of 0.25 We
find even better fits if we include all analytic terms through NNLO.
We do not include the NNLO logarithms because they are unknown and would require a two-loop calculation.
The fit expression including all analytic NNLO terms is
\bea\label{eq:NNLO}
    \frac{h^\text{NNLO}_{A_1}(1)}{\eta_A} = 
        c_0 + \text{NLO}_{\rm logs} + c_1m^2_{X_P} + c_2(2m_{U_P}^2+m_{S_P}^2) + c_3 a^2,
\eea
where the subscript $P$ on the meson masses indicates the taste pseudoscalar mass.  The fit parameter $c_0$ represents the quantity $1+X_A(\Lambda_\chi)/m_c^2$ appearing on the right-hand side of Eq.~(\ref{eq:schpt}), while $\rm{NLO}_{\rm logs}$ is a short-hand expression for the last term on the right-hand side of Eq.~(\ref{eq:schpt}).
By heavy-quark symmetry, the $c_i$ are suppressed by a factor of $1/m_c^2$.
The one-loop corrections start at $\text{O}(\bar{\Lambda}^2/m_Q^2)$ so that one has to go to NNLO to find
terms of $\text{O}[(\bar{\Lambda}^2/m_Q^2)p^2]$.
In order to estimate systematic errors we try adding a variety of even higher-order analytic terms to this
expression, as described in detail in Section~\ref{sec:chpt}.
We prefer to take a central value for the extrapolated form factor that is roughly in the middle of the range
of results from the various alternative fits used to estimate our central value.  The motivation for
this form is no greater than for the other fits that were tried.
Our preferred central value fit is to the form
\bea\label{eq:Standard}
    \frac{h^\text{NNLO}_{A_1}(1)}{\eta_A} = 
        c_0 + \text{NLO}_{\rm logs} + c_1m^2_{X_P} + c_2(2m_{U_P}^2+m_{S_P}^2) + c_3 a^2 + c_4 m^4_{X_P},
\eea
which, in addition to the analytic NNLO terms of Eq.~(\ref{eq:NNLO}), includes an NNNLO term proportional to
$m^4_{X_P}$.
Because the various fit Ans\"atze for $h_{A_1}(1)$ considered have at most six free parameters, we do not
need to impose constraints on any of the unknown coefficients.
The coefficients are of the size expected from power counting in heavy-meson chiral perturbation theory.

\begin{figure}[b]
    \centering
    \includegraphics[scale=.55]{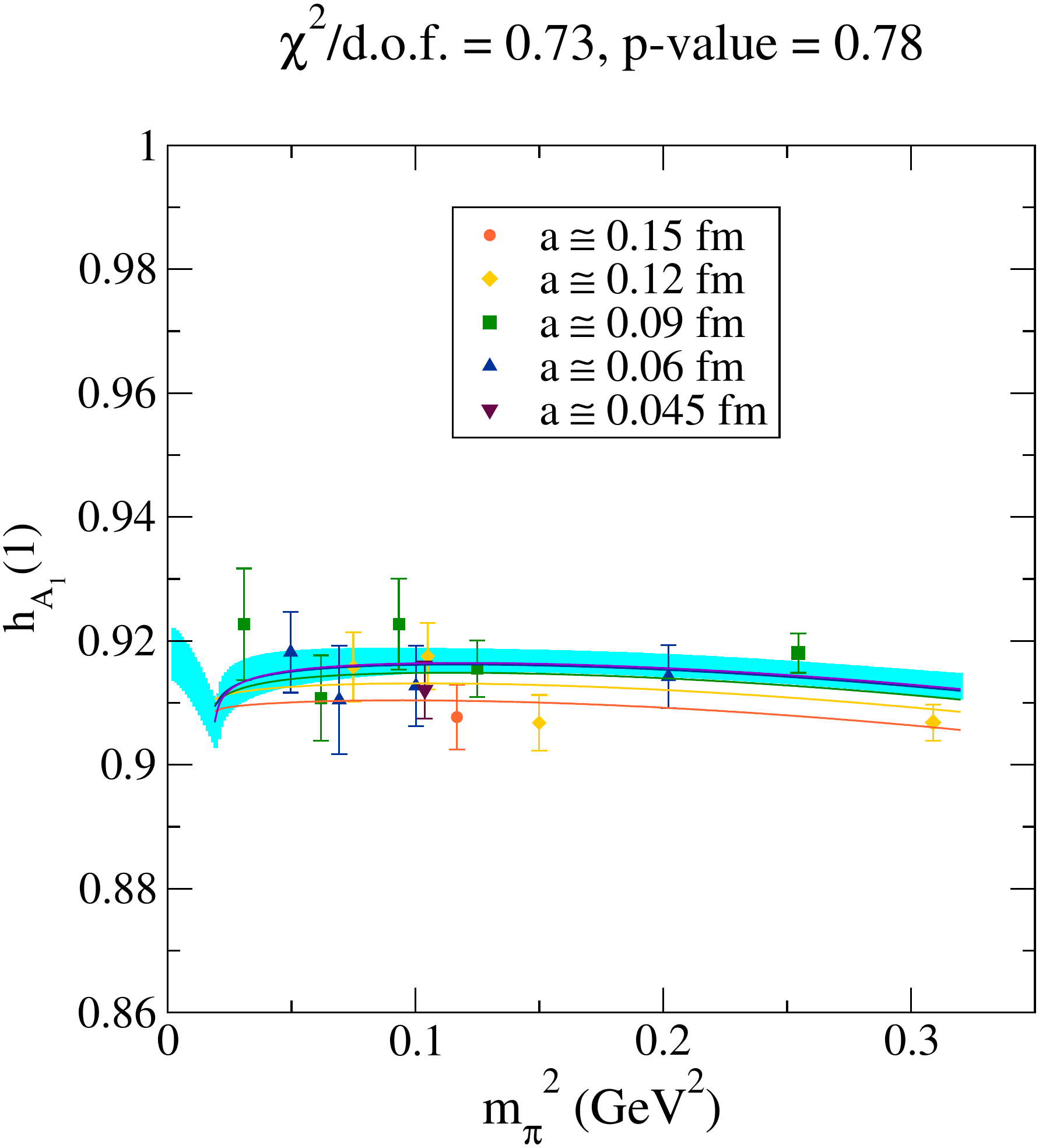}
    \caption{The full QCD points for $h_{A_1}(1)$ versus $m^2_\pi$ at five lattice spacings are 
        shown in comparison to the continuum curve and the various fit curves.  Fit curves at each lattice spacing are shown, with the lowest corresponding to $a=0.15$ fm and increasing monotonically as $a$ decreases.}
    \label{fig:chiralFit}
\end{figure}

\begin{figure}[b]
    \centering
    \includegraphics[scale=.55]{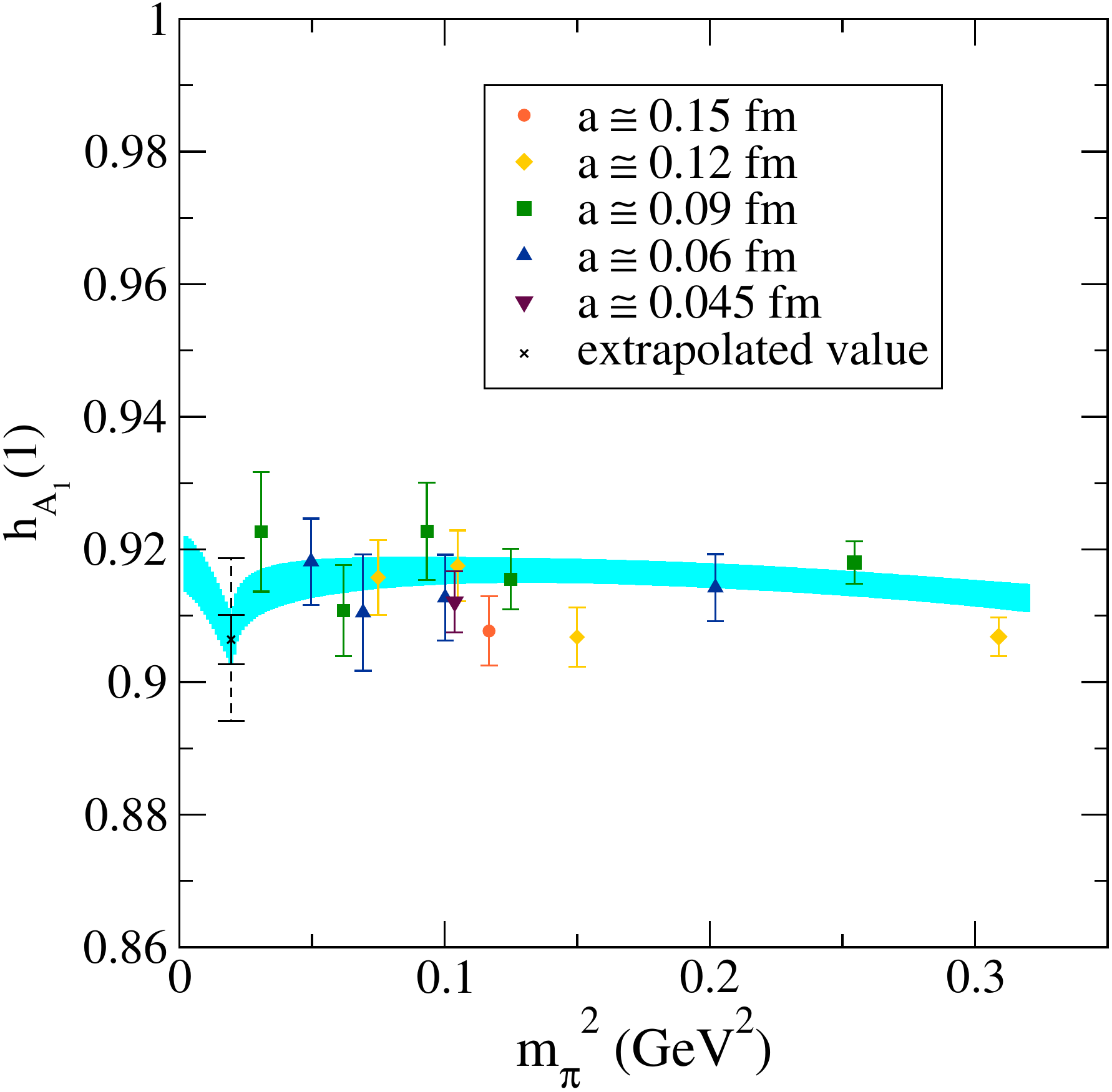}
    \caption{The full QCD points for $h_{A_1}(1)$ versus $m^2_\pi$ at five lattice spacings are 
        shown in comparison to the continuum curve.
        The cross is the extrapolated value, where the solid line is the statistical error, and the dashed 
        line is the total systematic error added to the statistical error in quadrature.}
    \label{fig:chiralFullQCD}
\end{figure}

Our preferred central value fit is shown in Fig.~\ref{fig:chiralFit}, where the curves show the light-quark
mass dependence at different lattice spacings.
The cyan band is the continuum extrapolated result.  A notable feature of the chiral extrapolation is a cusp that appears close to the physical pion mass.  The cusp is due to the presence of the $D\pi$ threshold and the fact that the $D$-$D^*$ splitting is very close to, but slightly larger than, the physical pion mass.
One can see from the curves in Fig.~\ref{fig:chiralFit} that the cusp is expected to be washed out by finite-lattice-spacing effects,
but is recovered in the continuum limit.   
The $p$~value for this fit is 0.78; the alternative fits that also include higher-order analytic
terms have similar $p$~values.
Figure~\ref{fig:chiralFullQCD} shows nearly the same plot, but with only the continuum curve displayed.
The extrapolated value for the form factor is also shown, including the full systematic error for our final
result.

\section{Systematic errors}
\label{sec:systematics}

In this section, we examine the uncertainties in our calculation in detail.
Statistical uncertainties are computed with a single elimination jackknife and fits use the full covariance
matrix to determine $\chi^2$.
We devote a subsection to each of the sources of uncertainty: fitting and excited states, the heavy-quark
mass and lattice-scale dependence, the chiral extrapolation of the light spectator-quark mass (in particular
the $D^*$-$D$-$\pi$ coupling), discretization errors, perturbation theory, and isospin effects.

\subsection{Fitting and excited states}

We determine plateau fits to the double ratio, Eq.~(\ref{eq:Rlatdef}).
The fits are done under a single elimination jackknife, after blocking the data by 4 on all ensembles.
The $\chi^2$ is defined using the full covariance matrix.
Statistical errors are determined in fits that include the full correlation matrix, which was remade for each
jackknife fit.
In order to correctly propagate the correlated statistical errors to the chiral/continuum extrapolation fits,
the jackknife data sets on different ensembles are combined into a larger block-diagonal jackknife data set.
The block size of 4 is chosen only to keep the combined data set to a manageable size for the chiral and
continuum extrapolation fits.
We find that the statistical errors do not grow with blocking, and that therefore the autocorrelation
errors are negligible even without blocking.
This was not true in our previous calculation~\cite{Bernard:2008dn}, although that calculation used many of
the same ensembles.
This is because in the current calculation, we move the source origin around the lattice randomly, whereas in
the previous calculation the source origin was fixed.

With several hundred configurations on each ensemble, and over two thousand configurations on some ensembles,
we do not have difficulty resolving the full covariance matrix in our correlator fits, and we do not need to
resort to a singular value decomposition cut on the eigenvalues of the covariance matrix.
We find that the averaged ratio data (constructed from our correlators using Eq.~(\ref{eq:avg})) on the
$0.09$~fm lattices are well-described by a fit to a constant over a range of 5 time slices, and that the fit
range where an acceptable fit is obtained is roughly the same in physical units across ensembles.
The correlated $\chi^2$/d.o.f.\ ranges from 0.08 to 0.85, with one exception.
On the 0.06~fm, 0.15$m_s$ ensemble, the $\chi^2$/d.o.f.\ is 1.71, a bit higher than one might expect, based
on fits to the same physical time range on other ensembles.
Also, the double ratio $R(t)$ appears somewhat asymmetric under the interchange of source and sink on this
ensemble, but this must be a statistical fluctuation, since $R(t)$ is symmetric by construction.
For this ensemble, we adopt the Particle Data Group (PDG) prescription and rescale the statistical error by
the square root of the $\chi^2$/d.o.f.
Time ranges for fits, their $p$~values, and the raw values for $h_{A_1}(1)$ are given in
Table~\ref{tab:fit_results}.
We take the good quality of our fits as evidence that systematic errors due to excited states are small
compared to other errors, and aside from the inflation of the error on one of our data points, we assign no
further error to fitting and excited states.

\subsection{Heavy-quark mass and lattice-scale dependence}

As discussed in Sec.~\ref{sec:kappa}, the simulation values for $\kappa_{b,c}$ differ from the best tuned
values for these quantities, since the initial tuning analysis was supplemented by additional data and
improved methodology.
We use Eq.~(\ref{eq:shift}) to perform the shift in the form factor given the tuned values of $\kappa_{b,c}$
in Table~\ref{tab:kappas}.
The dependence of $h_{A_1}$ on $\kappa$ (or $m_2$) can also be used to propagate the errors in $\kappa$ shown
in Table~\ref{tab:kappas} to the form factor.
This is done by inflating the difference from the mean under a jackknife for the data points on different
ensembles.
The inflation factor is the sum in quadrature of the statistical error and the parametric error in $h_{A_1}$
due to the $\kappa$ uncertainty labeled ``statistics and fitting'' only.  Thus, the statistical error in $h_{A_1}$ includes the ``statistics and fitting" error in the $\kappa$ tuning.  
The error in the determination of $\kappa_{b,c}$ coming from setting the lattice scale is treated separately
below.

This treatment of the heavy-quark mass tuning error assumes that the errors in $\kappa$ are independent for
each ensemble.
The error would be larger if the adjustment in the form factor varied systematically across multiple
ensembles.
To test the size of such a systematic error, we redo the central fit with all of the coarse ensembles shifted
together by 1$\sigma$ of the estimated errors in $\kappa_{b,c}$.
This leads to a small shift in the central value which is negligible compared to other errors.
The errors in $dh/d[1/(r_1m_b)]$ and in $dh/d[1/(r_1m_c)^2]$ are negligible compared to the other heavy-quark
mass tuning errors.

The relative lattice spacing between different ensembles is fixed in units of $r_1/a$.
The absolute lattice spacing is then fixed using the MILC determination of $r_1=0.3117(22)$~fm from~$f_\pi$
\cite{Bazavov:2011aa}.
Because the form factor is dimensionless, the error in setting the lattice scale mainly affects $h_{A_1}(1)$
by introducing an uncertainty in the determination of the bare $b$- and $c$-quark masses.
Changing $r_1$ within its error of approximately $0.7\%$ leads to an additional $0.1\%$ systematic error in
$h_{A_1}(1)$.

\subsection{Chiral extrapolation}
\label{sec:chpt}

We estimate the systematic error due to the chiral extrapolation by comparing various types of fits including
analytic terms of higher order than NLO in rS$\chi$PT, since the two-loop NNLO logarithms are unknown.
We also compare with continuum $\chi$PT, where staggered effects are removed from the one-loop logarithms.
Finally, we account for additional errors that appear due to the uncertainties in the parameters that enter
the NLO rS$\chi$PT expression.
The largest of these is the uncertainty in $\gDDp$, the coupling between the $D^*$, $D$, and $\pi$ in the
(continuum) chiral effective theory.
As emphasized in our previous calculation of the $B\to D^*\ell\nu$ form factor \cite{Bernard:2008dn}, the
chiral logarithms are of order $10^{-3}$ in the region where we have data, and the nonanalytic behavior is
only important near the physical pion mass.
In that region, $\chi$PT is expected to provide a good description of the physics.
This is important, because very near the physical pion mass there is a cusp in the form factor.
This is due to the presence of the $D\pi$ threshold and the fact that the $D$-$D^*$ splitting is so close to
the physical pion mass.
Because this cusp is a physical effect, it should be included in any version of the chiral extrapolation that
is used to estimate systematic errors.

Through NLO order (one-loop) in rS$\chi$PT there is only one free parameter, an overall constant.
The other parameters that appear in the continuum expression through one-loop are determined from either the
lattice or phenomenology.
They are $\gDDp$, $f_\pi$, $m_\pi$, and the $D$-$D^*$ mass splitting $\Delta^{(c)}$.
The constants $f_\pi$ and $\gDDp$ appear in an overall multiplicative factor
$\gDDp^2/48\pi^2f_\pi^2$ in front of the logarithmic term; see Eq.~(\ref{eq:schpt_long}).
The main uncertainty in the size of the cusp comes from the uncertainties of these one-loop input parameters.
The parameters $f_\pi$, $\Delta^{(c)}$, and the pion mass itself are all precisely determined from
experiment, and contribute only small errors to the overall determination of the size of the cusp.
The dominant error in the size of the cusp comes from the uncertainty in $\gDDp$.

There are additional parameters that enter the one-loop rS$\chi$PT expression due to lattice artifacts.
These are the taste splittings $a^2\Delta_{\xi}$ with $\xi=P,A,T,V,I$, and the taste-violating
hairpin-coefficients $a^2\delta'_A$ and $a^2\delta'_V$.
The former are well-determined from staggered meson spectrum calculations, and the latter are determined from
simultaneous rS$\chi$PT fits to $m_\pi^2/(m_x+m_y)$ and $f_\pi$.
Because the chiral logarithms are such a small contribution to the fit form in the region where we have data,
it makes essentially no difference whether we include the modifications for staggered fermions or not.
We see no difference in the extrapolated continuum result when comparing staggered and continuum $\chi$PT fit
results through 4 decimal places.
Thus, the uncertainties in the parameters specific to rS$\chi$PT are negligible in our extrapolation.

We find that a fit to the NLO expression supplemented by a term linear in $a^2$, does an adequate job of
fitting the data, with $\chi^2/\text{d.o.f.}=1.20$ corresponding to $p=0.25$.
The quality of the fit can be improved either by pruning the heaviest mass points or by adding higher-order
analytic terms to the fit function; we try both.
For our central value we choose a fit that falls around the middle of the range of all the fits that we have
tried.
For our error, we take the largest difference between the central value and the different alternatives.
Our preferred central value fit is to Eq.~(\ref{eq:Standard}), which, in addition to the analytic NNLO terms
of Eq.~(\ref{eq:NNLO}), includes an NNNLO term proportional to $m^4_{X_P}$.
Alternative fits with good $p$~values include the following: Eq.~(\ref{eq:Standard}) without the $c_4$
term, Eq.~(\ref{eq:Standard}) with an additional term $c_6 a^2(2m_{U_P}^2 + m_{S_P}^2)$, repeating these fits
but taking only the ensembles with $a\leq 0.09$~fm.
This cut on the lattice spacing also cuts out the data with the heaviest pion masses, as can be seen in
Table~\ref{tab:ChPT}.
We also considered a fit that tests for the presence of higher-order taste-breaking effects.
This fit is similar to the central value fit but with the taste-pseudoscalar pion mass in the analytic terms
replaced by the taste-tensor pion mass (which is close to the root-mean-square pion mass).
The largest variation from the central value of the form factor in all of these fits is 0.0049, or $0.5\%$.
Figure~\ref{fig:chiralFullQCD} shows all of the full QCD points in our calculation as a function of
(taste-Goldstone) pion mass, as well as the continuum extrapolated curve and the extrapolated value for 
$h_{A_1}(1)$ with the full systematic error.

The largest of the parametric uncertainties in our chiral extrapolation is that due to the chiral-Lagrangian
coupling~$\gDDp$, which sets the size of the cusp.
Our data do not constrain it, so we must take its value from elsewhere.
New lattice calculations of $\gDDp$ \cite{Becirevic:2012pf,Can:2012tx} have appeared since our previous work
on $B\to D^*\ell\nu$.
In Ref.~\cite{Becirevic:2012pf}, 2 light flavors of quarks were included in the sea, but otherwise the
systematic errors appear to be under control.
The authors find $\gDDp(N_f=2)=0.53(3)(3)$, where the first error is statistical and the second is systematic
error due to chiral extrapolation.
The calculation in Ref.~\cite{Can:2012tx} includes 2+1 light dynamical flavors, but only a single lattice
spacing.
The authors find $\gDDp=0.55(6)$, consistent with the 2-flavor calculation.
These results are also consistent with the values extracted from the experimental measurements of the $D^*$
decay width \cite{Anastassov:2001cw,Lees:2013uxa,Lees:2013zna}.
A new preliminary 2+1 flavor result for the analogous coupling in the $B$ system reports
$g_{B^*B\pi}=0.569(48)(59)$~\cite{Flynn:2013kwa}.
Finally, a 2+1 flavor calculation of the coupling in the static heavy-quark limit~\cite{Detmold:2012ge}
finds, after a careful study of systematic effects, $g_\text{static}=0.449(51)$.
Although the result of Ref.~\cite{Becirevic:2012pf} is a calculation directly at the charm quark mass, it
only has two flavors of sea quarks, so we take an error that encompasses that of the 2+1 flavor result in the
static limit in order to be conservative.
Thus, in our fits we take $\gDDp=0.53 \pm0.08$, leading to a parametric, systematic uncertainty in
$h_{A_1}(1)$ of $0.3\%$.

The size of the cusp is also expected to be modified by terms of higher order in the chiral expansion, \ie,
the two-loop chiral logarithms.
Although possible higher-order corrections are at least partially accounted for by our analytic terms in the
range of pion masses where we have data, the cusp is entirely determined by the chiral effective theory, so
it is important to consider how that prediction might be affected by higher-order corrections independent of
the analytic terms that we have added.
Because the effect occurs very near the physical pion mass, we expect the relevant power counting to be that
of $\text{SU}(2)_L \times \text{SU}(2)_R$ $\chi$PT.
We estimate the potential size of the two-loop corrections to the cusp by considering the size of the
one-loop corrections to $f_\pi$ compared to its $\text{SU}(2)$ chiral limit value $f_2$, since these one-loop
corrections to a parameter appearing in the coefficient of the one-loop term are expected to be typical of
the size of the other two-loop corrections.
We take the most recent value for $f_\pi/f_2=1.062(3)$ from the MILC Collaboration~\cite{Bazavov:2009ir} and
find that a $6\%$ change in $f_\pi$ leads to a $0.1\%$ change in $h_{A_1}(1)$.
Thus, for our chiral extrapolation error we include an additional $0.1\%$ systematic error due to higher-order chiral corrections to the cusp added in quadrature with the $0.5\%$ systematic error estimated from the spread in reasonable fits discussed above.

All other parametric uncertainties in the chiral formulas can be neglected.
The physical pion mass in the chiral extrapolation is taken from experiment, so the errors from the
uncertainties in the low-energy constant $B_0$ in Eq.~(\ref{eq:treelevel}) and in the light-quark masses are
negligible.
We take the charm meson mass splitting $\Delta^{(c)}$ from experiment, and the error due to its uncertainty
is also negligible.
Changing the (bare) strange quark mass within its error of approximately $2\%$ also has a negligible effect
on $h_{A_1}(1)$.

\subsection{Finite-volume effects}

The finite-volume effects can be estimated using heavy-light $\chi$PT, where the integrals are replaced by
discrete sums.
The corrections to the integrals in the formulas appearing for $B\to D^*$ decays were worked out by Arndt and Lin
\cite{Arndt:2004bg}.
Although the finite-volume effects would be large very near the cusp at the physical pion mass on the
ensembles we are using (ranging in size from 2.5--5.5~fm), for the values we have actually simulated, the
finite-size effects predicted by $\chi$PT are less than one part in $10^4$.
This is not a result of any particular cancellation, but rather due to the very small contribution of the
chiral logarithms to this quantity.
Thus, the finite-size effects are expected to be negligible for our calculation, and we do not assign any
additional error due to them.

\subsection{Discretization errors}

\begin{figure}[b]
\centering
    \includegraphics[scale=.5]{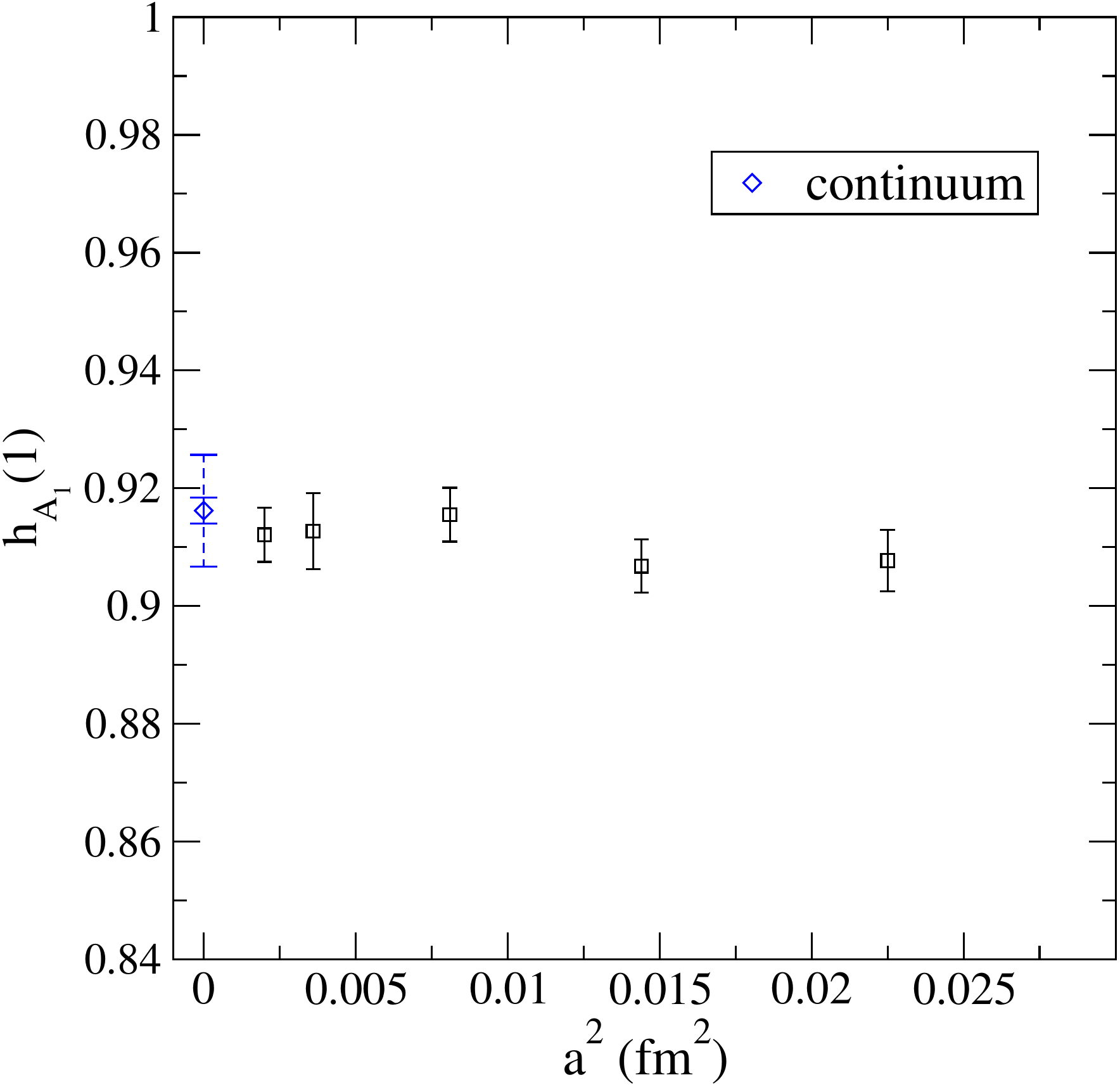}
    \caption{$h_{A_1}(1)$ versus $a^2$ for spectator mass $m_x=0.2\seaheavy$.
       The blue point at $a=0$ shows the extrapolated value for this~$m_x$ including the heavy-quark discretization error added in quadrature with the statistical error.}
    \label{fig:aSq}
\end{figure}

Figure~\ref{fig:aSq} shows the dependence of $h_{A_1}(1)$ as a function of $a^2$, for fixed spectator-quark
mass.
The observed lattice-spacing dependence is, at most, as large as the statistical error.
The HQET theory of heavy-quark discretization effects anticipates this small size but does not, however,
predict a simple power-series for the $a$ dependence, making a naive extrapolation problematic.
In Appendix~\ref{app:hqetdisc}, we present a detailed analysis for the expected $a$ dependence.
In short, we expect the overall size of heavy-quark discretization errors to be of order $a\bar\Lambda^2/m_c$
and $a^2\bar\Lambda^2$, but must choose a value of $\bar\Lambda$.
We compare the observed variation with $a^2$ of the data in Fig.~\ref{fig:aSq} with the
theory~\cite{Kronfeld:2000ck,Harada:2001fj}.
We find that if we choose $\bar\Lambda=450$~MeV, then the theoretical estimates are compatible with the
data's $a$ dependence.
In this way, we deduce that the discretization error on the superfine lattice ($a\approx0.060$~fm) is~1\%,
leading to the row labeled ``discretization errors'' in Table~\ref{tab:errors}.

\subsection{Perturbation theory}
\label{sec:PTerror}

The calculation of $\rhoA{j}$ defined in Eq.~(\ref{eq:rho_ratio}) is carried out at one-loop order in
perturbation theory, as discussed in Sec.~\ref{sec:PT}.
Because $\rhoA{j}$ is defined from a ratio of current renormalization factors, its deviation from unity is
expected to be small by construction.
Indeed, the one-loop corrections to $\rhoA{j}$ shown in Table~\ref{tab:rho} confirm our expectation.
They range from 0.05\% to 0.6\%.
In order to estimate the error due to the omitted higher-order corrections, we consider the variation of the
one-loop corrections to $\rhoA{j}$ with the quark masses used in this calculation.
We also consider the related renormalization factor $\rhoV{4}$, defined from the charm-bottom vector current
$V^4_{cb}$ analogously to the definition of $\rhoA{j}$ in Eq.~(\ref{eq:rho_ratio}).
We find $\rho^{[1]} \leq 0.1$ for both currents.
We then estimate the uncertainty as $\rho^{[1]}_\text{max} \cdot \alpha_s^2$ with $\rho^{[1]}_\text{max}=0.1$
and $\alpha_s=\alpha_V(2/a)$ evaluated at $a \approx 0.045$~fm, which yields a systematic error of $0.4\%$.

\subsection{Isospin Effects}

The experimental measurements of the branching fraction for $B\to D^*\ell\nu$ assume isospin symmetry, and
different isospin channels are averaged together \cite{Amhis:2012bh}.
We estimate the size of the effect of isospin corrections based on the chiral extrapolation.
One could explicitly include the difference between $u$ and $d$ quark masses in the chiral effective theory,
though this has not been worked out through one-loop for this process, to the best of our knowledge.
As a simple estimate of the size of isospin effects we vary the end point of our chiral extrapolation between
the physical $\pi^+$ and the $\pi^0$ mass.
We use the $\pi^+$ mass extrapolation for our central value, but shifting to the $\pi^0$ changes the result
by $0.1\%$.
Changing the charm mass splitting between the $D^{*0}$ and the $D^{*+}$ is a much smaller effect.
Thus, we quote an error of $0.1\%$ due to isospin effects.

\begin{table}
    \centering
    \caption{Final error budget for $h_{A_1}(1)$ where each error is discussed in the text.  
        Systematic errors are added in quadrature and combined in quadrature with the statistical error to 
        obtain the total error.}
    \label{tab:errors}
    \begin{tabular}{lc}
    \hline \hline
      Uncertainty & $h_{A_1}(1)$  \\
      \hline
      Statistics &  $0.4\%$   \\
      Scale ($r_1$) error & $0.1\%$  \\
      $\chi$PT fits \ &  $0.5\%$  \\
      $\gDDp$ &  $0.3\%$  \\
      Discretization errors &  $1.0\%$  \\
      Perturbation theory & $0.4\%$  \\
      Isospin & $0.1\%$ \\
    \hline 
       Total & $1.4\%$ \\
    \hline \hline
    \end{tabular}
\end{table}

\section{Electroweak Effects}
\label{sec:ew}

In this section, we discuss the electroweak and electromagnetic effects in the semileptonic rate,
Eq.~(\ref{eq:dGdw}).
They do not enter the lattice-QCD calculation but are needed, in addition to the hadronic form factor
$\mathcal{F}(1)=h_{A_1}(1)$, to obtain $|V_{cb}|$.
The factor $\eta_{\text{EW}}$ (written as $\eta_{\text{em}}$ in Ref.~\cite{Beringer:1900zz}) takes the
form~\cite{Sirlin:1981ie}
\begin{equation}
    \eta_{\text{EW}} = 1 + \frac{\alpha}{\pi} \left[ \ln\frac{M_W}{\mu} +
        \tan^2\theta_W \frac{M_W^2}{M_Z^2-M_W^2} \ln\frac{M_Z}{M_W} \right],
    \label{eq:sirlin}
\end{equation}
where the weak mixing angle is specified via $\cos\theta_W=g_2/(g_2^2+g_1^2)^{1/2}$; $g_2$ and $g_1$ are the 
gauge couplings of $\text{SU}(2)\times\text{U}(1)$.
The first (second) term stems from $W$-photon ($W$-$Z$) box diagrams plus associated parts from vertex and 
wavefunction renormalization.
This form assumes that $G_F$ in Eq.~(\ref{eq:dGdw}) is defined via the muon lifetime, which is the case for 
$G_F$ in Ref.~\cite{Beringer:1900zz}.
In the SM, $M_W=M_Z\cos\theta_W$, and the bracket simplifies to $\ln(M_Z/\mu)$.
With this assumption, taking the factorization scale $\mu=M_{B^\pm}$, and varying~$\mu$ by a factor of~2 to 
estimate the error, one finds
\begin{equation}
    \eta_{\text{EW,SM}} = 1.00662(16).
    \label{eq:SMsirlin}
\end{equation}
To reiterate, it is theoretically cleaner not to include this factor in $\mathcal{F}(w)$.
This way makes it more straightforward to study or remove the $\mu$ dependence in future work.

In the experiments~\cite{Amhis:2012bh}, the charged-lepton energy spectrum is corrected for bremsstrahlung
with the PHOTOS~\cite{Barberio:1993qi} generator.
For charged $B$ decay, this package has been shown~\cite{RichterWas:1992qb} to reproduce the exact
formula~\cite{Ginsberg:1966zz}.
For neutral $B$ decay, the charged $D^-$ and $l^+$ in the final state attract each other, which is reflected
in a slightly different formula for the radiation~\cite{Ginsberg:1968pz}.
Reference~\cite{Atwood:1989em} recommends treating this effect with a Coulomb correction,
$1+\alpha\pi/2=1.01146$ on the amplitude, which is larger than the electroweak correction and similar in size
to the uncertainties from experiment and from QCD.
Note, however, that a detailed study of radiative corrections in $K\to\pi l\nu$ finds that QCD-scale effects reduce the Coulomb effects, such that the total is closer to 1\% than 2\%~\cite{Cirigliano:2008wn}.
Already now, and certainly for any future determination of $|V_{cb}|$, a similar treatment is called for,
theoretically first and then in the combination of experimental measurements of neutral and charged decays.

\begin{table}
    \caption{Values of $|V_{cb}|$ implied by different choices of experimental inputs when accounting for
    electroweak and Coulomb corrections.
    The first column is the mode or combination of modes that is taken from experiment, the second and third
    columns give the experimental value for $10^3|V_{cb}||\bar{\eta}_\text{EW}|\mathcal{F}(1)$ and its 
    source, the fourth column is our estimate of the correction factor $|\bar{\eta}_\text{EW}|$, the last 
    column is the resulting $10^3|V_{cb}|$ using the result in Eq.~(\ref{eq:hA1(1)}).}
    \label{tab:Vcb}
    \begin{tabular*}{\textwidth}{@{\extracolsep{\fill}}ccccc}
    \hline\hline
    Mode    & \multicolumn{1}{c}{$10^3|V_{cb}||\bar{\eta}_\text{EW}|\mathcal{F}(1)$} & Ref. &
        $|\bar{\eta}_\text{EW}|$ & $10^3|V_{cb}|$ \\
    \hline
    $B^0$   & $35.60 \pm 0.57$ &       \cite{Schwanda:2013se} & $1.0182\pm 0.0016$ & 
        $38.59 \pm 0.62_\text{expt} \pm 0.52_\text{QCD} \pm 0.06_\text{QED}$  \\
    $B^\pm$ & $35.14 \pm 1.45$ & BaBar \cite{Aubert:2007qs}   & $1.0066\pm 0.0016$ & 
        $38.53 \pm 1.60_\text{expt} \pm 0.52_\text{QCD} \pm 0.06_\text{QED}$  \\
    Both    & $40.00 \pm 2.04$ & CLEO  \cite{Adam:2002uw}     & $1.0124\pm 0.0058$ & 
        $43.61 \pm 2.22_\text{expt} \pm 0.59_\text{QCD} \pm 0.25_\text{QED}$  \\
    Both    & $35.83 \pm 1.12$ & BaBar \cite{Aubert:2008yv}   & $1.0124\pm 0.0058$ & 
        $39.06 \pm 1.22_\text{expt} \pm 0.53_\text{QCD} \pm 0.22_\text{QED}$  \\
    Both    & $35.90 \pm 0.45$ & HFAG  \cite{Amhis:2012bh}    & $1.015 \pm 0.005$  & 
        $39.04 \pm 0.49_\text{expt} \pm 0.53_\text{QCD} \pm 0.19_\text{QED}$  \\
    \hline\hline
    \end{tabular*}
\end{table}

The current experiments do not take the Sirlin~\cite{Sirlin:1981ie} and Coulomb effects into account.  Further, to our knowledge a study of QCD-scale photons, analogous to Ref.~\cite{Cirigliano:2008wn}, is not available for heavy-meson decays.
In particular, charged and neutral decays are analyzed and combined without different radiative corrections.
The quantity reported to be $|V_{cb}|\mathcal{F}(1)$ is really
$|V_{cb}||\bar{\eta}_\text{EW}|\mathcal{F}(1)$, where $\bar{\eta}_\text{EW}$ is a suitably charge-weighted
average of Eq.~(\ref{eq:sirlin}) and the Coulomb effect.
Table~\ref{tab:Vcb} shows results for $|V_{cb}|$ from different choices for the experimental input and the
corresponding estimate of $\bar{\eta}_\text{EW}$.
The first entry shows an average with HFAG methods from $B^0$ decays only~\cite{Schwanda:2013se}, while the
second shows the $B^{\pm}$-only measurement from BaBar~\cite{Aubert:2007qs}; then $\bar{\eta}_\text{EW}$ is
simply Eq.~(\ref{eq:SMsirlin}) with and without the Coulomb factor, respectively.
The third and fourth entries are the results from single experiments, CLEO~\cite{Adam:2002uw} and
BaBar~\cite{Aubert:2008yv}, in which both modes were combined; here, we compute $\bar{\eta}_\text{EW}$ by
assuming a 50-50 split, varying between 100-0 and 0-100 to estimate the error.
This range is extreme, but with one experiment, the QCD and QED errors are smaller than the experimental
error.
The last row in Table~\ref{tab:Vcb} shows the 2012 result from HFAG~\cite{Amhis:2012bh} with our estimate of
the appropriate charge-weighted average for $\bar{\eta}_\text{EW}$.
The neutral data carry greater weight in the HFAG average~\cite{Schwanda:2013se}, so we take a value of
$\bar{\eta}_\text{EW}$ slightly larger than a 50-50 split, with generous error range, to allow for other
effects, such as photons at the QCD~scale.

\section{Results and Conclusions}
\label{sec:end}

We have improved on our previous calculation of the zero-recoil form factor for $B\to D^*\ell\nu$ decay by
increasing statistics, going to lighter quark masses at correspondingly larger volumes, and going to finer
lattice spacings.
Our final result, given the error budget in Table~\ref{tab:errors}, is
\begin{equation}
    \mathcal{F}(1) = h_{A_1}(1) = 0.906(4)
(1)
(5)
(3)
(9)
(4)
(1)
, \label{eq:hA1(1)}
\end{equation}
where the errors are statistical, 
scale uncertainty, 
chiral extrapolation errors,
parametric uncertainty in $g_{D^*D\pi}$, 
heavy-quark discretization errors, 
perturbative matching, 
and isospin effects.
Adding all systematic errors in quadrature, we obtain $h_{A_1}(1) = 0.906(4)(12)$, which is consistent with
our previous published result $h_{A_1}(1)=0.921(13)(20)$ \cite{Bernard:2008dn}, but with a significantly
smaller error.
The data added since our preliminary report~\cite{Bailey:2010gb} have reduced the $\chi$PT and $\gDDp$ errors
moderately.

From Table~\ref{tab:Vcb}, we choose the HFAG average of all data, with our conservative estimate of the QED 
correction, as our preferred way of obtaining $|V_{cb}|$.
Thus, we find 
\begin{equation} 
    |V_{cb}|=(39.04 \pm 0.49_\text{expt} \pm 0.53_\text{QCD} \pm 0.19_\text{QED})\times 10^{-3}.
\end{equation}
The QCD error is now commensurate with the experimental error.
This result is in agreement with our previous published result \cite{Bernard:2008dn}, but differs by
$3.0\sigma$ from the inclusive determination $|V_{cb}|=(42.42\pm0.86)\times10^{-3}$~\cite{Gambino:2013rza}.

The largest error in our determination of $h_{A_1}(1)$ is the systematic error due to heavy-quark
discretization effects.
We have made a detailed study of the expected $a$ dependence using HQET at finite lattice spacing.
A value of $\bar{\Lambda}$ is needed to compute this dependence; our choice of $\bar{\Lambda} \approx$ 450
MeV is consistent with the size of the discretization effects seen in the numerical data and can reproduce
the behavior of these effects over the five lattice spacings included in our calculation.
We could reduce this error by going to finer lattice spacings or by using a more improved Fermilab action,
\eg, the Oktay-Kronfeld action \cite{Oktay:2008ex}.
When using this action, it would be necessary to improve the currents to the same order.

Several subleading errors appear in our calculation at the 0.4-0.6$\%$ level.
They would be nontrivial to improve.
Reducing the error from the QED Coulomb correction would require a detailed study of electromagnetic effects
within HQET, and reducing the QCD matching error would require a two-loop lattice perturbation theory
calculation or nonperturbative matching.
The chiral extrapolation error would not necessarily be reduced by a straightforward simulation at the
physical light-quark masses because the $D^*$ would become unstable apart from finite-volume effects.
At the current level of precision, it is important to extend the calculation to nonzero recoil.
This would provide a useful cross-check of the method used to extrapolate the experimental form factor to
zero recoil \cite{Caprini:1997mu}.
Another important cross-check is our companion calculation of $|V_{cb}|$ using the $B\to D\ell\nu$ decay, has
been reported in Ref.~\cite{Qiu:2013ofa}.
Full details, including its determination of $|V_{cb}|$, will be presented in a forthcoming paper.

\acknowledgments

We thank Vincenzo Cirigliano, Christoph Schwanda, and Zbigniew W\c{a}s for useful correspondance.  
A.X.K. thanks the Fermilab Theory Group for hospitality while this work was finalized.
Computations for this work were carried out with resources provided by 
the USQCD Collaboration, the Argonne Leadership Computing Facility, 
the National Energy Research Scientific Computing Center, and the Los Alamos National Laboratory, which
are funded by the Office of Science of the United States Department of Energy; and with resources
provided by the National Institute for Computational Science, the Pittsburgh Supercomputer Center, 
the San Diego Supercomputer Center, and the Texas Advanced Computing Center, 
which are funded through the National Science Foundation's Teragrid/XSEDE Program. 
This work was supported in part by the U.S. Department of Energy under Grants 
No.~DE-FG02-91ER40628 (C.B.), 
No.~DE-FC02-06ER41446 (C.D., J.F., L.L.), 
No.~DE-SC0010120 (S.G.),
No.~DE-FG02-91ER40661 (S.G., R.Z.),
No.~DE-FC02-06ER41443 (R.Z.), 
No.~DE-FG02-13ER42001 (D.D., A.X.K.), 
No.~DE-FG02-13ER41976 (D.T.); 
by the National Science Foundation under Grants 
No.~PHY-1067881, No.~PHY-0757333, No.~PHY-0703296 (C.D., J.F., L.L.), 
No.~PHY-1212389 (R.Z.),
No.~PHY-1316748 (R.S.); 
by the URA Visiting Scholars' program (C.M.B., D.D., A.X.K.);
by the Science and Technology Facilities Council and the Scottish Universities Physics Alliance (J.L.);
by the MINECO (Spain) under Grants FPA2010-16696, FPA2006-05294, and \emph{Ram\'on y Cajal} program (E.G.); 
by the Junta de Andaluc\'{\i}a (Spain) under Grants FQM-101 and FQM-6552 (E.G.); by European Commission 
under Grant No. PCIG10-GA-2011-303781 (E.G.);
and by the Creative Research Initiatives program (3348-20090015) of the NRF grant funded by the Korean 
government (MEST) (J.A.B.).
This manuscript has been co-authored by an employee of Brookhaven Science Associates, LLC,
under Contract No.~DE-AC02-98CH10886 with the U.S. Department of Energy. 
Fermilab is operated by Fermi Research Alliance, LLC, under Contract No.~DE-AC02-07CH11359 with
the U.S. Department of Energy.

\appendix

\section{Staggered Chiral Perturbation Theory for $B\to D^*\ell\nu$ at zero-recoil}
\label{app:chpt}

The partially quenched expression for $h_{A_1}/\eta_{A}$ at zero-recoil through NLO in staggered chiral
perturbation theory was derived in Ref.~\cite{Laiho:2005ue}.
For completeness, it is given here.
The result is
\begin{eqnarray}\label{eq:schpt_long}  
    \frac{h_{A_1}^{(B_x)PQ,2+1}(1)}{\eta_A}&=& 1 + \frac{X_{A}(\Lambda_\chi)}{m_c^2}+\frac{g^2_{DD^*\pi}}{48
\pi^2f^2}\Bigg\{\frac{1}{16} \sum_{\begin{subarray}{c}j=xu,xu,xs\\ \Xi=I,P,4V,4A,6T \end{subarray}} \!\!\!\!\!\!\! \overline{F}_{j_\Xi} \nonumber\\
 &+& \frac{1}{3}\bigg[R^{[2,2]}_{X_I}\big(\{M^{(5)}_{X_I}\}
;\{\mu_I\}\big)\left(\frac{d\overline{F}_{X_I}}{dM^2_{X_I}}\right)
-\sum_{j \in
\{M^{(5)}_I\}}D^{[2,2]}_{j,X_I}\big(\{M^{(5)}_{X_I}\};\{\mu_I\}\big)\overline{F}_{j}
\bigg] \nonumber \\
  &+& a^2\delta'_{V}\bigg[R^{[3,2]}_{X_I}\big(\{M^{(7)}_{X_V}\}
;\{\mu_V\}\big)\left(\frac{d\overline{F}_{X_V}}{dM^2_{X_V}}\right)
-\sum_{j \in
\{M^{(7)}_V\}}D^{[3,2]}_{j,X_V}\big(\{M^{(7)}_{X_V}\};\{\mu_V\}\big)\overline{F}_{j}
 \Big] \nonumber \\
 &+& \big(V\rightarrow A\big)\Bigg\}, 
\end{eqnarray}
\noindent where
\begin{eqnarray}\label{eq:F}
F\left(M_j,z_j\right) &=& \frac{M_j^2}{z_j}\bigg\{z_j^3
\ln\frac{M_j^2}{\Lambda_\chi^2}-\frac{2}{3}z_j^3 -4z_j+2\pi \nonumber \\
&&
-\sqrt{z_j^2-1}(z_j^2+2)\left(\ln\Big[1-2z_j(z_j-\sqrt{z_j^2-1})\Big]-i\pi\right)\bigg\}
\nonumber
\\ && \longrightarrow (\Delta^{(c)})^2\ln\left(\frac{M_j^2}{\Lambda_\chi^2}\right)+{\cal
O}[(\Delta^{(c)})^3],
\end{eqnarray}
\noindent with $\overline{F}(M_j,z_j) = F(M_j,-z_j)$, and
$z_j=\Delta^{(c)}/M_j$, where $\Delta^{(c)}$ is the $D$-$D^*$ mass
splitting. The residues $R^{[n,k]}_j$ and $D^{[n,k]}_{j,i}$ are
defined in Refs.~\cite{Aubin:2003mg, Aubin:2003uc}.
\noindent These residues are a function of two sets of masses, the
numerator masses, $\{M\}=\{M_1, M_2,...,M_n \}$ and the
denominator masses, $\{\mu\}=\{ \mu_1, \mu_2,...,\mu_k \}$.  In
our 2+1 flavor case, we have
\begin{eqnarray} \{M_X^{(5)}\} &\equiv& \{M_\eta, M_X \}, \nonumber \\
    \{M_X^{(7)}\}&\equiv& \{M_\eta, M_{\eta'}, M_X \}, \nonumber \\
    \{\mu\}&\equiv&\{M_U,M_S\}. \end{eqnarray}
\noindent The expressions for the masses $M_{\eta_I}$, $M_{\eta_V}$, $M_{\eta'_V}$ in terms of the parameters
of the rooted staggered effective theory are given in Ref.~\cite{Aubin:2003mg}.

\section{Heavy-quark Discretization Effects}
\label{app:hqetdisc}

Let us define the various discretization errors in $\rhoA{j}\sqrt{R_{A_1}}$ via
\begin{equation}
    \rhoA{j}\sqrt{R_{A_1}} = h_{A_1}(1) + \mathrm{O}(\alpha_s^{1+\ell_\rho}) 
        + \mathrm{O}(\alpha_s^{1+\ell_c}a\bar{\Lambda}^2/m_c)
        + \mathrm{O}(\alpha_s^{1+\ell_d}a^2\bar{\Lambda}^2),
    \label{eq:rhoR=hA1}
\end{equation}
where $\bar{\Lambda}\approx M_B-m_b$ is a measure of nonperturbative QCD effects in heavy-light mesons.
These stem, respectively, from the truncation of the perturbative series for~$\rhoA{j}$, truncation of the
perturbative series for~$c_{\text{SW}}$ (\ie, improvement of the action), and from mismatches in the improved
lattice currents.
That the power-law effects in Eq.~(\ref{eq:rhoR=hA1}) start with $\bar\Lambda^2$ is a special property of
zero recoil, established below.
As written, Eq.~(\ref{eq:rhoR=hA1}) holds for general, multi-loop matching; for the calculation described in
this paper, we have one-loop matching for $\rhoA{j}$, so $\ell_\rho=1$, and we have tree-level improvement
for the action and current, so $\ell_c=\ell_d=0$.

We now assemble the formulae needed to prove the appearance of $\bar\Lambda^2$.
The discretization effects are estimated with the heavy-quark effective field theory (HQET)
\cite{Kronfeld:2000ck,Harada:2001fj}.
Wilson fermions exhibit heavy-quark symmetries for small $\kappa$, so HQET provides a suitable description.
For the lattice gauge theory (LGT) Lagrangian,
\begin{equation}
    \mathcal{L}_{\text{LGT}} \doteq \bar{h}(iv\cdot{D} - m_1)h +
        \frac{\bar{h}D_\perp^2h}{2m_2} +
        \frac{\bar{h}s\cdot Bh}{2m_B} + 
        \frac{\bar{h}[D_\perp^\alpha,iE_\alpha]h}{8m_D^2} +
        \frac{\bar{h}s_{\alpha\beta}\{D_\perp^\alpha,iE^\beta\}h}{4m_E^2} + \cdots,
    \label{eq:L-HQET}
\end{equation}
where $\doteq$ can be read ``has the same matrix elements as.''
Here, $v$ is a four vector specifying the rest-frame of the heavy-light meson, such that $v^2=-1$;
the heavy-quark field $h$ satisfies $-i\vslash h=h$,
and $s_{\alpha\beta}=-i\sigma_{\alpha\beta}/2$.
Then, $D_\perp^\mu=D^\mu+v^\mu\,v\!\cdot\!D$ is the covariant derivative orthogonal to $v$, 
$B^{\alpha\beta}=(\delta^\alpha_\mu+v^\alpha v_\mu)F^{\mu\nu}(\delta^\beta_\nu+v^\beta v_\nu)$ is the 
chromomagnetic field (in the $v$ frame), and $E^\beta=-v_\alpha F^{\alpha\beta}$  is the 
chromoelectric field (in the $v$ frame).
The HQET description for continuum QCD has the same structure
\begin{equation}
    \mathcal{L}_{\text{QCD}} \doteq \bar{h}(iv\cdot{D} - m)h +
        \frac{\bar{h}D_\perp^2h}{2m} +
        \frac{z_B\bar{h}s\cdot Bh}{2m} + 
        \frac{z_D\bar{h}[D_\perp^\alpha,iE_\alpha]h}{8m^2} +
        \frac{z_E\bar{h}s_{\alpha\beta}\{D_\perp^\alpha,iE^\beta\}h}{4m^2} + \cdots.
    \label{eq:L-HQET-QCD}
\end{equation}
In matrix elements, the rest mass $m_1$ does not enter, so one tunes $\kappa$ so that
\begin{equation}
    \frac{1}{2m_2}=\frac{1}{2m},
\end{equation}
and $c_{\text{SW}}$ so that
\begin{equation}
    \frac{1}{2m_B}=\frac{z_B}{2m} = \frac{1 + \text{O}(\alpha_s)}{2m} ,
\end{equation}
where the second equality follows because $z_B=1+\text{O}(\alpha_s)$.
In practice, we tune $\kappa$ via the heavy-strange \emph{meson} mass, as discussed in
Appendix~\ref{app:kappa-tuning}, and we choose $c_\text{SW}$ at the tadpole-improved tree level, which brings
in the second error exhibited in Eq.~(\ref{eq:rhoR=hA1}).

The LGT currents can also be described in the HQET, and the full description entails many operators
\cite{Kronfeld:2000ck,Harada:2001fj}.
Here, however, we need only the temporal vector current:
\begin{eqnarray}
    Z_{V_{cb}} V^4 = -Z_{V_{cb}} v\cdot V \doteq \bar{C}_{V^{cb}_\parallel} \bar{c}b & + & 
        \eta^{(0,2)}_{V^{cb}D_\perp^2} \frac{\bar{c}D_\perp^2b}{8m_{D_\perp^2b}^2} +
        \eta^{(0,2)}_{V^{cb}sB} \frac{\bar{c}s\cdot Bb}{8m_{sBb}^2} -
        \eta^{(0,2)}_{V^{cb}\alpha E} \frac{\bar{c}i\Eslash b}{4m_{\alpha Eb}^2} 
    \nonumber \\ & + &
        \eta^{(2,0)}_{V^{cb}D_\perp^2} \frac{\bar{c}\loarrow{D}_\perp^2b}{8m_{D_\perp^2c}^2} +
        \eta^{(2,0)}_{V^{cb}sB} \frac{\bar{c}s\cdot Bb}{8m_{sBc}^2} +
        \eta^{(2,0)}_{V^{cb}\alpha E} \frac{\bar{c}i\Eslash b}{4m_{\alpha Ec}^2} 
    \label{eq:V-HQET} \\ & + &
        z^{(1,1)}_{V^{cb}1} \frac{\bar{c}\loarrow{D}_\perp\cdot D_\perp b}{2m_{3c}\;2m_{3b}} +
        z^{(1,1)}_{V^{cb}s} \frac{\bar{c}\loarrow{D}_\perp^\alpha s_{\alpha\beta}D_\perp^\beta b}{2m_{3c}\;2m_{3b}}
    , \nonumber 
\end{eqnarray}
and the spatial axial vector current ($\epsilon$ is the $D^*$ polarization vector):
\begin{eqnarray} \hspace*{-2em}
    Z_{A_{cb}} \epsilon\cdot A \doteq \bar{C}_{A^{cb}_\perp} \bar{c}\eslash_\perp\gamma^5b & + & 
        \eta^{(0,2)}_{A^{cb}D_\perp^2} \frac{\bar{c}\eslash_\perp\gamma^5D_\perp^2b}{8m_{D_\perp^2b}^2} +
        \eta^{(0,2)}_{A^{cb}sB} \frac{\bar{c}\eslash_\perp\gamma^5s\cdot Bb}{8m_{sBb}^2} -
        \eta^{(0,2)}_{A^{cb}\alpha E} \frac{\bar{c}\eslash_\perp\gamma^5i\Eslash b}{4m_{\alpha Eb}^2} 
    \nonumber \\ & + &
        \eta^{(2,0)}_{A^{cb}D_\perp^2} \frac{\bar{c}\loarrow{D}_\perp^2\eslash_\perp\gamma^5b}{8m_{D_\perp^2c}^2} +
        \eta^{(2,0)}_{A^{cb}sB} \frac{\bar{c}s\cdot B\eslash_\perp\gamma^5b}{8m_{sBc}^2} +
        \eta^{(2,0)}_{A^{cb}\alpha E} \frac{\bar{c}i\Eslash\eslash_\perp\gamma^5b}{4m_{\alpha Ec}^2} 
    \label{eq:A-HQET} \\ & + &
        z^{(1,1)}_{A^{cb}1} \frac{\bar{c}(\loarrow{D}_\perp\eslash_\perp\gamma^5D_\perp)_1b}{2m_{3c}\;2m_{3b}} +
        z^{(1,1)}_{A^{cb}s} \frac{\bar{c}(\loarrow{D}_\perp\eslash_\perp\gamma^5D_\perp)_sb}{2m_{3c}\;2m_{3b}}
    .  \nonumber
\end{eqnarray}
The continuum currents enjoy the same description, but with different short-distance coefficients.
The matching factors~$Z_V$ and~$Z_A$ are defined so that the leading operators on the right-hand sides of
Eqs.~(\ref{eq:V-HQET}) and~(\ref{eq:A-HQET}) share the normalization with the corresponding continuum
currents.
With the one-loop calculation of $\rhoA{j}$, explained in Sec.~\ref{sec:PT}, the matching leads to 
$\ell_\rho=1$ in Eq.~(\ref{eq:rhoR=hA1}).
For the currents defined in Sec.~\ref{sec:corrs}, as well as for the continuum currents, the 
$\eta$-coefficients and $z$-coefficients all take the form $1+\text{O}(\alpha_s)$.
The rotation in Eq.~(\ref{eq:Psi=psi}) ensures that
\begin{equation}
    \frac{1}{2m_3} = \frac{1}{2m_2} + \text{O}(\alpha_sa),
\end{equation}
\ie, $\ell_d=0$ in Eq.~(\ref{eq:rhoR=hA1}).
The other masses in Eqs.~(\ref{eq:V-HQET}) and~(\ref{eq:A-HQET}) deviate from $m_2$ when $m_2a\not\ll1$ but 
all collapse to $m_2$ as $m_2a\to0$ \cite{ElKhadra:1996mp,Oktay:2008ex}.
These properties of the coefficients are crucial to the proof that the discretization effects start with 
$\bar\Lambda^2$ in Eq.~(\ref{eq:rhoR=hA1}).

Note that no dimension-four currents arise, which would describe discretization errors starting at 
order~$a\bar{\Lambda}$.
At nonzero recoil, such currents do appear, and their discretization errors are shown in detail in 
Eqs.~(2.37)--(2.44) of Ref.~\cite{Harada:2001fj}.
At zero recoil, the heavy-quark symmetry enlarges from
$\text{SU}_{b\text{-spin}}(2)\times\text{SU}_{c\text{-spin}}(2)$ to $\text{SU}_{\text{spin-flavor}}(4)$, and
a generalization of Luke's theorem requires the leading discretization/heavy-quark effects to vanish.
The discretization effects then stem from second-order breaking of heavy-quark symmetry, as explained in
Ref.~\cite{Kronfeld:2000ck}, leading to the extra suppression of $\bar{\Lambda}/m_c$ or $a\bar\Lambda$ in 
Eq.~(\ref{eq:rhoR=hA1}).
Luke's theorem also ensures that single insertions of chromoelectric interactions (spin-orbit and Darwin
terms) drop out at zero recoil.

We proceed by collecting results from Ref.~\cite{Kronfeld:2000ck} for the zero-recoil discretization 
errors in matrix elements of the currents in Eqs.~(\ref{eq:Vlat}) and~(\ref{eq:Alat})
and combining them into a formula for the discretization error in $\rhoA{j}\sqrt{R_{A_1}}$.
(Note that in Ref.~\cite{Kronfeld:2000ck} $\rho_A\sqrt{R_{A_1}}$ stands for a different double ratio.) %
The discretization errors stem from all higher-dimension terms on the right-hand sides of 
Eqs.~(\ref{eq:L-HQET}), (\ref{eq:V-HQET}), and~(\ref{eq:A-HQET}), but always take the form
\begin{equation}
    \texttt{error}_i = \left(\mathcal{C}^{\text{LGT}}_i-\mathcal{C}^{\text{QCD}}_i\right)
        \left\langle\mathcal{O}_i\right\rangle,
\end{equation}
where the $\mathcal{C}_i$ denote the short-distance coefficients, which are different for the lattice and
continuum, and the $\mathcal{O}_i$ denotes the HQET operators on the right-hand sides of
Eqs.~(\ref{eq:L-HQET}), (\ref{eq:V-HQET}), and~(\ref{eq:A-HQET}).
To get the errors, we then combine asymptotic forms of
$\mathcal{C}^{\text{LGT}}_i-\mathcal{C}^{\text{QCD}}_i$ with power-counting estimates of
$\langle\mathcal{O}_i\rangle$.
The former have been derived in Refs.~\cite{ElKhadra:1996mp,Oktay:2008ex}, and the latter are guided by the
data and some theoretical considerations to arrive at concrete error estimates.

\subsection{Second-order formulas at zero recoil}

From Eqs.~(7.20) and (7.30) of Ref.~\cite{Kronfeld:2000ck}, the HQET expansions through
$\textrm{O}(\bar{\Lambda}^2)$ of the matrix elements are
\begin{eqnarray}
    \langle B|\ZVbb V^4|B\rangle & = & 1 + W^{(2)}_{00},
    \label{eq:BVBW} \\
    \langle D^*(\bm{\epsilon})|\ZVcc V^4|D^*(\bm{\epsilon})\rangle & = & 1 + W^{(2)}_{11},
    \label{eq:DVDW} \\
    \langle D^*(\bm{\epsilon})|\ZAcb \bm{\epsilon}\cdot\bm{A}%
        |B\rangle & = & \bar{C}_{A^{cb}_\perp} W^{(0)}_{01} + W^{(2)}_{01},
    \label{eq:DABW}
\end{eqnarray}
where $\bar{C}_{A^{cb}_\perp}=1+\text{O}(\alpha_s)$ is a short-distance coefficient in
Eq.~(\ref{eq:A-HQET}), and $W^{(2)}_{01}$ is written $\bar{W}^{(2)}_{01}+\delta W^{(2)}_{01}$ in 
Ref.~\cite{Kronfeld:2000ck}.
The subscripts on $W^{(i)}_{JJ'}$ indicate the meson spins ($J=0$ for $B$ and $J=1$ for $D^*$), and the
superscript denotes the order in the heavy-quark expansion of the currents.
The expressions for the vector-current matrix elements have been simplified by noting 
$\bar{C}_{V^{hh}_\parallel}=1$ for the flavor-diagonal vector current,
and $W^{(0)}_{JJ}=1$ for $h\to h$ transitions.
Combining Eqs.~(\ref{eq:BVBW})--(\ref{eq:DABW}), one finds the $\textrm{O}(\bar{\Lambda}^2)$ expansion
\begin{equation}
    \rhoA{j}\sqrt{R_{A_1}} = \bar{C}_{A^{cb}_\perp} W^{(0)}_{01} + W^{(2)}_{01}
        -\half \bar{C}_{A^{cb}_\perp} \left(W^{(2)}_{00}+W^{(2)}_{11}\right).
    \label{eq:disc}
\end{equation}
We must obtain more explicit expressions for the terms on the right-hand side and compare them to the
analagous terms in the HQET expansion of~$h_{A_1}(1)$ in continuum QCD.

Let us start with $W^{(0)}_{01}$.
From Eq.~(7.31) of Ref.~\cite{Kronfeld:2000ck}
\begin{equation}
    W^{(0)}_{01} = 1 - \half\Delta_2(\Delta_2D-2\Theta_BE)
        - \half\Delta_B(\Delta_BR_1-\Theta_BR_2)
        - \frac{1}{2m_{Bc}2m_{Bb}}(\case{4}{3}R_1+2R_2),
    \label{eq:W001}
\end{equation}
where $D$, $E$, $R_1$, and $R_2$ are HQET matrix elements of order~$\bar{\Lambda}^2$, and
\begin{eqnarray}
    \Delta_I & = & \frac{1}{2m_{Ic}} - \frac{1}{2m_{Ib}},\quad I=2,B, \\
    \Theta_I & = & \frac{1}{2m_{Ic}} + \frac{3}{2m_{Ib}}
\end{eqnarray}
are combinations of the mass coefficients in Eq.~(\ref{eq:L-HQET}).
Beyond the leading~1, the terms in $W^{(0)}_{01}$ come from double insertions of the kinetic and
chromomagnetic interactions.
Equation~(\ref{eq:W001}) makes clear that we are working through $\textrm{O}(\bar{\Lambda}^2)$ in the
heavy-quark expansion, although it accommodates, in principle, all orders in perturbation theory
in~$\alpha_s$.

To obtain the analogous expression for Eq.~(\ref{eq:W001}) in continuum QCD, simply replace $m_{2h}\to m_h$
(because that is how the hopping parameter is tuned in the Fermilab method) and $1/m_{Bh}\to z_B/m_h$
[compare Eqs.~(\ref{eq:L-HQET}) and~(\ref{eq:L-HQET-QCD})].
Taking the difference, one sees that the error in $W^{(0)}_{01}$ stems from 
\begin{equation}
    \frac{1}{2m_{Bh}} - \frac{z_B}{2m_{2h}} = af_{Bh}.
\end{equation}
We have chosen $c_{\text{SW}}$ such that $f_{Bh}$ is of order~$\alpha_s$, and the mismatches in 
$W^{(0)}_{01}$ lead to errors of order~$\alpha_sa\bar{\Lambda}^2/m_h$.

Now let us turn to the error in the other terms in Eq.~(\ref{eq:disc}) and combine them into
\begin{equation}
    \ddot{W}^{(2)}_{01} = W^{(2)}_{01} - \half \bar{C}_{A^{cb}_\perp} \left(W^{(2)}_{00} + W^{(2)}_{11}\right).
    \label{eq:ddot}
\end{equation}
The right-hand side comes from the matrix elements of the dimension-five terms in Eqs.~(\ref{eq:V-HQET}) 
and~(\ref{eq:A-HQET}).
The matrix elements of $\Eslash$ vanish, and the others lead to 
\begin{eqnarray}
    W^{(2)}_{JJ} & = & - \left( \frac{1}{4m^2_{D_\perp^2h}} -
        \frac{z_{V^{hh}1}^{(1,1)}}{(2m_{3h})^2} \right) \mu_\pi^2
        + d_J \left( \frac{1}{4m^2_{sBh}} -
        \frac{z_{V^{hh}s}^{(1,1)}}{(2m_{3h})^2} \right) 
        \frac{\mu_G^2}{3},
    \label{eq:W2JJ} \\
    W^{(2)}_{01} & = & - \left(
        \frac{\eta_{A^{cb}D^2_\perp}^{(2,0)}}{8m^2_{D_\perp^2c}} +
        \frac{\eta_{A^{cb}D^2_\perp}^{(0,2)}}{8m^2_{D_\perp^2b}} +
        \third\frac{z_{A^{cb}1}^{(1,1)}}{2m_{3c}2m_{3b}} \right) \mu_\pi^2 \nonumber \\ & & { } 
         - \left( \frac{\eta_{A^{cb}sB}^{(2,0)}}{8m^2_{sBc}} -
        3\frac{\eta_{A^{cb}sB}^{(0,2)}}{8m^2_{sBb}} -
        \frac{z_{A^{cb}s}^{(1,1)}}{2m_{3c}2m_{3b}} \right) 
        \frac{\mu_G^2}{3}, 
    \label{eq:W201}
\end{eqnarray}
as in Eqs.~(7.22), (7.33) and~(7.34) of Ref.~\cite{Kronfeld:2000ck}.
Here, $\mu_\pi^2$ is the heavy-quark kinetic energy, and $\mu_G^2$ is known from the $B^*$-$B$ splitting.
Both $\mu_\pi^2$ and $\mu_G^2$ are of order~$\bar{\Lambda}^2$.
(Ref.~\cite{Kronfeld:2000ck} used another notation with $\mu_\pi^2=-\lambda_1$ and $\mu_G^2=3\lambda_2$.) %
We choose to define $m^2_{D_\perp^2h}$ and $m^2_{sBh}$ to all orders in $\alpha_s$ via the degenerate-mass
vector current, so $\eta_{V^{hh}D^2_\perp}^{(2,0)}\equiv1$, etc., so no $\eta$-like coefficients appear in
Eq.~(\ref{eq:W2JJ}).

At the tree level, the coefficients written as inverse masses are the same for all currents.
By construction, $\eta_{A^{cb}D_\perp^2}^{(2,0)}$, $\eta_{A^{cb}D_\perp^2}^{(0,2)}$,
$\eta_{A^{cb}sB}^{(2,0)}$, and $\eta_{A^{cb}sB}^{(0,2)}$, take the form $1+\text{O}(\alpha_s)$.
Furthermore, an analogous all-orders definition of $m_{3h}$ ensures that the $z_{J\bullet}^{(1,1)}$ take the
form $1+\text{O}(\alpha_s)$ too.
As $a\to0$, the right-hand sides of Eqs.~(\ref{eq:W2JJ}) and~(\ref{eq:W201}) approach continuum QCD.
In particular, the quantities inside large parentheses in Eq.~(\ref{eq:W2JJ}) must vanish as $a\to0$.

Combining Eqs.~(\ref{eq:W2JJ}) and~(\ref{eq:W201}) as specified in Eq.~(\ref{eq:ddot}),
\begin{eqnarray}
    \ddot{W}^{(2)}_{01} & = & - \left(
        \frac{\eta_{A^{cb}D^2_\perp}^{(2,0)}-\bar{C}_{A^{cb}_\perp}}{8m^2_{D_\perp^2c}} +
        \frac{\bar{C}_{A^{cb}_\perp}z_{V^{cc}1}^{(1,1)}}{8m^2_{3c}} +
        \frac{\eta_{A^{cb}D^2_\perp}^{(0,2)}-\bar{C}_{A^{cb}_\perp}}{8m^2_{D_\perp^2b}} +
        \frac{\bar{C}_{A^{cb}_\perp}z_{V^{bb}1}^{(1,1)}}{8m^2_{3b}} +
        \third\frac{z_{A^{cb}1}^{(1,1)}}{2m_{3c}2m_{3b}} \right) \mu_\pi^2
        \nonumber \\ & - &
        \left( \frac{\eta_{A^{cb}sB}^{(2,0)}-\bar{C}_{A^{cb}_\perp}}{8m^2_{sBc}} +
        \frac{\bar{C}_{A^{cb}_\perp}z_{V^{cc}s}^{(1,1)}}{8m^2_{3c}} -
        3\frac{\eta_{A^{cb}sB}^{(0,2)}-\bar{C}_{A^{cb}_\perp}}{8m^2_{sBb}} -
        3\frac{\bar{C}_{A^{cb}_\perp}z_{V^{bb}s}^{(1,1)}}{8m^2_{3b}} -
        \frac{z_{A^{cb}s}^{(1,1)}}{2m_{3c}2m_{3b}} \right) 
        \frac{\mu_G^2}{3},
    \hspace*{3em}
\end{eqnarray}
Once again, the analagous expression in continuum QCD can be obtained from $\ddot{W}^{(2)}_{01}$ by 
changing the short-distance coefficients accordingly.
The errors in $\ddot{W}^{(2)}_{01}$ stem from the mismatches
\begin{eqnarray} 
    a^2f_{D_\perp^2c} = 
    \frac{\eta_{A^{cb}D^2_\perp}^{(2,0)}(m_{0c}a,m_{0b}a)}{8m^2_{D_\perp^2c}} -
        \frac{\bar{C}_{A^{cb}_\perp}}{8m^2_{D_\perp^2c}} +
        \frac{\bar{C}_{A^{cb}_\perp}z_{V^{cc}1}^{(1,1)}(m_{0c}a,m_{0b}a)}{8m^2_{3c}}    
    & - & \frac{\eta_{A^{cb}D^2_\perp}^{(2,0)}(m_c/m_b)}{8m^2_{2c}},
    \label{diff:ccD2} \hspace*{2.5em} \\
    a^2f_{D_\perp^2b} = 
    \frac{\eta_{A^{cb}D^2_\perp}^{(0,2)}(m_{0c}a,m_{0b}a)}{8m^2_{D_\perp^2b}} -
        \frac{\bar{C}_{A^{cb}_\perp}}{8m^2_{D_\perp^2b}} +
        \frac{\bar{C}_{A^{cb}_\perp}z_{V^{bb}1}^{(1,1)}(m_{0c}a,m_{0b}a)}{8m^2_{3b}} 
        & - & \frac{\eta_{A^{cb}D^2_\perp}^{(0,2)}(m_c/m_b)}{8m^2_{2b}},
    \label{diff:bbD2} \\
    a^2f_{sBc} = 
    \frac{\eta_{A^{cb}sB}^{(2,0)}(m_{0c}a,m_{0b}a)}{8m^2_{sBc}} -
        \frac{\bar{C}_{A^{cb}_\perp}}{8m^2_{sBc}} +
        \frac{\bar{C}_{A^{cb}_\perp}z_{V^{cc}s}^{(1,1)}(m_{0c}a,m_{0b}a)}{8m^2_{3c}} 
        & - & \frac{\eta_{A^{cb}sB}^{(2,0)}(m_c/m_b)}{8m^2_{2c}}, 
    \label{diff:ccsB} \\
    a^2f_{sBb} = 
    \frac{\eta_{A^{cb}sB}^{(0,2)}(m_{0c}a,m_{0b}a)}{8m^2_{sBb}} -
        \frac{\bar{C}_{A^{cb}_\perp}}{8m^2_{sBb}} +
        \frac{\bar{C}_{A^{cb}_\perp}z_{V^{bb}s}^{(1,1)}(m_{0c}a,m_{0b}a)}{8m^2_{3b}}
        & - & \frac{\eta_{A^{cb}sB}^{(0,2)}(m_c/m_b)}{8m^2_{2b}},
    \label{diff:bbsB} \\
    a^2f_{3c3b1} = 
    \frac{z_{A^{cb}1}^{(1,1)}(m_{0c}a,m_{0b}a)}{2m_{3c}2m_{3b}}
        & - & \frac{z_{A^{cb}1}^{(1,1)}(m_c/m_b)}{2m_{2c}2m_{2b}},
    \label{diff:cbD2} \\
    a^2f_{3c3bs} = 
    \frac{z_{A^{cb}s}^{(1,1)}(m_{0c}a,m_{0b}a)}{2m_{3c}2m_{3b}}
        & - & \frac{z_{A^{cb}s}^{(1,1)}(m_c/m_b)}{2m_{2c}2m_{2b}},
    \label{diff:cbsB}
\end{eqnarray}
where the right-most terms are those stemming from continuum QCD.
Because the Fermilab method is based on Wilson fermions (as opposed to lattice NRQCD), the continuum limit of
the $\eta$s and $z$s must tend as $a\to0$ to the analogous coefficients for continuum QCD:
\begin{eqnarray}
    \lim_{a\to0}\eta^{(\bullet)}_{J\bullet}(m_{0c}a,m_{0b}a) & = & \eta^{(\bullet)}_{J\bullet}(m_c/m_b) \\
    \lim_{a\to0}z^{(1,1)}_{J\bullet}(m_{0c}a,m_{0b}a) & = & z^{(1,1)}_{J\bullet}(m_c/m_b) 
\end{eqnarray}
with $m_c/m_b=m_{0c}a/m_{0b}a$ fixed.
Therefore, in Eqs.~(\ref{diff:ccD2})--(\ref{diff:bbsB}), the first and fourth should cancel against each
other, and so should the second and third.
At nonzero lattice spacing, even when $m_{0h}a\sim1$, the difference between the first and second terms is of
order~$\alpha_s$, and similarly for the difference between the third and fourth terms.
This complicated pattern of cancellation ensures that the right-hand sides of
Eqs.~(\ref{diff:ccD2})--(\ref{diff:bbsB}) is of order~$\alpha_s a^2$.
Similarly, the cancellation on the right-hand sides of Eqs.~(\ref{diff:cbD2}) and~(\ref{diff:cbsB}) also
leaves mismatches of order~$\alpha_s a^2$.

This completes the demonstration that the heavy-quark discretization effects in Eq.~(\ref{eq:rhoR=hA1}) 
start with $\bar\Lambda^2$.
Note especially that the discretization effects of order~$a$ from the clover term mistuning are
suppressed by an additional (small) factor $\bar{\Lambda}/m_h$.
The discretization errors from the currents are, owing to the double-ratio, of order~$a^2$.
Note that to extend Eq.~(\ref{eq:rhoR=hA1}) beyond $\ell_d=0$, we would need not only one-loop matching of
the rotation in Eq.~(\ref{eq:Psi=psi}) but further rotations of the form $D_\perp^2\psi$
and~$s\!\cdot\!B\psi$.
In practice, we have $\ell_d=0$, so this complication is not needed for now.

\subsection{Discretization errors}
\label{sec:subleading}

We now turn to explicit estimates of the total discretization error.
Each term of Eq.~(\ref{eq:rhoR=hA1}) introduces an error into our calculation, which we address in turn.
The error of order~$\alpha_s^2$ from the one-loop computation of the matching factor~$\rhoA{j}$ is discussed
in Sec.~\ref{sec:PTerror}.

\subsubsection{Errors of order $\alpha_s a\bar{\Lambda}^2/m_h$}

This discretization error stems from the one-loop mismatch of the chromomagnetic masses $1/2m_{Bh}$ appearing
in $W^{(0)}_{01}$.
From Eq.~(\ref{eq:W001}), it is
\begin{eqnarray}
    \texttt{error}_B & = & a \frac{f_{Bb}}{2m_{2c}} 4E - a \frac{f_{Bc}}{2m_{2c}} [R_1-(R_2+E)] 
    \nonumber \\ & & { }-
        \frac{a}{3} \left[ \frac{f_{Bb}}{2m_{2c}} + 
            \frac{f_{Bc}+3f_{Bb}}{2m_{2b}}
            \right] [R_1+3(R_2+E)],
    \label{eq:errB}
\end{eqnarray}
where $f_{Bh}=f_B(m_{0h}a)$ is the mismatch function for heavy quark~$h$.
The reason for grouping the HQET matrix elements this way is explained below.
The mismatch function $f_B(m_0a)$ starts at order~$\alpha_s$, and we do not have an explicit expression for
it.
(The calculation is what one needs to match $c_{\text{SW}}$ at the one-loop level.) %
We shall take unimproved tree-level coefficients as a guide to the combinatoric factors, leading to the
Ansatz
\begin{equation}
	f_B(m_0a) = \frac{\alpha_s}{2(1+m_0a)}.
	\label{eq:fB}
\end{equation}
The relative signs in Eq.~(\ref{eq:errB}) are meaningful once one has chosen a coherent Ansatz for the mass
dependence of $f_B$, such as Eq.~(\ref{eq:fB}), and if, as argued in Sec.~\ref{sec:BPS}, we know the relative
signs of the HQET matrix elements $E$, $R_1-(R_2+E)$, and $R_1+3(R_2+E)$.
If we assume nothing about the latter, then the three terms on the right-hand side of Eq.~(\ref{eq:errB})
should be treated as independent and added in quadrature.

\subsubsection{Errors of order $\alpha_s a^2\bar{\Lambda}^2$}

These discretization errors stem from the differences in Eqs.~(\ref{diff:ccD2})--(\ref{diff:cbsB}).
Let us start with the first two terms in Eqs.~(\ref{diff:ccD2})--(\ref{diff:bbsB}).
The numerator differences are of order~$\alpha_s$ and the denominators can be deduced from Eqs.~(A17)
and~(A19) of Ref.~\cite{ElKhadra:1996mp}.
When $c_B=r_s$ they share the same coefficient
\begin{equation}
    \frac{1}{8m_{D_\perp^2}^2} = \frac{1}{8m_{sB}^2} = \frac{1}{8m_2^2} + a^2f_X(m_0a),
    \label{eq:mX2}
\end{equation}
where~\cite{ElKhadra:1996mp,Oktay:2008ex}
\begin{equation}
    f_X(m_0a) = \frac{1}{4(1+m_0a)} - \frac{1}{2}\left( \frac{m_0a}{2(2+m_0a)(1+m_0a)} \right)^2.
    \label{eq:fX}
\end{equation}
These errors can thus be estimated to be 
\begin{eqnarray}
    \texttt{error}_{X_1} & = & \alpha_s \left[
        \frac{1}{2(2m_{2c})^2} + a^2f_{Xc} +
        \frac{1}{2(2m_{2b})^2} + a^2f_{Xb} \right] \mu_\pi^2 
    \nonumber \\ & + &
        \alpha_s \left[\third
        \frac{1}{2(2m_{2c})^2} + a^2\third f_{Xc} -
        \frac{1}{2(2m_{2b})^2} - a^2f_{Xb} \right] \mu_G^2
    \label{eq:errmX}
\end{eqnarray}
where the relative signs and combinatorial factors have been retained.
We do not, however, know the sign and size of the (omitted) one-loop coefficients multiplying the two
brackets.
In Eq.~(\ref{eq:errmX}), $f_{Xh}$ means to evaluate Eq.~(\ref{eq:mX2}) with the $m_0a$ of quark~$h=c,b$.

In Eqs.~(\ref{diff:ccD2})--(\ref{diff:bbsB}), the cancellation of the third and fourth terms lead to
discretization effects correlated with the right-hand side of Eq.~(\ref{eq:errmX}).
Because the tree-level matches exactly, we have
\begin{equation}
    \texttt{error}_{X_2} = \alpha_s \left[
        \frac{1}{2(2m_{2c})^2} +
        \frac{1}{2(2m_{2b})^2} \right] \mu_\pi^2 +
        \alpha_s \left[\third
        \frac{1}{2(2m_{2c})^2} -
        \frac{1}{2(2m_{2b})^2} \right] \mu_G^2
    \label{eq:errm2}
\end{equation}
As $a\to0$, however, $\texttt{error}_{X_2}$ has to cancel 
the $1/(2m_2a)^2$ parts of $\texttt{error}_{X_1}$.
On the other hand, for $m_0a\gg 1$, the $f_X$ terms
dominate all others.
It seems safe, therefore, to combine these errors into
\begin{equation}
    \texttt{error}_X = \alpha_s a^2 (f_{Xc} + f_{Xb}) \mu_\pi^2
        + \alpha_s a^2 (\third f_{Xc} - f_{Xb}) \mu_G^2 .
    \label{eq:errX}
\end{equation}
Here, the relative sign and size of the two terms is unknown, owing to the unknown one-loop coefficients of 
the various $\eta$s.

The last discretization errors of order~$\alpha_s a^2\bar{\Lambda}^2$ stem from
Eqs.~(\ref{diff:cbD2})--(\ref{diff:cbsB}).
At the tree level, the numerators are 1, and in the denominators $m_3=m_2$.
At the one-loop level, mismatches appear
\begin{equation}
    \texttt{error}_{33} = 
        - a^2 \third (\mu_\pi^2 - \mu_G^2) f_{33}(m_{0c}a,m_{0b}a),
    \label{eq:err33}
\end{equation}
where $f_{33}$ is of order~$\alpha_s$.
Because, on the one hand, the mismatch vanishes as $a\to0$, yet, on the other, the lattice contribution
freezes out as the masses become large, we propose the following Ansatz:
\begin{equation}
    f_{33}(m_{0c}a,m_{0b}a) = \frac{\alpha_s}{2(1+m_{0c}a)2(1+m_{0b}a)}.
    \label{eq:f33}
\end{equation}
This error is likely to be smaller than the others, because $\mu_\pi^2-\mu_G^2$ is small; cf.\
Sec.~\ref{sec:BPS}.

\subsubsection{HQET matrix elements}
\label{sec:BPS}

We have good estimates for $\mu_\pi^2$ and $\mu_G^2$, because they appear in the heavy-quark expansions of
the meson masses and of kinematic distributions of inclusive semileptonic decays.
From the pseudoscalar-vector-meson mass difference
\begin{equation}
    \mu_G^2 = \case{3}{4}(M_{B^*}^2 - M_B^2) = 0.364~\text{GeV}^2 = (603~\text{MeV})^2,
    \label{eq:muG2}
\end{equation}
which can be taken to be exact.
Recent fits to inclusive $B\to X_cl\nu$ and $B\to X_s\gamma$ distributions yield a value for the kinetic
energy (in the ``kinetic'' scheme)~\cite{Antonelli:2009ws}
\begin{equation}
    \mu_\pi^2(1~\textrm{GeV}) = 0.424\pm0.042~\text{GeV}^2 = (651\pm32~\text{MeV})^2.
    \label{eq:mupi2}
\end{equation}
Thus, we have $\texttt{error}_{33}\approx0.0015$ (on lattices with $a\approx0.09$~fm).
We do not have estimates for $D$, $E$, $R_1$, and $R_2$ as good as Eqs.~(\ref{eq:muG2}) and~(\ref{eq:mupi2}),
but they satisfy sum rules such that $D>0$, $R_1>\max(R_2,-3R_2)$.

\subsection{Error estimation}

We would now like to combine the sources of heavy-quark discretization errors into a total
\begin{equation}
	\texttt{error} = \bigoplus_i \texttt{error}_i(m_0a),
\end{equation}
where $\bigoplus$ denotes sum in quadrature over independent terms in $\texttt{error}_B$, $\texttt{error}_X$,
and $\texttt{error}_{33}$.
With the error function $f_X$ derived and reasonable Ans\"atze for $f_B$ and $f_{33}$, the crucial ingredient
in these estimates is the value chosen for $\bar{\Lambda}$, estimating the needed HQET matrix elements to be
of order~$\bar{\Lambda}^2$.
Below we study our data and choose $\bar\Lambda$ to reproduce the observed lattice-spacing dependence.
We follow the detailed derivation given above and use $\mu_\pi^2$ and $\mu_G^2$ for
$\texttt{error}_X$ and $\texttt{error}_{33}$.
On the fine lattices ($a\approx0.09$~fm), we take the typical $\alpha_V(q^*)$ to be $0.261$, as in
Table~\ref{tab:rho}, and we use one-loop running to obtain $\alpha_V(q^*)$ at the other lattice spacings.

\input 450f.tex 
%
\input 450.tex
The discretization formulas can be re-applied to estimate the difference between $\rhoA{j}\sqrt{R_{A_1}}$ on
a lattice of spacing $a$ vs.\ the value on a reference lattice.
Table~\ref{tbl:diffs} shows such differences with $\bar{\Lambda}=450$~MeV and the fine ($a\approx0.090$~fm)
lattice as the reference.
The variation is similar to, albeit slightly larger than, the observed lattice-spacing dependence in
Fig.~\ref{fig:aSq}, as one can see by comparing the columns labeled ``Total'' and ``Data'' in 
Table~\ref{tbl:diffs}.
Guided in this way, Table~\ref{tbl:errors} shows the total error with $\bar{\Lambda}=450$~MeV.
On the superfine lattice, the error is 1\%, which we quote in Sec.~\ref{sec:systematics} as the heavy-quark
discretization error on~$h_{A_1}(1)$.
This estimate is neither overly cautious ($\bar\Lambda$ is justified by the data) nor aggressive (we could
have pushed $\bar\Lambda$ to be a small as the data would tolerate, or taken the error estimate of $0.7\%$
from the ultrafine lattice spacing).

\section{Heavy-quark Mass Tuning and Hyperfine Splitting}
\label{app:kappa-tuning}

Our method for tuning $\kappa$ for charm and bottom quarks closely follows that of
Refs.~\cite{Bernard:2010fr,Bazavov:2011aa}, where further details can be found.
Here, however, we use a mass-independent scale-setting scheme, determining $r_1/a$, for each $a$, at the
physical sea-quark masses $\hat{m} = m_s/27$ and $m_s$.
Before we used a mass-dependent set up, taking $r_1/a$ on each ensemble at the simulation sea masses 
$\sealight$ and $\seaheavy$.
The new method compensates for mistunings in the sea-quark masses.
We also use a new method for smoothing the lattice-spacing dependence that reduces errors, particularly at
smaller lattice spacings.
Finally, these second-generation tunings also have higher statistical precision than was available in 
Refs.~\cite{Bernard:2010fr,Bazavov:2011aa}.

We start with the dispersion relation for a
heavy-light meson on the lattice~\cite{ElKhadra:1996mp}
\begin{equation}
    E^2(\bm{p})= M_1^2 + \frac{M_1}{M_2}\bm{p}^2 + 
        \frac{1}{4} A_4\,(a\bm{p}^2)^2 + 
        \frac{1}{3} A_{4'} a^2 \sum_{j=1}^3 |p_j|^4 + \ldots ,  
    \label{eq:disprel}
\end{equation}
where  
\begin{equation}
    M_1 \equiv E(\bm{0})
\end{equation}
is called the rest mass, and the kinetic mass is given by
\begin{equation}
    M_2^{-1} \equiv 2 \left.\frac{\partial E(\bm{p})}{\partial p_j^2} 
        \right|_{\bm{p}=\bm{0}} .
\end{equation}
These meson masses $M_1$ and $M_2$ differ from corresponding quark masses, $m_1$ and 
$m_2$, by binding-energy effects.
The bare mass or, equivalently, the hopping parameter~$\kappa$ must be 
adjusted so that these masses reproduce an experimental charmed or 
$b$-flavored meson mass.
When $M_1$ and $M_2$ differ, as they do when $m_Qa\not\ll1$, one must choose.
Weak matrix elements are unaffected by the heavy-quark rest mass 
$m_1$~\cite{Kronfeld:2000ck}, so it does not make sense to adjust the 
bare mass to $M_1$.  On the other hand, as seen in Appendix~\ref{app:hqetdisc}, the analysis of discretization effects using HQET makes $M_2$ the natural choice.
We therefore focus on $M_2$,  adjusting $\kappa$ to the strange 
pseudoscalars $D_s$ and $B_s$, extrapolated to physical sea-quark masses, 
both because the signal degrades for lighter 
valence-quark masses and because this avoids introducing an unnecessary systematic 
uncertainty due to a chiral extrapolation in the valence-quark mass.  

\subsection{Tuning from the dispersion relation on the $\sealight/\seaheavy = 0.2$ ensembles}

We outline tuning the charm and bottom $\kappa$ values with the following steps, which are described in more
detail below.
We work at all available lattice spacings with the $\sealight/\seaheavy = 0.2$ ensembles. 

\begin{figure}[b]
    \centering
    \includegraphics[clip=true,width=0.48\textwidth]{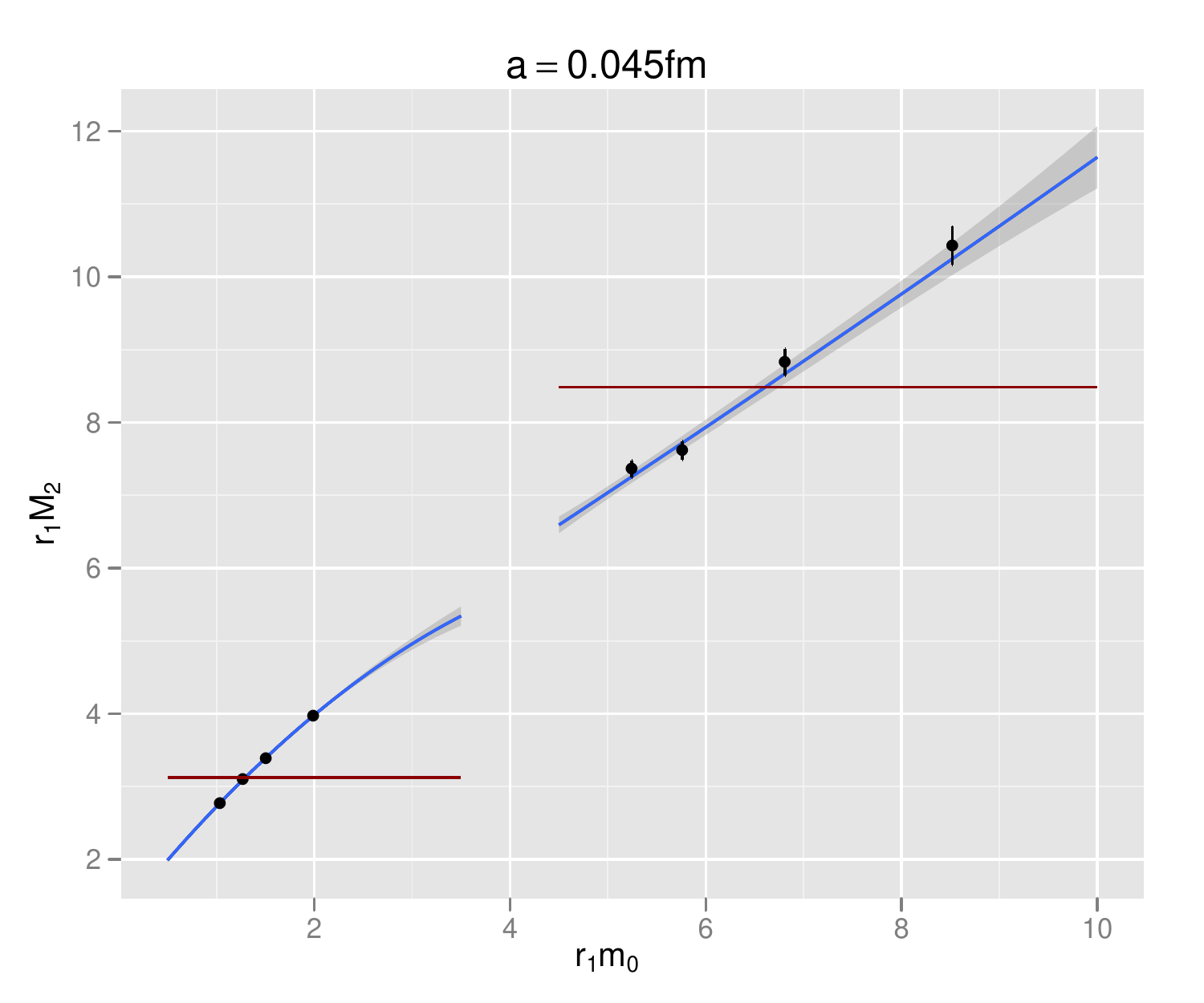} \hfill
    \includegraphics[clip=true,width=0.48\textwidth]{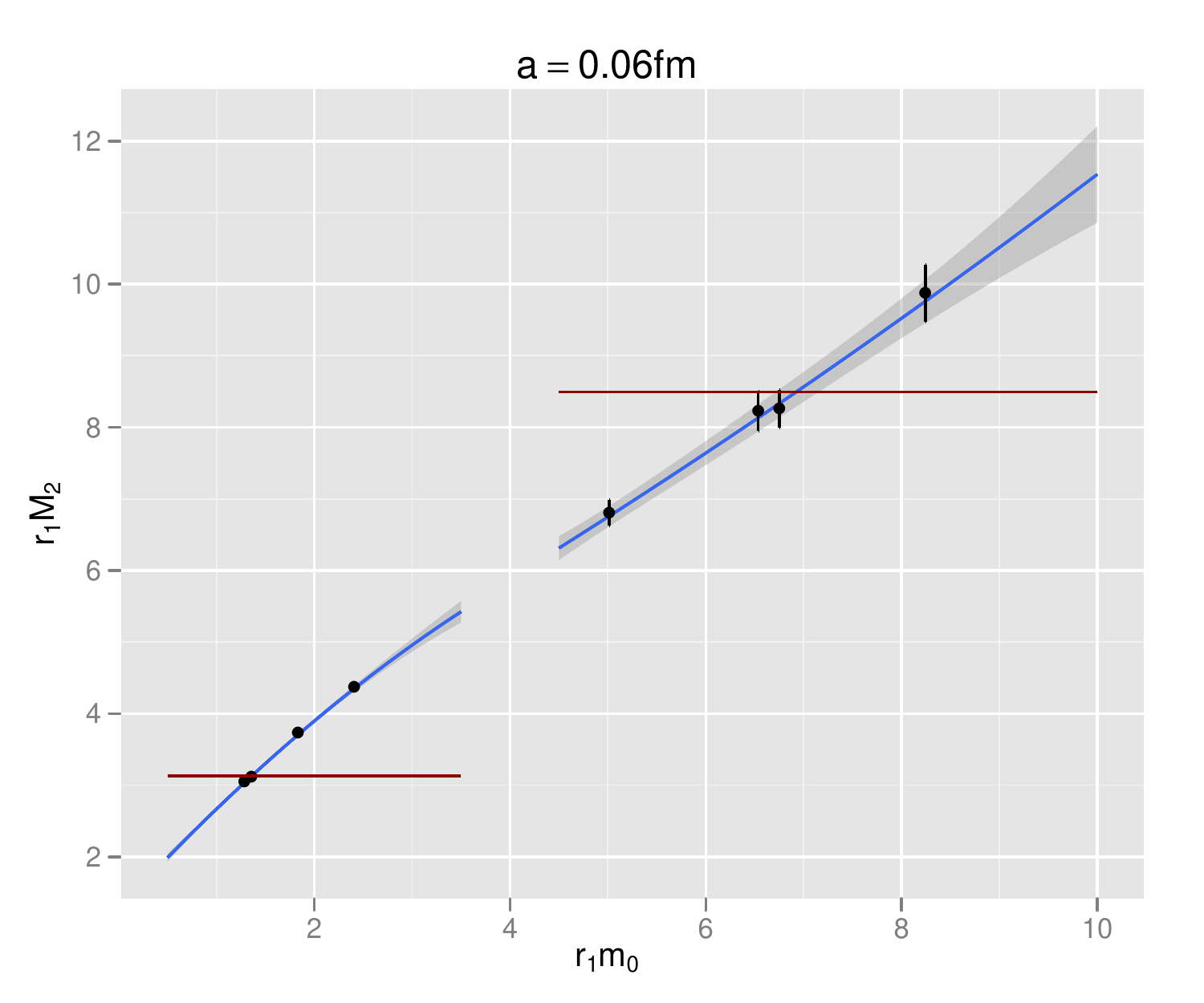} \\
    \includegraphics[clip=true,width=0.48\textwidth]{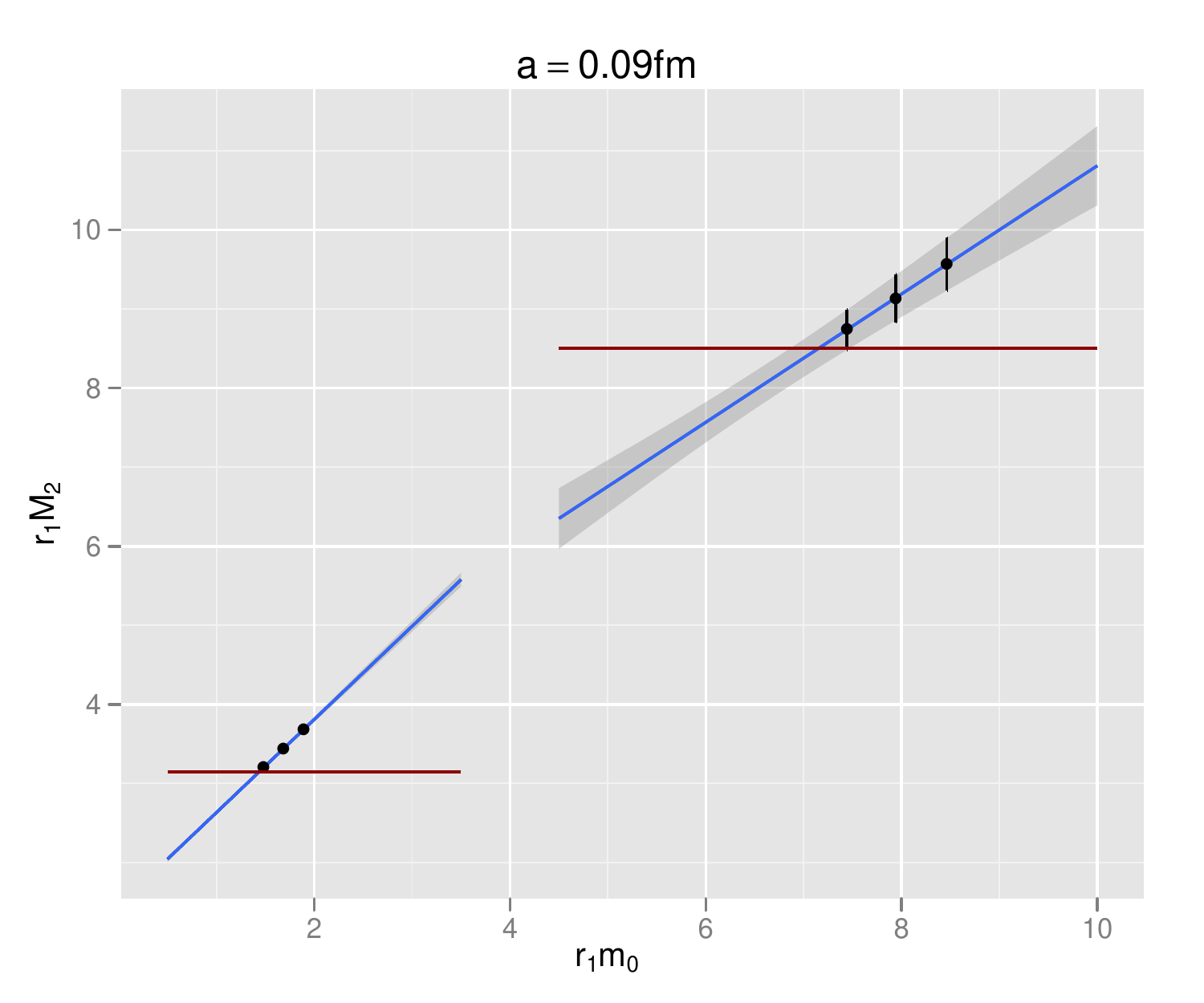} \hfill
    \includegraphics[clip=true,width=0.48\textwidth]{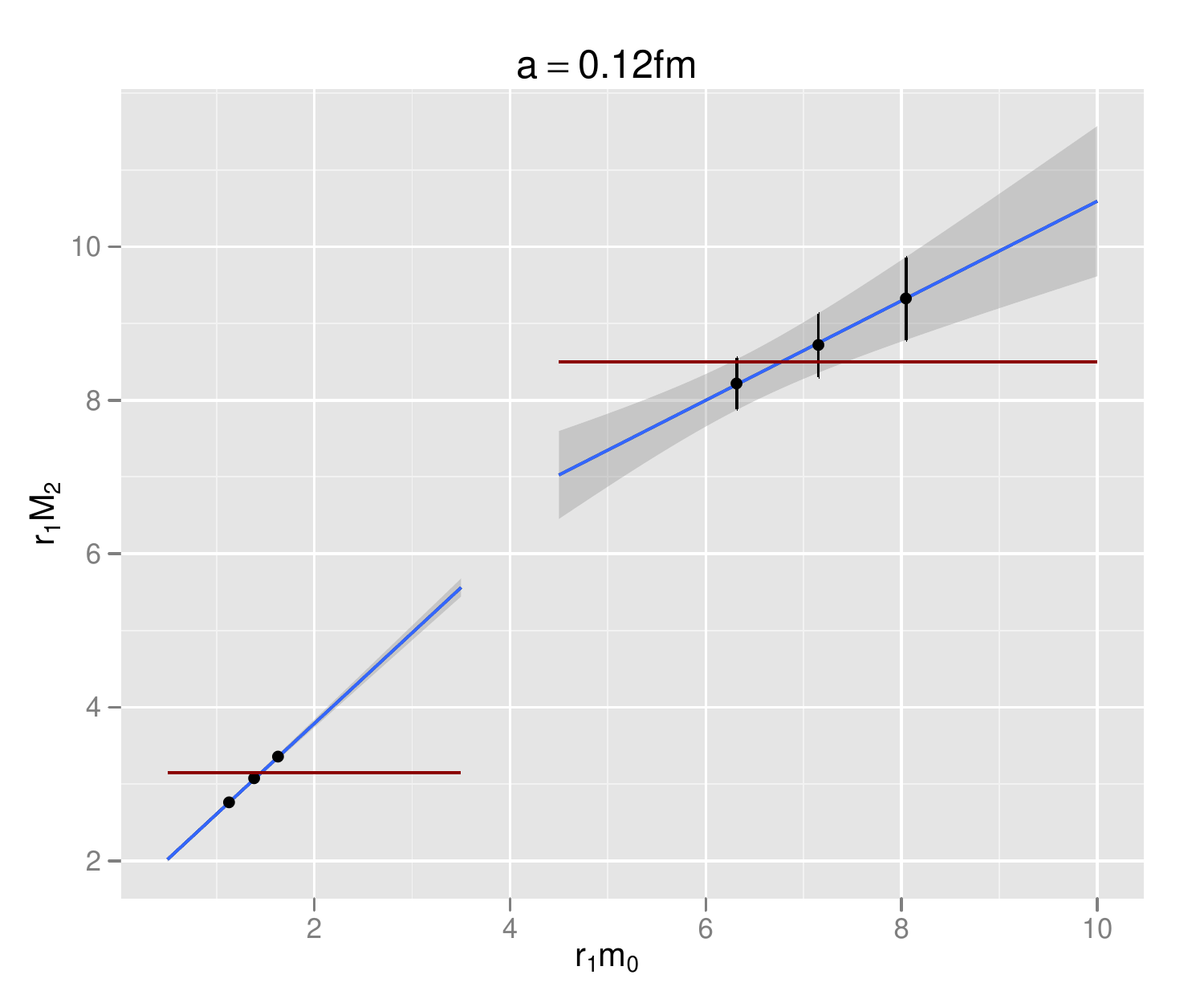} 
    \caption{Interpolation of $r_1M_2(\kappa)$ to the corresponding physical $D_s$ and $B_s$ meson masses
    (indicated by the horizonal lines).
    Separate (quadratic or linear) interpolations are performed for charm and bottom.
    These results are from Analysis~B.
    The figure for the $a\approx0.15~\fm$ lattice is not shown.}
    \label{fig:kappaInterepAnalyB}
\end{figure}

\begin{itemize}
\item
We have generated correlators for heavy-light pseudoscalar mesons at multiple $\kappa$
values and with light-quark masses bracketing the tuned strange quark
mass on the ensemble with $\sealight/\seaheavy = 0.2$ at each of the
lattice spacings $a\approx 0.045$, $0.06$, $0.09$, $0.12$ and
$0.15\,\fm$.  The charm- and bottom-quark mass regions are bracketed with at least three $\kappa$ values each.  In general, the available two-point data are a mix
of results from $\kappa$ tuning only production runs and results
from full analysis production runs. 
\item
Ground-state energies $aE(a\bm{p})$ for a range of $a\bm{p}$ were
determined by (constrained) chi-square minimization fits including
local-local, smeared-local and smeared-smeared (source-sink) two-point
functions.
\item
The energies $E$ are fit to the dispersion relation in
Eq.~(\ref{eq:disprel}) in constrained chi-square minimizations
using prior distributions for the coefficients $M_1/M_2$, $A_4$ and
$A_{4'}$ motivated by the tree-level dispersion relation for a
clover heavy quark with estimated corrections for binding energy
effects in a heavy-light meson \cite{Bazavov:2011aa}.
\item
We linearly adjust each meson kinetic mass
\begin{equation}
    M_2\left(m_q\right) = M_2\left(m_s\right) + C_v \left(m_q - m_s\right)/m_s
\end{equation}
to get the value corresponding to the physical valence strange quark $m_q=m_s$
listed in Table~\ref{tab:physMass} for each ensemble.
The $C_v$ are determined either by interpolation of the lattice results
or estimated from the experimental meson masses and the physical quark masses.
\item
On the asqtad ensembles, the mass $\seaheavy$ of the heaviest sea-quark flavor
can differ significantly from the physical strange quark mass, $m_s$.
We correct linearly for this sea-quark mass variation:
\begin{equation}
    M_2\left(\sealight,\seaheavy\right) = M_2\left(\hat{m},m_s\right) + 
        C_s \left(2\ratlight + \ratheavy\right)
    \label{eq:sea-qk-variation}
\end{equation}
where $M_2(\hat{m},m_s)$ is the meson mass in the limit of physical sea-quark masses,
$\ratlight=(\sealight-\hat{m})/m_s$, and $\ratheavy=(\seaheavy-m_s)/m_s$.
The average physical mass is $\hat{m}=(m_u+m_d)/2$ while $\sealight$ is the sea-quark mass used in
simulations.
We estimate $r_1C_s\approx 0.02$ for the $D_s$ and $r_1C_s\approx 0.012$ for the $B_s$ based upon an analysis
of the sea-quark mass dependence on the $a\approx0.12\,\fm$ lattice, and we take the physical mass $\hat{m}$
from Table~\ref{tab:physMass}.
\item
On each ensemble, the lattice masses $M_2(\kappa_h,m_s)$, adjusted to
the correct (valence and sea) strange quark mass, must be fit to an
interpolating function prior to implicitly solving for the $\kappa$ value
needed to match the lattice $M_2$ to the experimental value of the
$D_s$ or $B_s$ meson mass.  We have tested two different interpolating
functions, finding negligible difference in the resulting tuned $\kappa$ values.
For the first fit-function we use the same HQET-inspired form as in our previous tuning analyses:
\begin{equation}
M_2(\kappa) = \Lambda + m_2(\kappa) + \frac{\lambda_1}{m_2(\kappa)}
\end{equation}
where quark mass $m_2$ is computed to tree level. The parameters
$\Lambda$ and $\lambda_1$ are determined in a chi-square minimization.
The set of two parameters are determined separately for charm and bottom.
The second fit function is quadratic or linear in the tree-level bare quark mass $am_0$.
Again, the coefficients of the best fit are determined separately
for charm and bottom. In Fig.~\ref{fig:kappaInterepAnalyB} we show
examples of polynomial interpolations of $r_1 M_2(\kappa)$ and indicate
values corresponding to the known $D_s$ and $B_s$ masses. 
\item 
We use MILC's smoothed $r_1/a$ measurements and the value $r_1 =
0.3117(22)~\fm$ \cite{Bazavov:2011aa} to set the lattice spacing in
our determinations of $\kappa_c$ and $\kappa_b$.
\end{itemize}

\begin{figure}
    \centering
    \includegraphics[clip=true,width=0.97\textwidth]{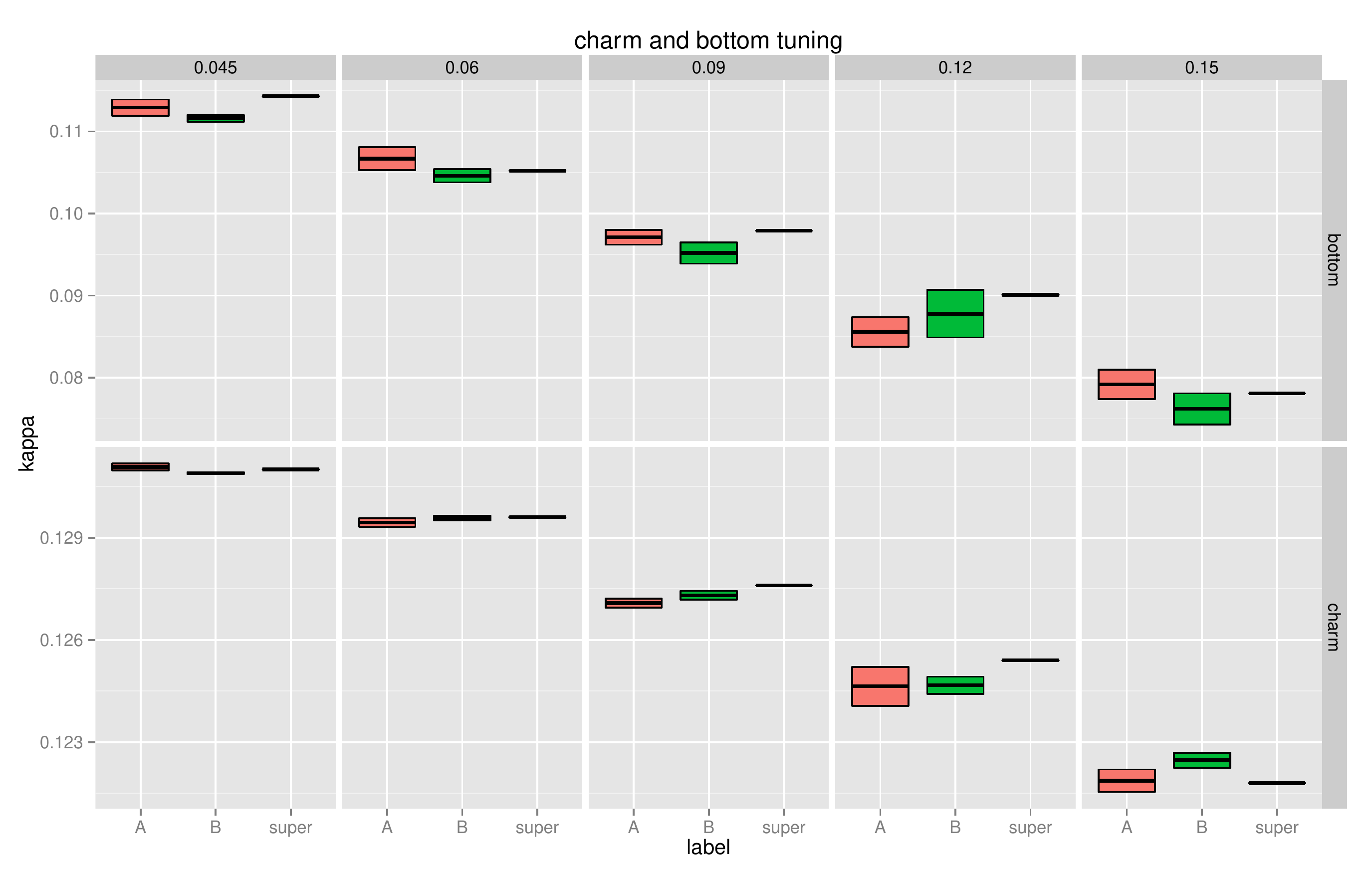}
    \caption{Comparison of charm and bottom $\kappa$ values from analyses A (red) and B(green) together with the $\kappa$ values used in the
        production campaign.}
    \label{fig:KappaDR}
\end{figure}

The process outlined above is used in two separate analyses. Analysis~A
is based on the two-point functions listed in Table~\ref{tab:AkTuneEnsembles}
and a block-elimination jackknife with block sizes ranging from 5 to 32
to estimate statistical errors. Analysis~B uses
the two-point functions listed in Table~\ref{tab:BkTuneEnsembles} together
with a bootstrap procedure in the error analysis. For several
ensembles, Analysis~B adds two additional (a charm-like and a
bottom-like) $\kappa$ values from the two-points generated in our
full analysis campaign.  The rest and kinetic masses from the two different analyses for the five $\hat m' / m'_s =0.2$ ensembles with different lattice spacing are listed in Tables~\ref{tab:RestKinetic0p15}-\ref{tab:RestKinetic0p045}.
The charm and bottom $\kappa$ values obtained in the
two analyses are tabulated, with statistical errors, in
Table~\ref{tab:KappaDRsummary}. The table also shows a comparison of
$\kappa$ values obtained from the HQET-inspired interpolation versus an
interpolation quadratic in $m_0$.  The tabulated (quadratic) results
are plotted in Figure~\ref{fig:KappaDR} for comparison. The results from the two
analyses are statistically consistent (with highly correlated
statistical errors). We take a weighted average from the two analyses
(see Table~\ref{tab:KappaDRsummary}) and use the resulting charm and bottom
$\kappa$ values in subsequent steps of the analysis.

\begin{table}
    \centering
    \caption{Ensembles with sea-quark $\sealight/\seaheavy=0.2$ that are used in $\kappa$ tuning, smoothed values of 
        $r_1/a$ and the physical quark masses, $m_s$ and $\hat{m}=(m_u+m_d)/2$ obtained from the 
        analysis of the light spectrum and decay constants \protect\cite{Bazavov:2010hj}.}
    \label{tab:physMass}
    \begin{tabular}{lllllll}
    \hline \hline
    $\approx a$ (fm) &~$r_1/a$  &$\beta$ &$a\seaheavy$  & $a\sealight$ & $am_s$   &$a\hat{m}$ \\ \hline
    0.15             &2.221530  &6.572   &0.0484  &0.0097   &0.04185  &0.001508   \\
    0.12             &2.738591  &6.76    &0.05    &0.01     &0.03357  &0.001215   \\
    0.09             &3.788732  &7.09    &0.031   &0.0062   &0.02446  &0.0008922  \\
    0.06             &5.353063  &7.47    &0.018   &0.0036   &0.01751  &0.0006401  \\
    0.045            &7.208234  &7.81    &0.014   &0.0028   &0.01298  &0.0004742  \\
    \hline \hline
    \end{tabular}
\end{table}

\begin{table}
    \centering
    \caption{Analysis A of $\sealight/\seaheavy=0.2$ ensembles, configurations$\times$sources and two-point valence 
        masses and $\kappa$ values.
        In all cases we use local-local, smeared-local, and smeared-smeared (source-sink) two-point functions in 
        fits. 
        The number of states (+ opposite parity states) and time range fit are shown.
        Where three fit ranges are shown, the first refers to the smeared-smeared correlator, the second, the
        smeared-local correlator, and the third, the local-local correlator.
        Where only one range is shown, all three correlators are fit to the same range.
        Two-point functions with momenta $|\bm{p}|\le|2|2\pi/L$ are included in the analysis.}
    \label{tab:AkTuneEnsembles}
    \begin{tabular}{l  l  c  c  l l}
    \hline \hline
    $a$ (fm)   & cfgs$\times$srcs & $am_q$       & $\kappa$                 & states & $t$ range \\
    \hline
    0.15     & 631$\times$8  & 0.0387, 0.0484 & 0.070, 0.076, 0.080      & $2 + 2$ & $[5,17]$ \\
             &         & 0.0387, 0.0484 & 0.090, 0.100, 0.115      & $2 + 2$ & $[6,18]$ \\
             &	       & 0.0387, 0.0484 & 0.115, 0.122, 0.125      & $2 + 2$ & $[8,20]$ \\  [0.5em]
    0.12     & 2259$\times$4   & 0.340,  0.370  & 0.074, 0.086, 0.098      & $2 + 2$ & $[9, 16]$ \\
             &         & 0.340,  0.370  & 0.1175, 0.1200, 0.1225   & $2 + 2$ & $[11, 21]$ \\  [0.5em]
    0.09     & 1912$\times$8  & 0.0250, 0.0270 & 0.090, 0.092, 0.094      & $2 + 2$ & $[10,20]$ \\
             &         & 0.0250, 0.0270 & 0.1240, 0.1255, 0.1270   & $3 + 3$ & $[12,24]$ \\
             &         & 0.0261, 0.0310 & 0.1276, 0.979            & $3 + 3$ & $[12,20]$ \\  [0.5em]
    0.06     & 670$\times$4  & 0.0188         & 0.100, 0.106, 0.122      & $2 + 2$ & $[15, 31 ]$ \\
             &         & 0.0188         & 0.124, 0.127, 0.130      & $2 + 2$ & $[20, 30]; [24, 34]; [28, 38 ]$ \\  [0.5em]
    0.045    & 801$\times$4   & 0.130,  0.135  & 0.106, 0.111, 0.116      & $2 + 2$ & $[18, 36]$  \\
             &         & 0.130,  0.135  & 0.128                    & $2 + 2$ & $[19, 35]; [20, 36]; [20, 36]$  \\
             &         & 0.130,  0.135  & 0.130, 0.132             & $2 + 2$ & $[20, 36]$  \\
    \hline \hline
    \end{tabular}
\end{table}

\begin{table}
    \centering
    \caption{Analysis B of $\sealight/\seaheavy=0.2$ ensembles, configurations$\times$sources and two-point valence 
        masses and $\kappa$ values.
        In all cases we use local-local, smeared-local, and smeared-smeared (source-sink) 2-pt functions in 
        fits. 
        The number of states (+ opposite parity states) and time range fit are shown.
        Two-point functions with momenta $|\bm{p}|\le|3|2\pi/L$ are fit.}
    \label{tab:BkTuneEnsembles}
    \begin{tabular}{l  l  c c   l l}
    \hline \hline
    $a$ (fm) & cfgs $\times$ srcs & $am_q$ & $\kappa$ & states & $t$ range \\
    \hline
    0.15   & 631 $\times$ 8   &  0.0484        & 0.070, 0.076, 0.080              &$3+3$ & $[6,22]$ \\
           &                  &  0.0484        & 0.085, 0.090,  0.094,  0.110     &$3+3$ & $[6,22]$ \\
           &                  &  0.0484        & 0.115, 0.122, 0.125              &$3+3$ & $[6,22]$ \\
           & 631 $\times$ 24  &  0.0484        & 0.0781, 0.1218                   &$3+3$ & $[6,22]$ \\  [0.5em]
    0.12   & 2259 $\times$ 4  &  0.349         & 0.0820, 0.0860, 0.0901           &$3+3$ & $[6,24]$ \\
           &                  &  0.349         & 0.1230, 0.1254, 0.1280           &$3+3$ & $[6,28]$ \\  [0.5em]
    0.09   & 1912 $\times$ 8  &  0.0270        & 0.090, 0.092, 0.094              &$3+3$ & $[12,36]$ \\
           & 1931 $\times$ 4  &  0.0261        & 0.0979                           &$3+3$ & $[12,36]$ \\
           & 1912 $\times$ 8  &  0.0270        & 0.1240, 0.1255, 0.1270           &$3+3$ & $[12,40]$ \\
           & 1931 $\times$ 4  &  0.0261        & 0.1276                           &$3+3$ & $[12,40]$ \\  [0.5em]
    0.06   & 670 $\times$ 4   &  0.0188        & 0.100, 0.106, 0.122              &$3+3$ & $[26,48]$ \\
           & 673 $\times$ 8   &  0.0188        & 0.1052                           &$3+3$ & $[26,48]$ \\
           & 670 $\times$ 4   &  0.0188        & 0.124, 0.127, 0.130              &$3+3$ & $[22,52]$ \\
           & 673 $\times$ 8   &  0.0188        & 0.1296                           &$3+3$ & $[22,52]$ \\  [0.5em]
    0.045  & 801 $\times$ 4   &  0.130         & 0.106, 0.111, 0.1143, 0.116      &$3+3$ & $[19,60]$  \\
           &                  &  0.130         & 0.128, 0.130, 0.1310, 0.132      &$3+3$ & $[19,70]$  \\
    \hline \hline
    \end{tabular}
\end{table}

\input tablesM1M2

\begin{table}
    \centering
\caption{Charm ($\kappa_c$) and bottom ($\kappa_b$) results from an analysis of the
  energy-momentum dispersion relation on the $\sealight/\seaheavy=0.2$ ensembles.
  Results from analyses A and B are listed with statistical errors.
  Under analysis A we tabulate $\kappa$ values found using an
  HQET-inspired interpolating function.  The weighted average of A:poly and
  B:poly results and the results after the smoothing fit are listed.  The
  third column shows $\kappa$ values used in the production campaign.}
\label{tab:KappaDRsummary}
\begin{tabular}{lllllllll}\hline\hline
   a (fm)  &system  &production   & A:HQET        & A:poly   & B:poly        & wt. avg.     & smoothed \\ 
\hline    
   0.15    &charm   &0.1218        &0.12210(30)  &0.12187(33) &0.12247(22)  & 0.12229(26)  & 0.12237(26) \\
   0.12    &        &0.1254        &0.12452(47)  &0.12464(57) &0.12467(25)  & 0.12467(32)  & 0.12423(15) \\
   0.09    &        &0.1276        &0.12721(14)  &0.12708(13) &0.12731(13)  & 0.12720(13)  & 0.12722(9)  \\
   0.06    &        &0.1296        &0.12959(12)  &0.12944(13) &0.12957(07)  & 0.12954(09)  & 0.12960(4)  \\
   0.045   &        &0.1310        &0.13124(10)  &0.13107(10) &0.13089(03)  & 0.13090(04)  & 0.130921(16)\\
   &&&&&&& \\                                                                 
   0.15    &bottom  &0.0781        &0.0803(11)   &0.0792(18)  &0.0762(19)   & 0.0778(18)   & 0.0775(16)  \\
   0.12    &        &0.0901        &0.0864(12)   &0.0856(18)  &0.0878(29)   & 0.0862(22)   & 0.0868(9)   \\
   0.09    &        &0.0979        &0.0971( 8)   &0.0971(09)  &0.0952(13)   & 0.0965(10)   & 0.0967(7)   \\
   0.06    &        &0.1052        &0.1064(15)   &0.1067(14)  &0.1046(08)   & 0.1051(10)   & 0.1052(5)   \\
   0.045   &        &0.1143        &0.1125(10)   &0.1129(10)  &0.1116(04)   & 0.1118(05)   & 0.1116(3)   \\ 
\hline\hline
\end{tabular}
\end{table}

\begin{table}
    \centering
    \caption{Rest masses of $D_s$ and $B_s$ interpolated to the
      respective weighted average $\kappa_c$ and $\kappa_b$ values for
      each of the $\sealight/\seaheavy = 0.2$ ensembles in the tuning
      sample. $\sigma_{\rm tune}$ is the error propagated from the
      uncertainty in the weighted averages of $\kappa_c$ and
      $\kappa_b$ and $\delta M_{\rm sea}$ is the correction applied to
      the rest-masses due to the extrapolation to the physical
      sea-quark masses.  The rest-masses listed in columns 2 and 5 are
      the results obtained before applying the sea-quark mass correction.
      Masses are in MeV.}
    \label{tab:M1-vs-a}
    \begin{tabular}{llrrlrr}
    \hline\hline
    $a/r_1$ &  $M_1(D_s)$ & $\sigma_\text{tune}$ &  $\delta M_\text{sea}$ & 
        $M_1(B_s)$ & $\sigma_\text{tune}$ &  $\delta M_\text{sea}$ \\
    \hline
       0.4501 & 1721.2 & 10.4   & $-11.0$ & 3190.0 & 57.2 &  $-6.7$ \\
       0.3652 & 1779.9 & 16.9   & $-20.7$ & 3411.9 & 87.0 & $-12.6$ \\  
       0.2639 & 1878.7 & 10.5   & $-14.1$ & 3801.2 & 58.5 &  $-8.6$ \\  
       0.1868 & 1927.5 & 11.3   &  $-7.3$ & 4280.3 & 82.8 &  $-4.4$ \\  
       0.1387 & 1950.2 &  7.4   &  $-9.0$ & 4622.1 & 59.1 &  $-5.5$ \\  
      \hline\hline
  \end{tabular}
\end{table}

\subsection{Smoothing and extending $\kappa$ tuning to other ensembles}
\label{sec:smooth}

In the second step of our $\kappa$ tuning analysis we improve the raw tuned results by smoothing them as a function of lattice
spacing and by adding the constraint that the {\em rest} masses $M_1$
extrapolate to their physical values at zero lattice spacing.  This
treatment gives the small adjustments in the central values and the
reduction in error, shown in the last column of Table~\ref{tab:KappaDRsummary}.
The improvement in error gets progressively
better as the lattice spacing is decreased.

The continuum extrapolation of the rest masses $M_1$ adds a useful constraint to the $\kappa$ tuning analysis, since the rest masses are determined to much higher statistical
accuracy than the kinetic masses $M_2$.  On each ensemble with fixed
lattice spacing $a/r_1$, their dependence on heavy valence quark
$\kappa$ can be described accurately with an interpolating function
$M_1(\kappa,a/r_1)$, which we take to be quadratic in the bare heavy-quark 
mass and which we determine separately for charm-like and bottom-like
masses.  Thus, on each ensemble, a tuned value of $\kappa$ and its
error implies, through interpolation, an inferred value of
$M_1(a/r_1)$ with appropriately propagated error.  (The errors from
the interpolation were negligible compared with the errors arising
from uncertainties in the tuned values of $\kappa$ themselves.)  The
inferred rest masses are shown in Table~\ref{tab:M1-vs-a} for the
$D_s$ and $B_s$ on the ensembles with $\sealight/\seaheavy=0.2$, and are 
uncorrected for unphysical sea-quark masses.  We determine
the sea-quark mass correction following
Eq.~(\ref{eq:sea-qk-variation}), but with a coefficient $C_s^\prime$
appropriate for the rest mass. The resulting sea-quark-mass correction
is shown in Table~\ref{tab:M1-vs-a}. Our smoothing procedure then fits
the inferred, adjusted values of $M_1(a/r_1)$ to a smooth function of
lattice spacing $a/r_1$, with the constraint that the intercept
$M_1(0)$ agrees with the physical mass.

\begin{figure}
    \centering
    \includegraphics[clip=true,width=0.49\textwidth]{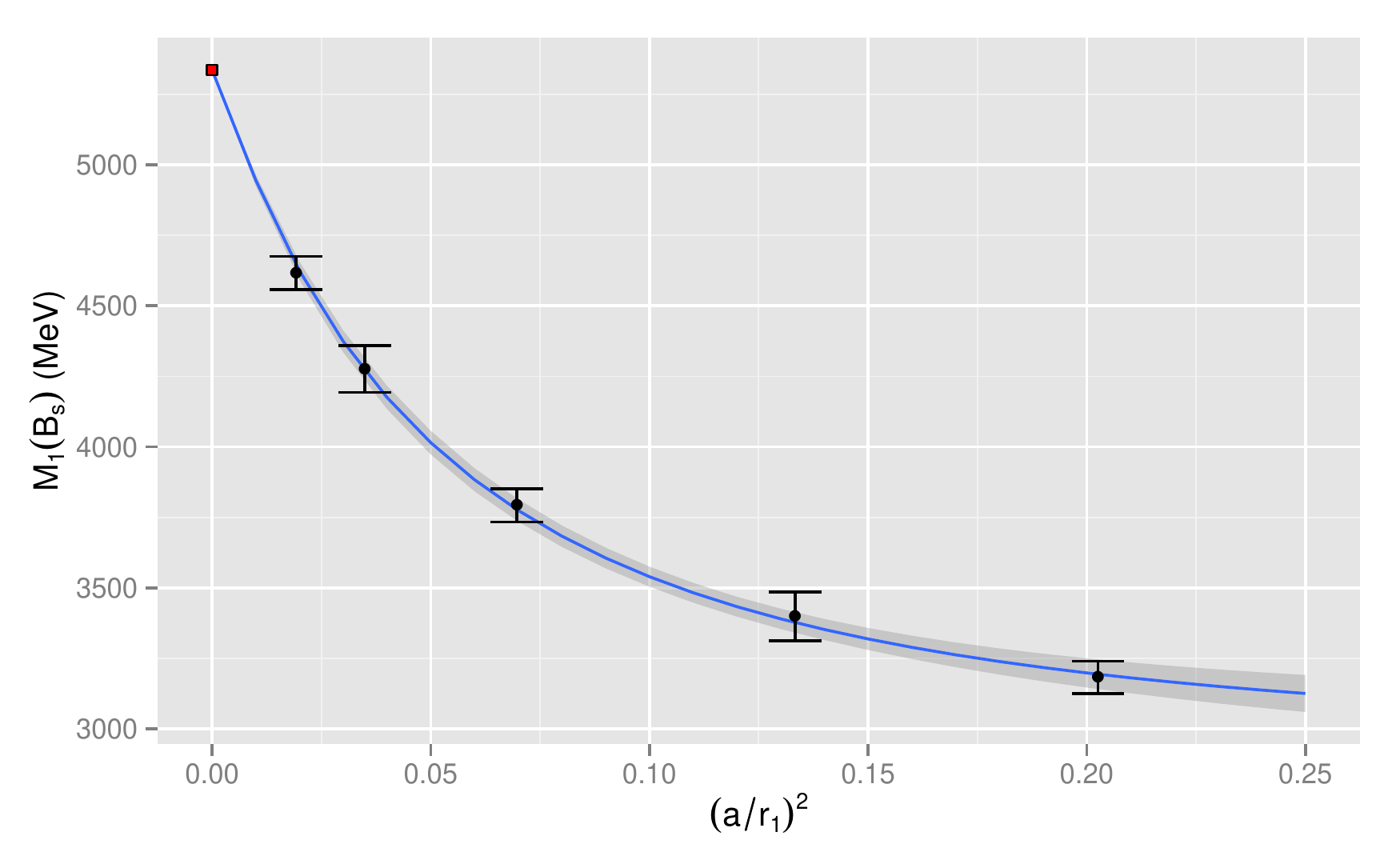} \hfill
    \includegraphics[clip=true,width=0.49\textwidth]{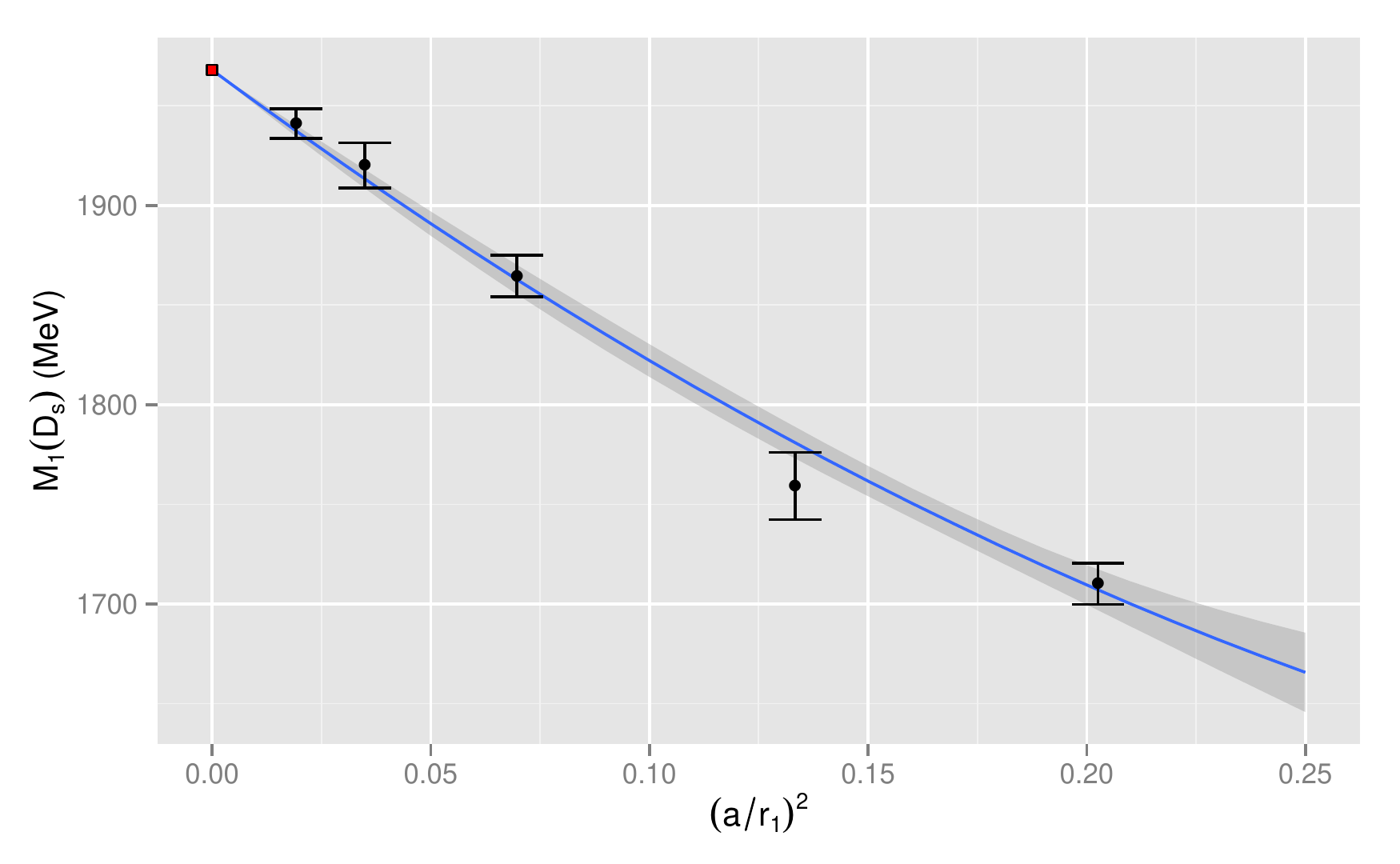} 
    \caption{Lattice rest masses (in MeV) for the $B_s$ (left) and $D_s$ (right) at physical valence and sea-quark 
        masses as a function of $(a/r_1)^2$.
    \label{fig:M1-vs-a}}
\end{figure}

For the $B_s$ we use the empirically chosen form
\begin{equation}
  M_1(a^2;B_s) = M(B_s)_\text{phys} + b_1 x + b_2 x^2
\end{equation}
where $x = (a/r_1)^2/[0.1 + (a/r_1)^2]$.  In units of the physical
$B_s$ meson mass $M$ this parameter becomes $x = (aM)^2/[7.3 +
  (aM)^2]$, which reduces the model to a quadratic in $a^2$ for $aM \ll
3$.  The resulting fit is shown in the left panel of Fig.~\ref{fig:M1-vs-a}
($\chi^2/\text{d.o.f.} = 0.4/3$, $p = 0.94$).

For charm-like masses, evidently, the value of $aM(D_s)$ is sufficiently 
small that a simple quadratic in $(a/r_1)^2$ suffices:
\begin{equation}
  M_1(a^2;D_s) = M(D_s)_\text{phys} + c_1 \,\frac{a^2}{r_1^2} + c_2 \,\frac{a^4}{r_1^4}
\end{equation}
The resulting fits are shown in the right panel of
Fig.~\ref{fig:M1-vs-a} ($\chi^2/\text{d.o.f.} = 2.4/3$, $p=0.49$).

We then use the best fits to determine the smoothed values of $M_1$ at
each lattice spacing.  Ensemble by ensemble, through the valence quark
mass interpolation, these smoothed values, in turn, provide the
smoothed $\kappa$s for each $0.2 \seaheavy$ ensemble. They are
recorded in Table~\ref{tab:KappaDRsummary}. 

Finally, we need to extend our determination of $\kappa_c$ and
$\kappa_b$ to predict their values for ensembles with values of
$\sealight/\seaheavy$ other than 0.2.  Because we are using a
mass-independent scheme, we interpolate only in $\beta$, where we note
that the variation of $\beta$ with sea quark mass (at approximately
constant lattice spacing) is very slight.  Because $\kappa_c$ and
$\kappa_b$ are tuned to masses adjusted to the physical sea-quark
masses, this mass-independent scheme is based on physical hadron
($\pi$, $K$, $D_s$, and $B_s$) masses and physical $f_\pi$ at all
lattice spacings.  To predict the $\kappa$ values at other $\beta$'s the
functions $\kappa_c(\beta)$ and $\kappa_b(\beta)$ are fit to a cubic
spline.  The spline is used {\it only} to determine the derivatives
$d\kappa_c/d\beta$ and $d\kappa_b/d\beta$ at the $\beta_i$'s for the
five $0.2 \seaheavy$ ensembles.  The derivatives are, in turn, used
to obtain $\kappa_c$ and $\kappa_b$ at the slightly shifted $\beta$
values for each of the four lattice spacings where we need them.  The
results are the final smoothed $\kappa$ values listed in
Table~\ref{tab:predict}.

\subsection{Scale error}
\label{sec:scale}

As noted above, we take $r_1 = 0.3117(22)~\fm$~\cite{Bazavov:2011aa}.  The error in the scale
determination introduces an error in converting the experimental mass to
$aM_2$, which propagates, in turn, to the tuned $\kappa$s.  The systematic error on the 
tuned $\kappa$s due to the uncertainty in the lattice scale determination is therefore 
obtained by changing $r_1$ from its central value by one standard deviation and propagating 
this change through our $\kappa$ tuning analysis.  Our final tuned $\kappa_b$ and 
$\kappa_c$ results including both statistical and $r_1$ systematic errors 
are shown in Table~\ref{tab:predict}.  We note that the derivative of the $\kappa$s with respect
to $r_1$ is negative.  So increasing $r_1$ by 0.0022 causes $\kappa$
to decrease by the amount shown. This exercise was done only on the
$0.2 \seaheavy$ ensembles.  We assume that the errors are the same for
ensembles at nearby $\beta$ (nearly same lattice spacing).

The correct way to propagate the scale error to the dimensionful quantities 
that we calculate is first to compute the
physical quantity for a fixed $r_1$, propagating only the statistical
error in $\kappa$ (\ie, not first combining statistical and
scale errors in some way), and then to recompute the same quantity
with the shifted $\kappa$ and shifted $r_1$.  The difference in the
central values of the final result is, then, the $r_1$ systematic
error.

\begin{table}
    \centering
    \caption{Final, smoothed mass-independent $\kappa_c$ and $\kappa_b$ values and production values for 
        various ensembles.
        The second error reflects the uncertainty in the $r_1$ determination.
        See Sec.~\protect\ref{sec:scale} for the preferred way to handle it.
        Smoothing is discussed in Sec.~\ref{sec:smooth}.}
    \begin{tabular}{lllllcc}
    \hline\hline
    \multicolumn{3}{c}{Ensemble}  & \multicolumn{2}{c}{Tuned (final)} & \multicolumn{2}{c}{Production} \\
    $\approx a$ (fm) & ~~$\beta$ & \multicolumn{1}{c}{$\sealight/\seaheavy$} &  \multicolumn{1}{c}{$\kappa_c$} & 
        \multicolumn{1}{c}{$\kappa_b$} & $\kappa_c$  &  $\kappa_b$ \\
    \hline
    0.15 & 6.566 & 0.1  		  &   0.12231(26)(20) &  0.0772(16)(3) & --    & --    \\
         & 6.572 & 0.2  		  &   0.12237(26)(20) &  0.0775(16)(3) & 0.1218 & 0.0781  \\
         & 6.586 & 0.4  		  &   0.12252(26)(20) &  0.0780(16)(3) & --    & --     \\
    \hline                                                    
    0.12 & 6.76  & 0.1            &   0.12423(15)(16) &  0.0868(9)(3) & 0.1254 & 0.0901  \\
         & 6.76  & 0.14           &   0.12423(15)(16) &  0.0868(9)(3) & 0.1254 & 0.0901  \\
         & 6.76  & 0.2            &   0.12423(15)(16) &  0.0868(9)(3) & 0.1254 & 0.0901  \\
         & 6.79  & 0.4  		  &   0.12452(15)(16) &  0.0879(9)(3) & 0.1259 & 0.0918  \\
    \hline                                                    
    0.09 & 7.075 & 0.05  		  &   0.12710(9)(14) &  0.0964(7)(3) & 0.1275 & 0.0976 \\
         & 7.08  & 0.1  		  &   0.12714(9)(14) &  0.0965(7)(3) & 0.1275 & 0.0976 \\
         & 7.085 & 0.14  		  &   0.12718(9)(14) &  0.0966(7)(3) & 0.1275 & 0.0977 \\
         & 7.09  & 0.2  		  &   0.12722(9)(14) &  0.0967(7)(3) & 0.1276 & 0.0979 \\
         & 7.10  & 0.3  		  &   0.12730(9)(14) &  0.0970(7)(3) & --     & --     \\
         & 7.11  & 0.4  		  &   0.12737(9)(14) &  0.0972(7)(3) & 0.1277 & 0.0982 \\
    \hline
    0.06 & 7.46  & 0.1  		  &   0.12955(4)(11) &  0.1050(5)(2)  & 0.1296 & 0.1052 \\
         & 7.465 & 0.14  		  &   0.12957(4)(11) &  0.1051(5)(2)  & 0.1296 & 0.1052 \\
         & 7.47  & 0.2  		  &   0.12960(4)(11) &  0.1052(5)(2)  & 0.1296 & 0.1052 \\
         & 7.475 & 0.3  		  &   0.12962(4)(11) &  0.1052(5)(2)  & --     & --     \\
         & 7.48  & 0.4  		  &   0.12964(4)(11) &  0.1054(5)(2)  & 0.1295 & 0.1048 \\
    \hline                                                    
    0.045 & 7.81  & 0.2  		  & 0.130921(16)(70) &  0.1116(3)(2)  & 0.1310 & 0.1143 \\
    \hline\hline
  \end{tabular}
\label{tab:predict}
\end{table}

\subsection{\boldmath$D^*_s$-$D_s$ and $B^*_s$-$B_s$ hyperfine splittings}
\label{app:hfs}

\begin{figure}[b]
    \includegraphics[clip=true,width=0.49\textwidth]{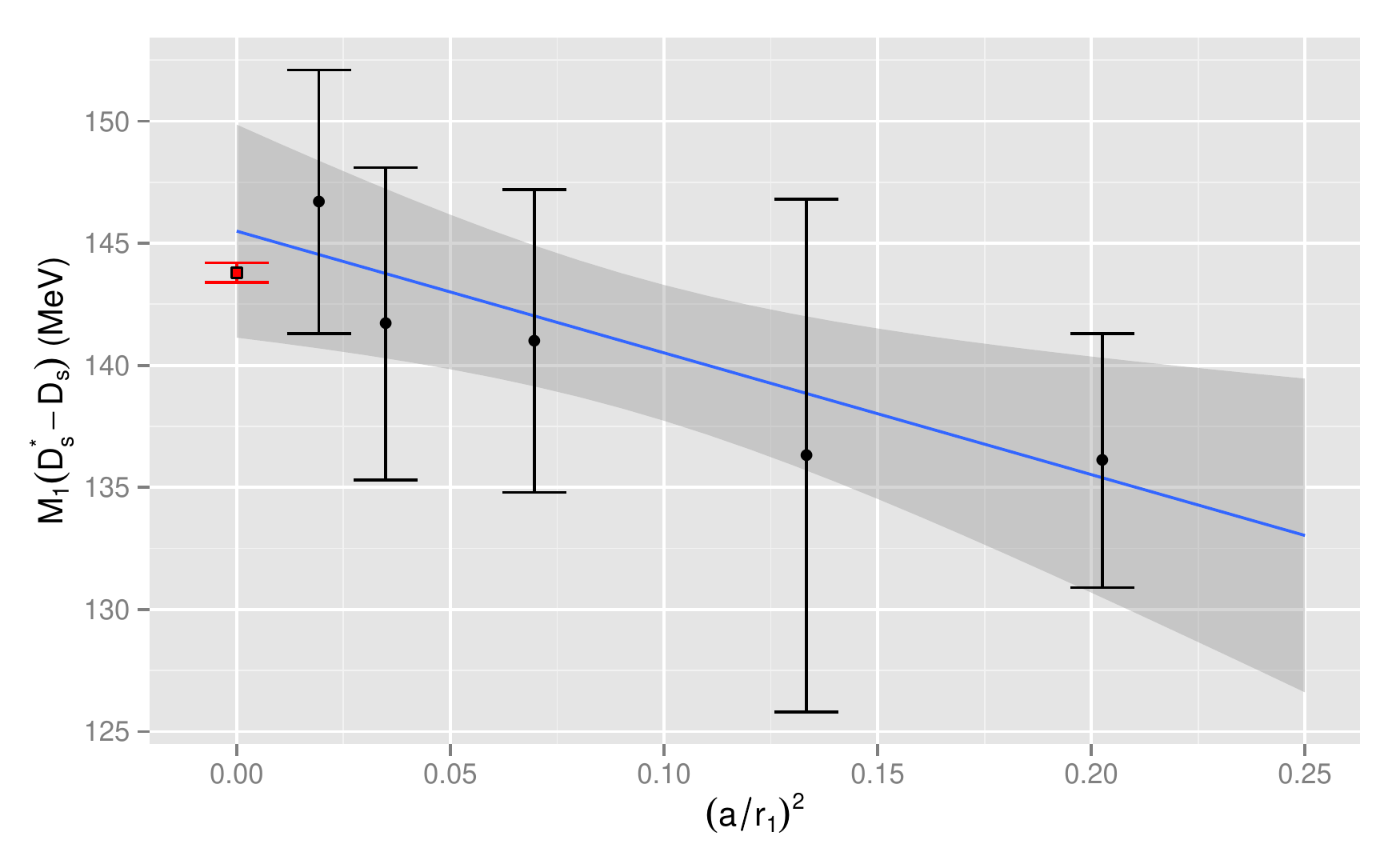} \hfill
    \includegraphics[clip=true,width=0.49\textwidth]{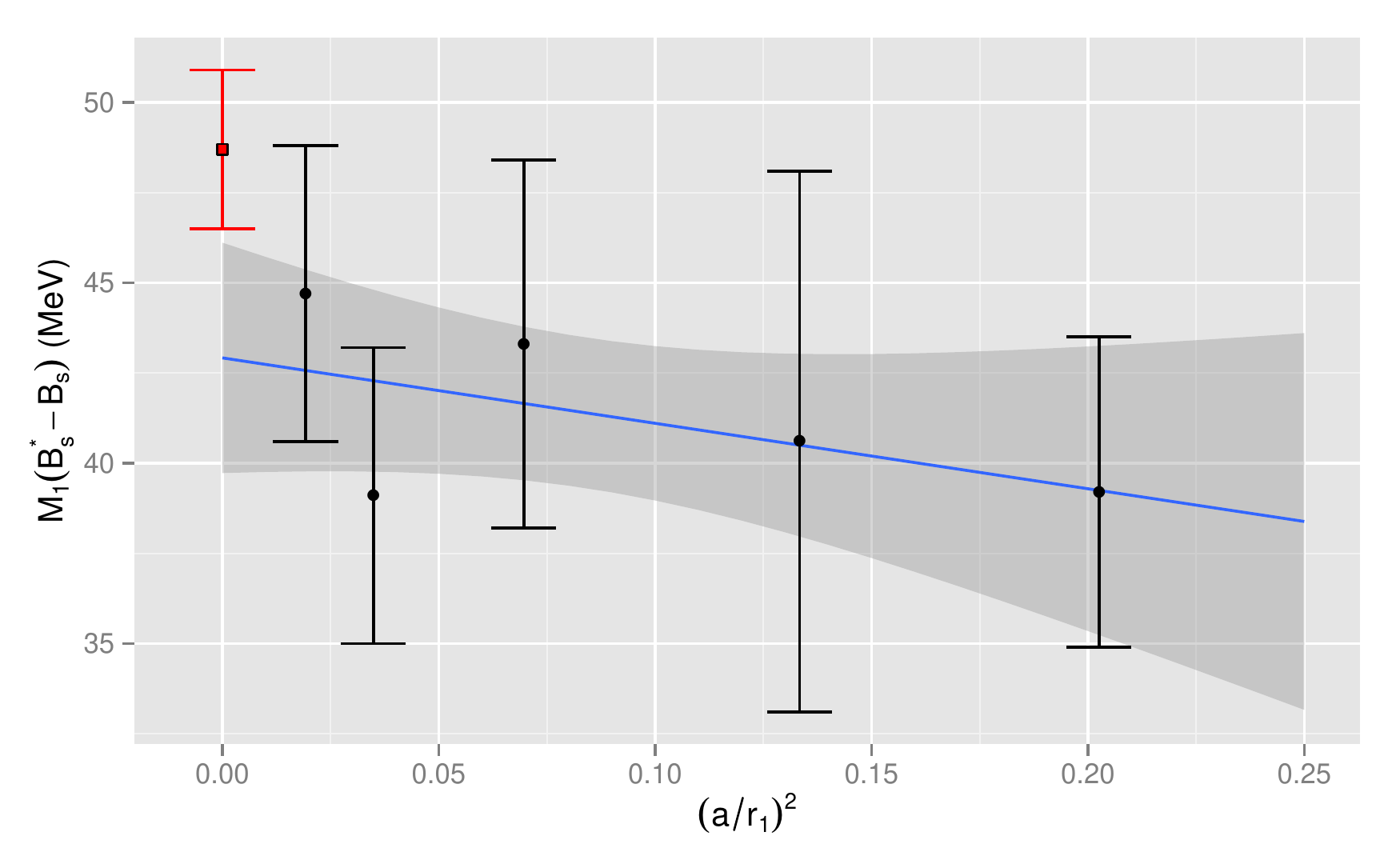} \\
    \includegraphics[clip=true,width=0.49\textwidth]{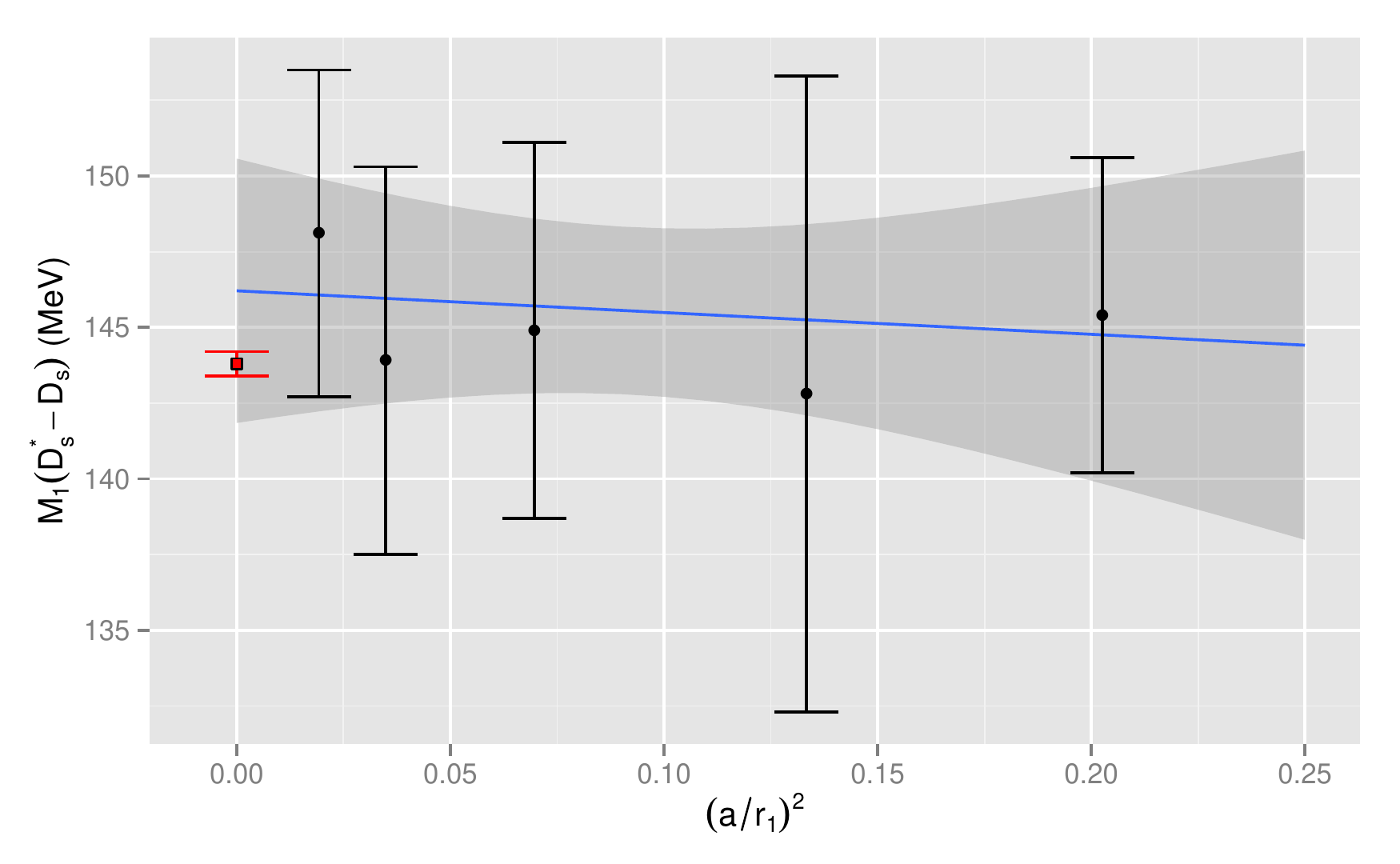} \hfill
    \includegraphics[clip=true,width=0.49\textwidth]{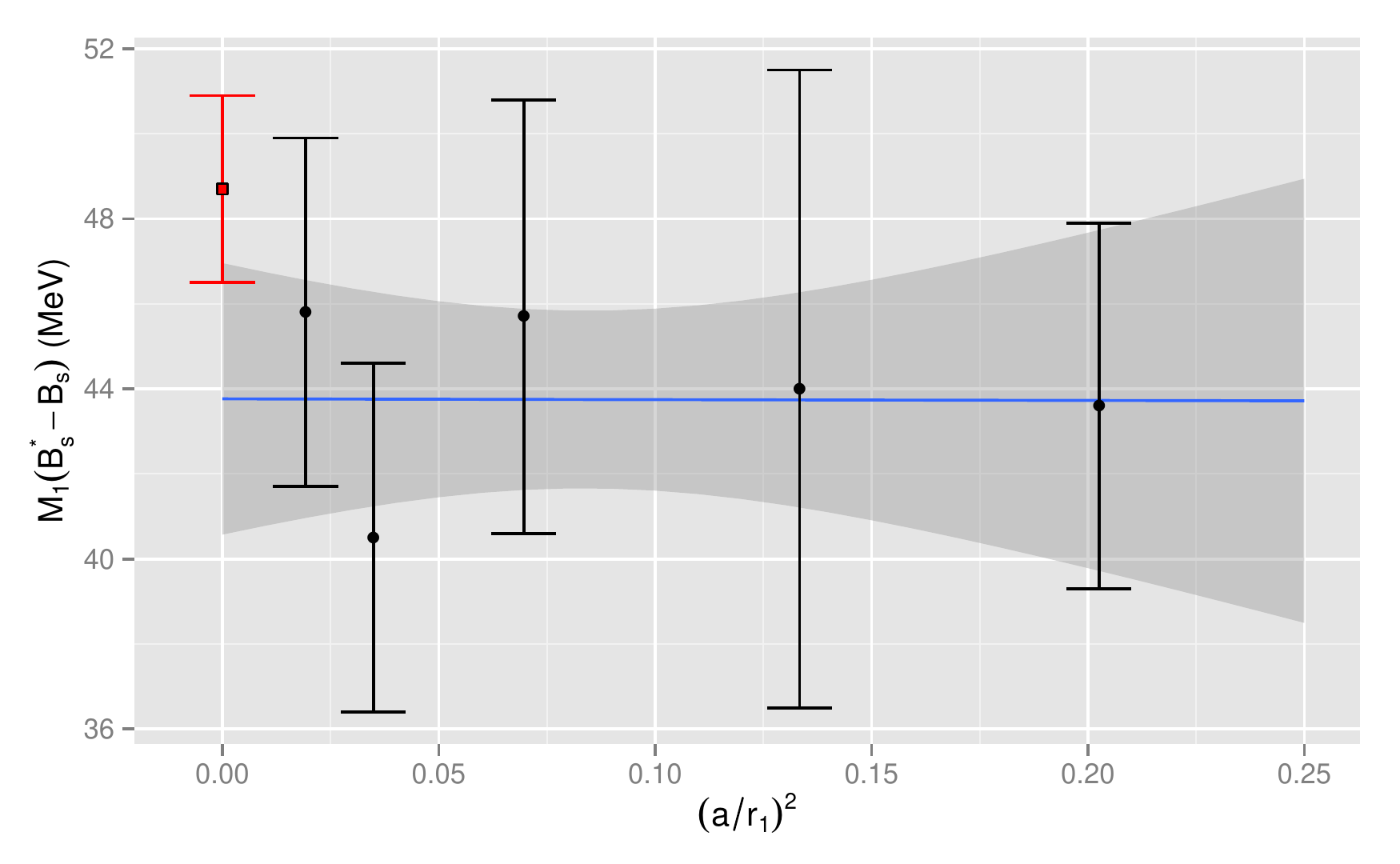} \\
    \caption{Hyperfine splittings for the $D_s$ (left) and $B_s$ (right) systems in MeV, shown with full 
    errors, extrapolated (linearly in $(a/r_1)^2$) to zero lattice spacing.
    Experimental values are indicated by the (red) points at $a=0$.
    Upper panels: before correction for leading heavy-quark discretization error.
    Lower panels: after correction.}
    \label{fig:DBhfs}
\end{figure}

The hyperfine splittings, $M(D_s^*)-M(D_s)$ and $M(B_s^*)-M(B_s)$, are
sensitive to the heavy-quark mass and to discretization effects, and
they therefore provide a good test of both our analysis of
discretization errors and of our $\kappa$-tuning analysis.  As with
the pseudoscalar mesons $D_s$ and $B_s$, we made sea-quark mass adjustments for the
vector mesons $D^*_s$ and $B^*_s$, as discussed above.  We computed
the hyperfine splitting at the physical strange quark mass over a
range of valence $\kappa$ values.  For purposes of interpolation we fit
the rest-mass splitting on each ensemble as a quadratic in $1/(am_0)$,
the inverse bare quark mass.  This fitting function works well over
the entire range of valence $\kappa$s from charm to bottom.  After
interpolation we apply a correction for heavy-quark discretization
errors to leading order in heavy-quark effective theory as described
in \cite{Bernard:2010fr}.  The resulting values are listed in
Tables~\ref{tab:Dshfs} and \ref{tab:Bshfs} and are shown in
Fig.~\ref{fig:DBhfs}. An error budget is also tabulated.  For the
remaining heavy-quark discretization error (beyond leading order), 
we used the full leading-order correction at 0.06~fm.  Error contributions are
combined in quadrature.  Our results for the splittings are extrapolated three
ways to zero lattice spacing:  The values corrected for heavy-quark
discretization errors are extrapolated linearly in $(a/r_1)^2$.
The uncorrected values are similarly extrapolated.  The corrected
values are simply averaged (extrapolated with slope fixed to zero).
All results are consistent.  They are compared with the experimental
values given in the last line of each table \cite{Beringer:1900zz}.  
The largest uncertainty comes from the adjustment
from the simulation sea-quark masses to the physical sea-quark masses.
For the $D_s$ hyperfine splitting, the prediction is well within $1 \sigma$ of the experimental value, and 
for the $B_s$, about $1.3\sigma$ below (lower panels of Fig.~\ref{fig:DBhfs}).  Without the
leading heavy-quark correction, the extrapolated result for the $D_s$
splitting is also well within $1\sigma$ of the experimental value and for
the $B_s$, slightly more than $1.3\sigma$ below (upper panels), but
the extrapolation model (linear in $(a/r_1)^2$) is then less reliable.

\begin{table}[t]
  \caption{Hyperfine splitting $\Delta M_\text{hfs}(D_s)=M_{D^*_s}-M_{D_s}$ in MeV at the physical
    valence- and sea-quark masses as a function of $a/r_1$.  The 
    splittings shown in the second column include a correction for 
    heavy-quark discretization errors to leading 
    order~\cite{Bernard:2010fr}, while the third column shows the
    uncorrected value.  The remaining columns give the error budget.
    Shown are the fit error, charm mass tuning error, sea-quark mass
    adjustment, the combination (in quadrature) of these three sources of
    statistical error, the systematic scale error, the systematic
    heavy-quark discretization error, and the combination (in quadrature)
    of the statistical and systematic errors. 
    The three rows at zero lattice spacing give, respectively,
    the value obtained by linear extrapolation of the corrected splittings in $(a/r_1)^2$,
    by similarly extrapolating the uncorrected splittings, and by taking the mean of the corrected splittings,
    and the last row gives the experimental value.}
\label{tab:Dshfs}
\begin{tabular}{cccccccccr}
\hline\hline
  $a/r_1$  & $\Delta M_\text{hfs}(D_s)$ & uncorrected & fit  & tune & sea quark & net stat. & $r_1$ scale & hvy.\ qk. & total \\
\hline
0.4501 & 145.4 & 136.1 & 1.4 & 1.0 & 4.1 &   4.5 & 2.1 & 1.8 &  5.2 \\
0.3652 & 142.8 & 136.3 & 6.6 & 0.9 & 7.7 &  10.1 & 1.8 & 1.8 & 10.5 \\
0.2639 & 144.9 & 141.0 & 1.9 & 0.7 & 5.2 &   5.6 & 1.9 & 1.8 &  6.2 \\
0.1868 & 143.9 & 141.7 & 3.4 & 3.7 & 2.8 &   5.8 & 2.0 & 1.8 &  6.4 \\
0.1387 & 148.1 & 146.7 & 2.3 & 2.3 & 3.4 &   4.7 & 2.1 & 1.8 &  5.4 \\
\hline
  0.0000 & 146(4) & corrected \\
  0.0000 & 145(4) & uncorrected \\
  0.0000 & 146(3) & mean \\
\hline
  expt   & 143.8(4) \\
\hline\hline
\end{tabular}
\end{table}

\begin{table}
  \caption{The same as Table~\ref{tab:Dshfs}, but for the hyperfine splitting 
  $\Delta M_\text{hfs}(B_s)=M_{B^*_s}-M_{B_s}$.}
\label{tab:Bshfs}
\begin{tabular}{cccccccccc}
\hline\hline
  $a/r_1$  & $\Delta M_\text{hfs}(B_s)$ & uncorrected & fit  & tune & sea quark & net stat. & $r_1$ scale & hvy.\ qk.& total \\
\hline
0.4501 & 43.6 & 39.2 & 1.4 & 1.3 &  3.5 &   4.0 &  0.6 & 1.4 & 4.3 \\
0.3652 & 44.0 & 40.6 & 2.8 & 0.9 &  6.8 &   7.4 &  0.6 & 1.4 & 7.5 \\
0.2639 & 45.7 & 43.3 & 1.2 & 0.7 &  4.7 &   4.9 &  0.7 & 1.4 & 5.1 \\
0.1868 & 40.5 & 39.1 & 2.8 & 0.7 &  2.5 &   3.8 &  0.6 & 1.4 & 4.1 \\
0.1387 & 45.8 & 44.7 & 2.4 & 0.5 &  3.0 &   3.9 &  0.6 & 1.4 & 4.1 \\
\hline
  0.0000 & 44(3) & corrected \\
  0.0000 & 43(3) & uncorrected \\
  0.0000 & 44(2) & mean \\
\hline
  expt   & $48.7^{+2.3}_{-2.1}$ \\
\hline\hline
\end{tabular}
\end{table}

%


\bibliography{SuperBib}

\end{document}

%% file: 450f.tex
\begin{table}[tp]
    \centering
    \caption[tbl:diffs]{Absolute difference of $h_{A_1}(1)$ from mismatches in the heavy-quark
        Lagrangian and current, estimating HQET quantities $E$, $R_1$, $R_2$ with $\Lambda^2$, 
        $\Lambda=450$~MeV, and taking $\mu_\pi^2=0.424~\textrm{GeV}^2$ and $\mu_G^2=0.364~\textrm{GeV}^2$.
        To obtain the totals, we use three uncorrelated $f_B$ terms and two $f_X$.
        The total difference is estimated using the $a=0.09$~fm lattice as a baseline.
        The right-most column shows the difference in the data between $h_{A_1}(1)$ on a given lattice and 
        the value at $a\approx0.09$~fm, computed at $m_x=0.2\seaheavy$ as in Fig.~\ref{fig:aSq}.}
    \label{tbl:diffs}
    \begin{tabular}{cccccccc}
    \hline\hline
    $a$~(fm) & $\alpha_V(q^*)$ & $m_{0b}a$ & $m_{0c}a$ &  $B$  &  $X$   & Total &    Data    \\
    \hline
     0.15\hs &       0.340     &   3.211   &   0.699   & 0.020 & 0.0102 & 0.022 & 0.0072(81) \\ 
     0.12\hs &       0.300     &   2.462   &   0.532   & 0.009 & 0.0044 & 0.010 & 0.0087(71) \\ 
     0.09\hs &       0.261     &   1.664   &   0.362   &   --  &   --   &   --  &   --       \\ 
     0.06\hs &       0.220     &   1.123   &   0.240   & 0.003 & 0.0035 & 0.005 & 0.0033(82) \\ 
     0.045   &       0.198     &   0.808   &   0.176   & 0.004 & 0.0046 & 0.006 & 0.0042(69) \\ 
    \hline\hline
    \end{tabular}
\end{table}
%

%% file: 450.tex
%
\begin{table}[tp]
    \centering
    \caption[tbl:errors]{Absolute error on $h_{A_1}(1)$ from mismatches
        in the heavy-quark Lagrangian and current, estimating HQET
        quantities $E$, $R_1$, $R_2$ with $\Lambda^2$,
        $\Lambda=450$~MeV, and taking
        $\mu_\pi^2=0.424~\textrm{GeV}^2$ and
        $\mu_G^2=0.364~\textrm{GeV}^2$.
        To obtain the totals, we use three uncorrelated $f_B$ terms
        and two $f_X$.}
    \label{tbl:errors}
    \begin{tabular}{ccccccc}
    \hline\hline
    $a$~(fm) & $\alpha_V(q^*)$ & $m_{0b}a$ & $m_{0c}a$ & $B$ & $X$ & Total \\
    \hline
     0.150   &       0.340     & 3.211 & 0.699 & 0.020 & 0.016 & 0.026 \\ 
     0.120   &       0.300     & 2.462 & 0.532 & 0.017 & 0.011 & 0.020 \\ 
     0.090   &       0.261     & 1.664 & 0.362 & 0.014 & 0.006 & 0.016 \\ 
     0.060   &       0.220     & 1.123 & 0.240 & 0.009 & 0.003 & 0.010 \\ 
     0.045   &       0.198     & 0.808 & 0.176 & 0.007 & 0.001 & 0.007 \\ 
    \hline\hline
    \end{tabular}
\end{table}

%% file: tablesM1M2.tex

\begin{table}
\renewcommand{\arraystretch}{1.15}
\caption{Results for rest and kinetic masses (in lattice units) on the $a\approx 0.15\,\fm$ ensemble. }
\label{tab:RestKinetic0p15}
\begin{tabular}{l@{\quad}lll@{\quad}lll}
\hline\hline
     &\multicolumn{3}{c}{Analysis A}   &\multicolumn{3}{c}{Analysis B} \\
$\kappa$ &$am_q$ &$aM_1$ &$aM_2$ &$am_q$ &$aM_1$ &$aM_2$  \\ \hline
0.125  &0.04213 & 1.1459(7)  & 1.284(14)  &0.0484 &1.1566(6)  &1.295(15) \\
0.122  &        & 1.2324(9)  & 1.406(24)  &       &1.2427(7)  &1.419(18) \\
0.115  &        & 1.4182(10) & 1.717(22)  &       &1.4282(9)  &1.719(28) \\
0.110  &        & --         & --         &       &1.5515(10) &1.938(37) \\
0.100  &        & 1.7759(14) & 2.524(56)  &       &-- &-- \\
0.090  &        & 1.9991(19) & 3.165(101) &       &2.0077(18) &3.003(100) \\
0.085  &        & --         & --         &       &2.1181(21) &3.290(123) \\
0.080  &        & 2.2193(22) & 3.764(131) &       &2.2287(23) &3.629(156) \\
0.076  &        & 2.3087(24) & 4.077(155) &       &2.3182(26) &3.901(182) \\
0.070  &        & 2.4444(29) & 4.654(232) &       &-- &-- \\
\hline \hline
\end{tabular}
\end{table}


\begin{table}
\renewcommand{\arraystretch}{1.15}
\caption{Results for rest and kinetic masses (in lattice units) on the $a\approx 0.12\,\fm$ ensemble. }
\label{tab:RestKinetic0p12}
\begin{tabular}{l@{\quad}lll@{\quad}lll}
\hline\hline
     &\multicolumn{3}{c}{Analysis A}   &\multicolumn{3}{c}{Analysis B} \\
$\kappa$ &$am_q$ &$aM_1$ &$aM_2$ &$am_q$ &$aM_1$ &$aM_2$  \\ \hline
0.1280  &0.03357 &        & --      &0.0349 &0.9239(3) &1.008(8) \\
0.1254  &      &        & --     &      &1.0066(3) &1.120(10) \\
0.1230  &      &        & --     &      &1.0787(4) &1.223(13) \\
0.1225  &      & 1.0918(4)  & 1.228(15)  &      &-- &-- \\
0.1200  &      & 1.1628(4)  & 1.327(17)  &      &-- &-- \\
0.1175  &      & 1.2309(5)  & 1.429(23)  &      &-- &-- \\
0.0980  &      & 1.7040(9)  & 2.378(69)  &      &-- &-- \\
0.0901  &      & --         & --         &      &1.8837(11) &3.000(123) \\
0.0860  &      & 1.9728(16) & 3.064(137) &      &1.9760(11) &3.181(152) \\
0.0820  &      & --         & --         &      &2.0651(12) &3.404(197) \\
0.0740  &      & 2.2419(24) & 4.037(261) &      &-- &-- \\
\hline \hline
\end{tabular}
\end{table}



\begin{table}
\renewcommand{\arraystretch}{1.15}
\caption{Results for rest and kinetic masses (in lattice units) on the $a\approx 0.09\,\fm$ ensemble. }
\label{tab:RestKinetic0p09}
\begin{tabular}{l@{\quad}lll@{\quad}lll}
\hline\hline
     &\multicolumn{3}{c}{Analysis A}   &\multicolumn{3}{c}{Analysis B} \\
$\kappa$ &$am_q$ &$aM_1$ &$aM_2$ &$am_q$ &$aM_1$ &$aM_2$  \\ \hline
0.1276  &0.02468 & 0.7698(3)  &  0.798(6)  &0.0261 &0.7720(2) &0.810(7) \\
0.1270  &        & 0.7900(3)  &  0.842(9)  &0.0270 &0.7940(2) &0.844(5) \\
0.1255  &        & 0.8392(3)  &  0.895(8)  &      &0.8428(2) &0.907(7) \\
0.1240  &        & 0.8862(4)  &  0.953(10) &      &0.8898(2) &0.971(8) \\
0.0979  &        & 1.5306(12) &  2.210(83) &0.0261 &1.5577(7) &1.975(45) \\
0.0940  &        & 1.6450(10) &  2.390(70) &0.0270 &1.6479(7) &2.306(67) \\
0.0920  &        & 1.6902(10) &  2.498(84) &      &1.6931(7) &2.411(79) \\
0.0900  &        & 1.7353(10) &  2.605(95) &      &1.7382(7) &2.525(88) \\
\hline \hline
\end{tabular}
\end{table}



\begin{table}
\renewcommand{\arraystretch}{1.15}
\caption{Results for rest and kinetic masses (in lattice units) on the $a\approx 0.06\,\fm$ ensemble. }
\label{tab:RestKinetic0p06}
\begin{tabular}{l@{\quad}lll@{\quad}lll}
\hline\hline
     &\multicolumn{3}{c}{Analysis A}   &\multicolumn{3}{c}{Analysis B} \\
$\kappa$ &$am_q$ &$aM_1$ &$aM_2$ &$am_q$ &$aM_1$ &$aM_2$  \\ \hline
0.130  &0.01777 & 0.5518(4)  & 0.563(5)   &0.0188 &0.5536(3) &0.570(4) \\
0.1296 &        & --         & --         &      &0.5693(2) &0.582(4) \\
0.127  &        & 0.6593(5)  & 0.678(11)  &      &0.6608(4) &0.696(7) \\
0.124  &        & 0.7568(7)  & 0.790(16)  &      &0.7581(5) &0.817(10) \\
0.122  &        & --         & --         &      &-- &-- \\
0.112  &        & 1.0924(13) & 1.325(56)  &      &1.0935(12) &1.271(33) \\
0.106  &        & 1.2412(18) & 1.621(95)  &      &1.2430(14) &1.536(52) \\
0.1052 &        & --         & --         &      &1.2640(9) &1.543(49) \\
0.100  &        & 1.3833(23) & 1.975(164) &      &1.3856(18) &1.845(75) \\
\hline \hline
\end{tabular}
\end{table}



\begin{table}
\renewcommand{\arraystretch}{1.15}
\caption{Results for rest and kinetic masses (in lattice units) on the $a\approx 0.045\,\fm$ ensemble. }
\label{tab:RestKinetic0p045}
\begin{tabular}{l@{\quad}lll@{\quad}lll}
\hline\hline
     &\multicolumn{3}{c}{Analysis A}   &\multicolumn{3}{c}{Analysis B} \\
$\kappa$ &$am_q$ &$aM_1$ &$aM_2$ &$am_q$ &$aM_1$ &$aM_2$  \\ \hline
0.132   &0.01298 & 0.3818(2) & 0.394(3)  &0.0130 &0.3819(2) &0.384(1) \\
0.1310  &        & --        & --        &      &0.4239(2) &0.429(2) \\
0.130   &        & 0.4631(3) & 0.484(5)  &      &0.4632(2) &0.470(2) \\
0.128   &        & 0.5368(4) & 0.564(7)  &      &0.5370(3) &0.550(3) \\
0.116   &        & 0.9025(6) & 1.056(32) &      &0.9021(6) &1.021(16) \\
0.1143  &        & --        & --        &      &0.9480(5) &1.056(18) \\
0.111   &        & 1.0336(6) & 1.266(48) &      &1.0331(8) &1.225(26) \\
0.106   &        & 1.1576(7) & 1.535(75) &      &1.1573(9) &1.446(37) \\
\hline \hline
\end{tabular}
\end{table}


%% file: B2Dprd.bbl
\begin{thebibliography}{91}%
\makeatletter
\providecommand \@ifxundefined [1]{%
 \@ifx{#1\undefined}
}%
\providecommand \@ifnum [1]{%
 \ifnum #1\expandafter \@firstoftwo
 \else \expandafter \@secondoftwo
 \fi
}%
\providecommand \@ifx [1]{%
 \ifx #1\expandafter \@firstoftwo
 \else \expandafter \@secondoftwo
 \fi
}%
\providecommand \natexlab [1]{#1}%
\providecommand \enquote  [1]{``#1''}%
\providecommand \bibnamefont  [1]{#1}%
\providecommand \bibfnamefont [1]{#1}%
\providecommand \citenamefont [1]{#1}%
\providecommand \href@noop [0]{\@secondoftwo}%
\providecommand \href [0]{\begingroup \@sanitize@url \@href}%
\providecommand \@href[1]{\@@startlink{#1}\@@href}%
\providecommand \@@href[1]{\endgroup#1\@@endlink}%
\providecommand \@sanitize@url [0]{\catcode `\\12\catcode `\$12\catcode
  `\&12\catcode `\#12\catcode `\^12\catcode `\_12\catcode `\%12\relax}%
\providecommand \@@startlink[1]{}%
\providecommand \@@endlink[0]{}%
\providecommand \url  [0]{\begingroup\@sanitize@url \@url }%
\providecommand \@url [1]{\endgroup\@href {#1}{\urlprefix }}%
\providecommand \urlprefix  [0]{URL }%
\providecommand \Eprint [0]{\href }%
\providecommand \doibase [0]{http://dx.doi.org/}%
\providecommand \selectlanguage [0]{\@gobble}%
\providecommand \bibinfo  [0]{\@secondoftwo}%
\providecommand \bibfield  [0]{\@secondoftwo}%
\providecommand \translation [1]{[#1]}%
\providecommand \BibitemOpen [0]{}%
\providecommand \bibitemStop [0]{}%
\providecommand \bibitemNoStop [0]{.\EOS\space}%
\providecommand \EOS [0]{\spacefactor3000\relax}%
\providecommand \BibitemShut  [1]{\csname bibitem#1\endcsname}%
\let\auto@bib@innerbib\@empty
\bibitem [{\citenamefont {Beringer}\ \emph {et~al.}(2012)\citenamefont
  {Beringer} \emph {et~al.}}]{Beringer:1900zz}%
  \BibitemOpen
  \bibfield  {author} {\bibinfo {author} {\bibfnamefont {J.}~\bibnamefont
  {Beringer}} \emph {et~al.} (\bibinfo {collaboration} {Particle Data Group}),\
  }\href {\doibase 10.1103/PhysRevD.86.010001} {\bibfield  {journal} {\bibinfo
  {journal} {Phys. Rev.}\ }\textbf {\bibinfo {volume} {D86}},\ \bibinfo {pages}
  {010001} (\bibinfo {year} {2012})}\BibitemShut {NoStop}%
\bibitem [{\citenamefont {Laiho}\ \emph {et~al.}(2010)\citenamefont {Laiho},
  \citenamefont {Lunghi},\ and\ \citenamefont {Van~de Water}}]{Laiho:2009eu}%
  \BibitemOpen
  \bibfield  {author} {\bibinfo {author} {\bibfnamefont {J.}~\bibnamefont
  {Laiho}}, \bibinfo {author} {\bibfnamefont {E.}~\bibnamefont {Lunghi}}, \
  and\ \bibinfo {author} {\bibfnamefont {R.~S.}\ \bibnamefont {Van~de Water}},\
  }\href {\doibase 10.1103/PhysRevD.81.034503} {\bibfield  {journal} {\bibinfo
  {journal} {Phys. Rev.}\ }\textbf {\bibinfo {volume} {D81}},\ \bibinfo {pages}
  {034503} (\bibinfo {year} {2010})},\ \Eprint {http://arxiv.org/abs/0910.2928}
  {arXiv:0910.2928 [hep-ph]} \BibitemShut {NoStop}%
\bibitem [{\citenamefont {Lunghi}\ and\ \citenamefont
  {Soni}(2010)}]{Lunghi:2009ke}%
  \BibitemOpen
  \bibfield  {author} {\bibinfo {author} {\bibfnamefont {E.}~\bibnamefont
  {Lunghi}}\ and\ \bibinfo {author} {\bibfnamefont {A.}~\bibnamefont {Soni}},\
  }\href {\doibase 10.1103/PhysRevLett.104.251802} {\bibfield  {journal}
  {\bibinfo  {journal} {Phys. Rev. Lett.}\ }\textbf {\bibinfo {volume} {104}},\
  \bibinfo {pages} {251802} (\bibinfo {year} {2010})},\ \Eprint
  {http://arxiv.org/abs/0912.0002} {arXiv:0912.0002 [hep-ph]} \BibitemShut
  {NoStop}%
\bibitem [{\citenamefont {Lenz}\ \emph {et~al.}(2011)\citenamefont {Lenz},
  \citenamefont {Nierste}, \citenamefont {Charles}, \citenamefont
  {Descotes-Genon}, \citenamefont {Jantsch}, \citenamefont {Kaufhold},
  \citenamefont {Lacker}, \citenamefont {Monteil}, \citenamefont {Niess},\ and\
  \citenamefont {T'Jampens}}]{Lenz:2010gu}%
  \BibitemOpen
  \bibfield  {author} {\bibinfo {author} {\bibfnamefont {A.}~\bibnamefont
  {Lenz}}, \bibinfo {author} {\bibfnamefont {U.}~\bibnamefont {Nierste}},
  \bibinfo {author} {\bibfnamefont {J.}~\bibnamefont {Charles}}, \bibinfo
  {author} {\bibfnamefont {S.}~\bibnamefont {Descotes-Genon}}, \bibinfo
  {author} {\bibfnamefont {A.}~\bibnamefont {Jantsch}}, \bibinfo {author}
  {\bibfnamefont {C.}~\bibnamefont {Kaufhold}}, \bibinfo {author}
  {\bibfnamefont {H.}~\bibnamefont {Lacker}}, \bibinfo {author} {\bibfnamefont
  {S.}~\bibnamefont {Monteil}}, \bibinfo {author} {\bibfnamefont
  {V.}~\bibnamefont {Niess}}, \ and\ \bibinfo {author} {\bibfnamefont
  {S.}~\bibnamefont {T'Jampens}} (\bibinfo {collaboration} {{CKMfitter}}),\
  }\href {\doibase 10.1103/PhysRevD.83.036004} {\bibfield  {journal} {\bibinfo
  {journal} {Phys. Rev.}\ }\textbf {\bibinfo {volume} {D83}},\ \bibinfo {pages}
  {036004} (\bibinfo {year} {2011})},\ \Eprint {http://arxiv.org/abs/1008.1593}
  {arXiv:1008.1593 [hep-ph]} \BibitemShut {NoStop}%
\bibitem [{\citenamefont {Lunghi}\ and\ \citenamefont
  {Soni}(2011)}]{Lunghi:2010gv}%
  \BibitemOpen
  \bibfield  {author} {\bibinfo {author} {\bibfnamefont {E.}~\bibnamefont
  {Lunghi}}\ and\ \bibinfo {author} {\bibfnamefont {A.}~\bibnamefont {Soni}},\
  }\href {\doibase 10.1016/j.physletb.2011.02.016} {\bibfield  {journal}
  {\bibinfo  {journal} {Phys. Lett.}\ }\textbf {\bibinfo {volume} {B697}},\
  \bibinfo {pages} {323} (\bibinfo {year} {2011})},\ \Eprint
  {http://arxiv.org/abs/1010.6069} {arXiv:1010.6069 [hep-ph]} \BibitemShut
  {NoStop}%
\bibitem [{\citenamefont {Hashimoto}\ \emph {et~al.}(2002)\citenamefont
  {Hashimoto}, \citenamefont {Kronfeld}, \citenamefont {Mackenzie},
  \citenamefont {Ryan},\ and\ \citenamefont {Simone}}]{Hashimoto:2001nb}%
  \BibitemOpen
  \bibfield  {author} {\bibinfo {author} {\bibfnamefont {S.}~\bibnamefont
  {Hashimoto}}, \bibinfo {author} {\bibfnamefont {A.~S.}\ \bibnamefont
  {Kronfeld}}, \bibinfo {author} {\bibfnamefont {P.~B.}\ \bibnamefont
  {Mackenzie}}, \bibinfo {author} {\bibfnamefont {S.~M.}\ \bibnamefont {Ryan}},
  \ and\ \bibinfo {author} {\bibfnamefont {J.~N.}\ \bibnamefont {Simone}},\
  }\href {\doibase 10.1103/PhysRevD.66.014503} {\bibfield  {journal} {\bibinfo
  {journal} {Phys. Rev.}\ }\textbf {\bibinfo {volume} {D66}},\ \bibinfo {pages}
  {014503} (\bibinfo {year} {2002})},\ \Eprint
  {http://arxiv.org/abs/hep-ph/0110253} {hep-ph/0110253} \BibitemShut {NoStop}%
\bibitem [{\citenamefont {Bernard}\ \emph {et~al.}(2009)\citenamefont {Bernard}
  \emph {et~al.}}]{Bernard:2008dn}%
  \BibitemOpen
  \bibfield  {author} {\bibinfo {author} {\bibfnamefont {C.}~\bibnamefont
  {Bernard}} \emph {et~al.} (\bibinfo {collaboration} {Fermilab Lattice and
  MILC}),\ }\href {\doibase 10.1103/PhysRevD.79.014506} {\bibfield  {journal}
  {\bibinfo  {journal} {Phys. Rev.}\ }\textbf {\bibinfo {volume} {D79}},\
  \bibinfo {pages} {014506} (\bibinfo {year} {2009})},\ \Eprint
  {http://arxiv.org/abs/0808.2519} {arXiv:0808.2519 [hep-lat]} \BibitemShut
  {NoStop}%
\bibitem [{\citenamefont {Bailey}\ \emph {et~al.}(2010)\citenamefont {Bailey}
  \emph {et~al.}}]{Bailey:2010gb}%
  \BibitemOpen
  \bibfield  {author} {\bibinfo {author} {\bibfnamefont {J.~A.}\ \bibnamefont
  {Bailey}} \emph {et~al.} (\bibinfo {collaboration} {Fermilab Lattice and
  MILC}),\ }\href@noop {} {\bibfield  {journal} {\bibinfo  {journal} {PoS}\
  }\textbf {\bibinfo {volume} {LATTICE2010}},\ \bibinfo {pages} {311} (\bibinfo
  {year} {2010})},\ \Eprint {http://arxiv.org/abs/1011.2166} {arXiv:1011.2166
  [hep-lat]} \BibitemShut {NoStop}%
\bibitem [{\citenamefont {Akeroyd}\ \emph {et~al.}()\citenamefont {Akeroyd}
  \emph {et~al.}}]{Aushev:2010bq}%
  \BibitemOpen
  \bibfield  {author} {\bibinfo {author} {\bibfnamefont {A.~G.}\ \bibnamefont
  {Akeroyd}} \emph {et~al.},\ }\href {http://belle2.kek.jp} {\enquote {\bibinfo
  {title} {{Physics at Super $B$ Factory}},}\ }\Eprint
  {http://arxiv.org/abs/1002.5012} {arXiv:1002.5012 [hep-ex]} \BibitemShut
  {NoStop}%
\bibitem [{\citenamefont {Sirlin}(1982)}]{Sirlin:1981ie}%
  \BibitemOpen
  \bibfield  {author} {\bibinfo {author} {\bibfnamefont {A.}~\bibnamefont
  {Sirlin}},\ }\href {\doibase 10.1016/0550-3213(82)90303-0} {\bibfield
  {journal} {\bibinfo  {journal} {Nucl. Phys.}\ }\textbf {\bibinfo {volume}
  {B196}},\ \bibinfo {pages} {83} (\bibinfo {year} {1982})}\BibitemShut
  {NoStop}%
\bibitem [{\citenamefont {Ginsberg}(1968)}]{Ginsberg:1968pz}%
  \BibitemOpen
  \bibfield  {author} {\bibinfo {author} {\bibfnamefont {E.~S.}\ \bibnamefont
  {Ginsberg}},\ }\href {\doibase 10.1103/PhysRev.171.1675} {\bibfield
  {journal} {\bibinfo  {journal} {Phys. Rev.}\ }\textbf {\bibinfo {volume}
  {171}},\ \bibinfo {pages} {1675} (\bibinfo {year} {1968})},\ \bibinfo {note}
  {(E) \href{http://dx.doi.org/10.1103/PhysRev.174.2169.3}{\emph{ibid.}
  \textbf{174}, 2169 (1968)},
  \href{http://dx.doi.org/10.1103/PhysRev.187.2280.3}{\textbf{187}, 2280
  (1969)}}\BibitemShut {NoStop}%
\bibitem [{\citenamefont {Atwood}\ and\ \citenamefont
  {Marciano}(1990)}]{Atwood:1989em}%
  \BibitemOpen
  \bibfield  {author} {\bibinfo {author} {\bibfnamefont {D.}~\bibnamefont
  {Atwood}}\ and\ \bibinfo {author} {\bibfnamefont {W.~J.}\ \bibnamefont
  {Marciano}},\ }\href {\doibase 10.1103/PhysRevD.41.1736} {\bibfield
  {journal} {\bibinfo  {journal} {Phys. Rev.}\ }\textbf {\bibinfo {volume}
  {D41}},\ \bibinfo {pages} {1736} (\bibinfo {year} {1990})}\BibitemShut
  {NoStop}%
\bibitem [{\citenamefont {{K\"orner}}\ and\ \citenamefont
  {Schuler}(1990)}]{Korner:1989qb}%
  \BibitemOpen
  \bibfield  {author} {\bibinfo {author} {\bibfnamefont {J.~G.}\ \bibnamefont
  {{K\"orner}}}\ and\ \bibinfo {author} {\bibfnamefont {G.~A.}\ \bibnamefont
  {Schuler}},\ }\href {\doibase 10.1007/BF02440838} {\bibfield  {journal}
  {\bibinfo  {journal} {Z. Phys.}\ }\textbf {\bibinfo {volume} {C46}},\
  \bibinfo {pages} {93} (\bibinfo {year} {1990})}\BibitemShut {NoStop}%
\bibitem [{\citenamefont {Hashimoto}\ \emph {et~al.}(1999)\citenamefont
  {Hashimoto}, \citenamefont {El-Khadra}, \citenamefont {Kronfeld},
  \citenamefont {Mackenzie}, \citenamefont {Ryan},\ and\ \citenamefont
  {Simone}}]{Hashimoto:1999yp}%
  \BibitemOpen
  \bibfield  {author} {\bibinfo {author} {\bibfnamefont {S.}~\bibnamefont
  {Hashimoto}}, \bibinfo {author} {\bibfnamefont {A.~X.}\ \bibnamefont
  {El-Khadra}}, \bibinfo {author} {\bibfnamefont {A.~S.}\ \bibnamefont
  {Kronfeld}}, \bibinfo {author} {\bibfnamefont {P.~B.}\ \bibnamefont
  {Mackenzie}}, \bibinfo {author} {\bibfnamefont {S.~M.}\ \bibnamefont {Ryan}},
  \ and\ \bibinfo {author} {\bibfnamefont {J.~N.}\ \bibnamefont {Simone}},\
  }\href {\doibase 10.1103/PhysRevD.61.014502} {\bibfield  {journal} {\bibinfo
  {journal} {Phys. Rev.}\ }\textbf {\bibinfo {volume} {D61}},\ \bibinfo {pages}
  {014502} (\bibinfo {year} {1999})},\ \Eprint
  {http://arxiv.org/abs/hep-ph/9906376} {hep-ph/9906376} \BibitemShut {NoStop}%
\bibitem [{\citenamefont {Bazavov}\ \emph
  {et~al.}(2010{\natexlab{a}})\citenamefont {Bazavov} \emph
  {et~al.}}]{Bazavov:2009bb}%
  \BibitemOpen
  \bibfield  {author} {\bibinfo {author} {\bibfnamefont {A.}~\bibnamefont
  {Bazavov}} \emph {et~al.},\ }\href {\doibase 10.1103/RevModPhys.82.1349}
  {\bibfield  {journal} {\bibinfo  {journal} {Rev. Mod. Phys.}\ }\textbf
  {\bibinfo {volume} {82}},\ \bibinfo {pages} {1349} (\bibinfo {year}
  {2010}{\natexlab{a}})},\ \Eprint {http://arxiv.org/abs/0903.3598}
  {arXiv:0903.3598 [hep-lat]} \BibitemShut {NoStop}%
\bibitem [{\citenamefont {Weisz}(1983)}]{Weisz:1982zw}%
  \BibitemOpen
  \bibfield  {author} {\bibinfo {author} {\bibfnamefont {P.}~\bibnamefont
  {Weisz}},\ }\href {\doibase 10.1016/0550-3213(83)90595-3} {\bibfield
  {journal} {\bibinfo  {journal} {Nucl. Phys.}\ }\textbf {\bibinfo {volume}
  {B212}},\ \bibinfo {pages} {1} (\bibinfo {year} {1983})}\BibitemShut
  {NoStop}%
\bibitem [{\citenamefont {Curci}\ \emph {et~al.}(1983)\citenamefont {Curci},
  \citenamefont {Menotti},\ and\ \citenamefont {Paffuti}}]{Curci:1983an}%
  \BibitemOpen
  \bibfield  {author} {\bibinfo {author} {\bibfnamefont {G.}~\bibnamefont
  {Curci}}, \bibinfo {author} {\bibfnamefont {P.}~\bibnamefont {Menotti}}, \
  and\ \bibinfo {author} {\bibfnamefont {G.}~\bibnamefont {Paffuti}},\ }\href
  {\doibase 10.1016/0370-2693(83)91043-2} {\bibfield  {journal} {\bibinfo
  {journal} {Phys. Lett.}\ }\textbf {\bibinfo {volume} {B130}},\ \bibinfo
  {pages} {205} (\bibinfo {year} {1983})}\BibitemShut {NoStop}%
\bibitem [{\citenamefont {Weisz}\ and\ \citenamefont
  {Wohlert}(1984)}]{Weisz:1983bn}%
  \BibitemOpen
  \bibfield  {author} {\bibinfo {author} {\bibfnamefont {P.}~\bibnamefont
  {Weisz}}\ and\ \bibinfo {author} {\bibfnamefont {R.}~\bibnamefont
  {Wohlert}},\ }\href {\doibase 10.1016/0550-3213(84)90543-1} {\bibfield
  {journal} {\bibinfo  {journal} {Nucl. Phys.}\ }\textbf {\bibinfo {volume}
  {B236}},\ \bibinfo {pages} {397} (\bibinfo {year} {1984})},\ \bibinfo {note}
  {(E) \href{http://dx.doi.org/10.1016/0550-3213(84)90563-7}{\emph{ibid.}\
  \textbf{B247}, 544 (1984)}}\BibitemShut {NoStop}%
\bibitem [{\citenamefont {L\"uscher}\ and\ \citenamefont
  {Weisz}(1985{\natexlab{a}})}]{Luscher:1984xn}%
  \BibitemOpen
  \bibfield  {author} {\bibinfo {author} {\bibfnamefont {M.}~\bibnamefont
  {L\"uscher}}\ and\ \bibinfo {author} {\bibfnamefont {P.}~\bibnamefont
  {Weisz}},\ }\href {\doibase 10.1007/BF01206178} {\bibfield  {journal}
  {\bibinfo  {journal} {Commun. Math. Phys.}\ }\textbf {\bibinfo {volume}
  {97}},\ \bibinfo {pages} {59} (\bibinfo {year} {1985}{\natexlab{a}})},\
  \bibinfo {note} {(E) \href{http://dx.doi.org/10.1007/BF01205792}{\emph{ibid.}
  \textbf{98}, 433 (1985)}}\BibitemShut {NoStop}%
\bibitem [{\citenamefont {L\"uscher}\ and\ \citenamefont
  {Weisz}(1985{\natexlab{b}})}]{Luscher:1985zq}%
  \BibitemOpen
  \bibfield  {author} {\bibinfo {author} {\bibfnamefont {M.}~\bibnamefont
  {L\"uscher}}\ and\ \bibinfo {author} {\bibfnamefont {P.}~\bibnamefont
  {Weisz}},\ }\href {\doibase 10.1016/0370-2693(85)90966-9} {\bibfield
  {journal} {\bibinfo  {journal} {Phys. Lett.}\ }\textbf {\bibinfo {volume}
  {B158}},\ \bibinfo {pages} {250} (\bibinfo {year}
  {1985}{\natexlab{b}})}\BibitemShut {NoStop}%
\bibitem [{\citenamefont {Hao}\ \emph {et~al.}(2007)\citenamefont {Hao},
  \citenamefont {von Hippel}, \citenamefont {Horgan}, \citenamefont {Mason},\
  and\ \citenamefont {Trottier}}]{Hao:2007iz}%
  \BibitemOpen
  \bibfield  {author} {\bibinfo {author} {\bibfnamefont {Z.}~\bibnamefont
  {Hao}}, \bibinfo {author} {\bibfnamefont {G.~M.}\ \bibnamefont {von Hippel}},
  \bibinfo {author} {\bibfnamefont {R.~R.}\ \bibnamefont {Horgan}}, \bibinfo
  {author} {\bibfnamefont {Q.~J.}\ \bibnamefont {Mason}}, \ and\ \bibinfo
  {author} {\bibfnamefont {H.~D.}\ \bibnamefont {Trottier}},\ }\href {\doibase
  10.1103/PhysRevD.76.034507} {\bibfield  {journal} {\bibinfo  {journal} {Phys.
  Rev.}\ }\textbf {\bibinfo {volume} {D76}},\ \bibinfo {pages} {034507}
  (\bibinfo {year} {2007})},\ \Eprint {http://arxiv.org/abs/0705.4660}
  {arXiv:0705.4660 [hep-lat]} \BibitemShut {NoStop}%
\bibitem [{\citenamefont {Aubin}\ \emph {et~al.}(2004)\citenamefont {Aubin}
  \emph {et~al.}}]{Aubin:2004fs}%
  \BibitemOpen
  \bibfield  {author} {\bibinfo {author} {\bibfnamefont {C.}~\bibnamefont
  {Aubin}} \emph {et~al.} (\bibinfo {collaboration} {MILC}),\ }\href {\doibase
  10.1103/PhysRevD.70.114501} {\bibfield  {journal} {\bibinfo  {journal} {Phys.
  Rev.}\ }\textbf {\bibinfo {volume} {D70}},\ \bibinfo {pages} {114501}
  (\bibinfo {year} {2004})},\ \Eprint {http://arxiv.org/abs/hep-lat/0407028}
  {hep-lat/0407028} \BibitemShut {NoStop}%
\bibitem [{\citenamefont {Blum}\ \emph {et~al.}(1997)\citenamefont {Blum},
  \citenamefont {DeTar}, \citenamefont {Gottlieb}, \citenamefont {Heller},
  \citenamefont {Hetrick}, \citenamefont {Rummukainen}, \citenamefont {Sugar},
  \citenamefont {Toussaint},\ and\ \citenamefont {Wingate}}]{Blum:1996uf}%
  \BibitemOpen
  \bibfield  {author} {\bibinfo {author} {\bibfnamefont {T.}~\bibnamefont
  {Blum}}, \bibinfo {author} {\bibfnamefont {C.}~\bibnamefont {DeTar}},
  \bibinfo {author} {\bibfnamefont {S.}~\bibnamefont {Gottlieb}}, \bibinfo
  {author} {\bibfnamefont {U.~M.}\ \bibnamefont {Heller}}, \bibinfo {author}
  {\bibfnamefont {J.~E.}\ \bibnamefont {Hetrick}}, \bibinfo {author}
  {\bibfnamefont {K.}~\bibnamefont {Rummukainen}}, \bibinfo {author}
  {\bibfnamefont {R.~L.}\ \bibnamefont {Sugar}}, \bibinfo {author}
  {\bibfnamefont {D.}~\bibnamefont {Toussaint}}, \ and\ \bibinfo {author}
  {\bibfnamefont {M.}~\bibnamefont {Wingate}},\ }\href {\doibase
  10.1103/PhysRevD.55.1133} {\bibfield  {journal} {\bibinfo  {journal} {Phys.
  Rev.}\ }\textbf {\bibinfo {volume} {D55}},\ \bibinfo {pages} {1133} (\bibinfo
  {year} {1997})},\ \Eprint {http://arxiv.org/abs/hep-lat/9609036}
  {hep-lat/9609036} \BibitemShut {NoStop}%
\bibitem [{\citenamefont {Orginos}\ and\ \citenamefont
  {Toussaint}(1999)}]{Orginos:1998ue}%
  \BibitemOpen
  \bibfield  {author} {\bibinfo {author} {\bibfnamefont {K.}~\bibnamefont
  {Orginos}}\ and\ \bibinfo {author} {\bibfnamefont {D.}~\bibnamefont
  {Toussaint}} (\bibinfo {collaboration} {MILC}),\ }\href {\doibase
  10.1103/PhysRevD.59.014501} {\bibfield  {journal} {\bibinfo  {journal} {Phys.
  Rev.}\ }\textbf {\bibinfo {volume} {D59}},\ \bibinfo {pages} {014501}
  (\bibinfo {year} {1999})},\ \Eprint {http://arxiv.org/abs/hep-lat/9805009}
  {hep-lat/9805009} \BibitemShut {NoStop}%
\bibitem [{\citenamefont {Laga\"e}\ and\ \citenamefont
  {Sinclair}(1999)}]{Lagae:1998pe}%
  \BibitemOpen
  \bibfield  {author} {\bibinfo {author} {\bibfnamefont {J.~F.}\ \bibnamefont
  {Laga\"e}}\ and\ \bibinfo {author} {\bibfnamefont {D.~K.}\ \bibnamefont
  {Sinclair}},\ }\href {\doibase 10.1103/PhysRevD.59.014511} {\bibfield
  {journal} {\bibinfo  {journal} {Phys. Rev.}\ }\textbf {\bibinfo {volume}
  {D59}},\ \bibinfo {pages} {014511} (\bibinfo {year} {1999})},\ \Eprint
  {http://arxiv.org/abs/hep-lat/9806014} {hep-lat/9806014} \BibitemShut
  {NoStop}%
\bibitem [{\citenamefont {Lepage}(1999)}]{Lepage:1998vj}%
  \BibitemOpen
  \bibfield  {author} {\bibinfo {author} {\bibfnamefont {G.~P.}\ \bibnamefont
  {Lepage}},\ }\href {\doibase 10.1103/PhysRevD.59.074502} {\bibfield
  {journal} {\bibinfo  {journal} {Phys. Rev.}\ }\textbf {\bibinfo {volume}
  {D59}},\ \bibinfo {pages} {074502} (\bibinfo {year} {1999})},\ \Eprint
  {http://arxiv.org/abs/hep-lat/9809157} {hep-lat/9809157} \BibitemShut
  {NoStop}%
\bibitem [{\citenamefont {Orginos}\ \emph {et~al.}(1999)\citenamefont
  {Orginos}, \citenamefont {Toussaint},\ and\ \citenamefont
  {Sugar}}]{Orginos:1999cr}%
  \BibitemOpen
  \bibfield  {author} {\bibinfo {author} {\bibfnamefont {K.}~\bibnamefont
  {Orginos}}, \bibinfo {author} {\bibfnamefont {D.}~\bibnamefont {Toussaint}},
  \ and\ \bibinfo {author} {\bibfnamefont {R.~L.}\ \bibnamefont {Sugar}}
  (\bibinfo {collaboration} {MILC}),\ }\href {\doibase
  10.1103/PhysRevD.60.054503} {\bibfield  {journal} {\bibinfo  {journal} {Phys.
  Rev.}\ }\textbf {\bibinfo {volume} {D60}},\ \bibinfo {pages} {054503}
  (\bibinfo {year} {1999})},\ \Eprint {http://arxiv.org/abs/hep-lat/9903032}
  {hep-lat/9903032} \BibitemShut {NoStop}%
\bibitem [{\citenamefont {Susskind}(1977)}]{Susskind:1976jm}%
  \BibitemOpen
  \bibfield  {author} {\bibinfo {author} {\bibfnamefont {L.}~\bibnamefont
  {Susskind}},\ }\href {\doibase 10.1103/PhysRevD.16.3031} {\bibfield
  {journal} {\bibinfo  {journal} {Phys. Rev.}\ }\textbf {\bibinfo {volume}
  {D16}},\ \bibinfo {pages} {3031} (\bibinfo {year} {1977})}\BibitemShut
  {NoStop}%
\bibitem [{\citenamefont {Sharatchandra}\ \emph {et~al.}(1981)\citenamefont
  {Sharatchandra}, \citenamefont {Thun},\ and\ \citenamefont
  {Weisz}}]{Sharatchandra:1981si}%
  \BibitemOpen
  \bibfield  {author} {\bibinfo {author} {\bibfnamefont {H.~S.}\ \bibnamefont
  {Sharatchandra}}, \bibinfo {author} {\bibfnamefont {H.~J.}\ \bibnamefont
  {Thun}}, \ and\ \bibinfo {author} {\bibfnamefont {P.}~\bibnamefont {Weisz}},\
  }\href {\doibase 10.1016/0550-3213(81)90200-5} {\bibfield  {journal}
  {\bibinfo  {journal} {Nucl. Phys.}\ }\textbf {\bibinfo {volume} {B192}},\
  \bibinfo {pages} {205} (\bibinfo {year} {1981})}\BibitemShut {NoStop}%
\bibitem [{\citenamefont {Hamber}\ \emph {et~al.}(1983)\citenamefont {Hamber},
  \citenamefont {Marinari}, \citenamefont {Parisi},\ and\ \citenamefont
  {Rebbi}}]{Hamber:1983kx}%
  \BibitemOpen
  \bibfield  {author} {\bibinfo {author} {\bibfnamefont {H.~W.}\ \bibnamefont
  {Hamber}}, \bibinfo {author} {\bibfnamefont {E.}~\bibnamefont {Marinari}},
  \bibinfo {author} {\bibfnamefont {G.}~\bibnamefont {Parisi}}, \ and\ \bibinfo
  {author} {\bibfnamefont {C.}~\bibnamefont {Rebbi}},\ }\href {\doibase
  10.1016/0370-2693(83)91412-0} {\bibfield  {journal} {\bibinfo  {journal}
  {Phys. Lett.}\ }\textbf {\bibinfo {volume} {B124}},\ \bibinfo {pages} {99}
  (\bibinfo {year} {1983})}\BibitemShut {NoStop}%
\bibitem [{\citenamefont {Prelov\v{s}ek}(2006)}]{Prelovsek:2005rf}%
  \BibitemOpen
  \bibfield  {author} {\bibinfo {author} {\bibfnamefont {S.}~\bibnamefont
  {Prelov\v{s}ek}},\ }\href {\doibase 10.1103/PhysRevD.73.014506} {\bibfield
  {journal} {\bibinfo  {journal} {Phys. Rev.}\ }\textbf {\bibinfo {volume}
  {D73}},\ \bibinfo {pages} {014506} (\bibinfo {year} {2006})},\ \Eprint
  {http://arxiv.org/abs/hep-lat/0510080} {hep-lat/0510080} \BibitemShut
  {NoStop}%
\bibitem [{\citenamefont {Bernard}(2006)}]{Bernard:2006zw}%
  \BibitemOpen
  \bibfield  {author} {\bibinfo {author} {\bibfnamefont {C.}~\bibnamefont
  {Bernard}},\ }\href {\doibase 10.1103/PhysRevD.73.114503} {\bibfield
  {journal} {\bibinfo  {journal} {Phys. Rev.}\ }\textbf {\bibinfo {volume}
  {D73}},\ \bibinfo {pages} {114503} (\bibinfo {year} {2006})},\ \Eprint
  {http://arxiv.org/abs/hep-lat/0603011} {hep-lat/0603011} \BibitemShut
  {NoStop}%
\bibitem [{\citenamefont {Bernard}\ \emph {et~al.}(2007)\citenamefont
  {Bernard}, \citenamefont {DeTar}, \citenamefont {Fu},\ and\ \citenamefont
  {Prelov\v{s}ek}}]{Bernard:2007qf}%
  \BibitemOpen
  \bibfield  {author} {\bibinfo {author} {\bibfnamefont {C.}~\bibnamefont
  {Bernard}}, \bibinfo {author} {\bibfnamefont {C.~E.}\ \bibnamefont {DeTar}},
  \bibinfo {author} {\bibfnamefont {Z.}~\bibnamefont {Fu}}, \ and\ \bibinfo
  {author} {\bibfnamefont {S.}~\bibnamefont {Prelov\v{s}ek}},\ }\href {\doibase
  10.1103/PhysRevD.76.094504} {\bibfield  {journal} {\bibinfo  {journal} {Phys.
  Rev.}\ }\textbf {\bibinfo {volume} {D76}},\ \bibinfo {pages} {094504}
  (\bibinfo {year} {2007})},\ \Eprint {http://arxiv.org/abs/0707.2402}
  {arXiv:0707.2402 [hep-lat]} \BibitemShut {NoStop}%
\bibitem [{\citenamefont {Aubin}\ \emph {et~al.}(2008)\citenamefont {Aubin},
  \citenamefont {Laiho},\ and\ \citenamefont {Van~de Water}}]{Aubin:2008wk}%
  \BibitemOpen
  \bibfield  {author} {\bibinfo {author} {\bibfnamefont {C.}~\bibnamefont
  {Aubin}}, \bibinfo {author} {\bibfnamefont {J.}~\bibnamefont {Laiho}}, \ and\
  \bibinfo {author} {\bibfnamefont {R.~S.}\ \bibnamefont {Van~de Water}},\
  }\href {\doibase 10.1103/PhysRevD.77.114501} {\bibfield  {journal} {\bibinfo
  {journal} {Phys. Rev.}\ }\textbf {\bibinfo {volume} {D77}},\ \bibinfo {pages}
  {114501} (\bibinfo {year} {2008})},\ \Eprint {http://arxiv.org/abs/0803.0129}
  {arXiv:0803.0129 [hep-lat]} \BibitemShut {NoStop}%
\bibitem [{\citenamefont {Bernard}\ \emph {et~al.}(2006)\citenamefont
  {Bernard}, \citenamefont {Golterman},\ and\ \citenamefont
  {Shamir}}]{Bernard:2006ee}%
  \BibitemOpen
  \bibfield  {author} {\bibinfo {author} {\bibfnamefont {C.}~\bibnamefont
  {Bernard}}, \bibinfo {author} {\bibfnamefont {M.}~\bibnamefont {Golterman}},
  \ and\ \bibinfo {author} {\bibfnamefont {Y.}~\bibnamefont {Shamir}},\ }\href
  {\doibase 10.1103/PhysRevD.73.114511} {\bibfield  {journal} {\bibinfo
  {journal} {Phys. Rev.}\ }\textbf {\bibinfo {volume} {D73}},\ \bibinfo {pages}
  {114511} (\bibinfo {year} {2006})},\ \Eprint
  {http://arxiv.org/abs/hep-lat/0604017} {hep-lat/0604017} \BibitemShut
  {NoStop}%
\bibitem [{\citenamefont {Adams}(2005)}]{Adams:2004mf}%
  \BibitemOpen
  \bibfield  {author} {\bibinfo {author} {\bibfnamefont {D.~H.}\ \bibnamefont
  {Adams}},\ }\href {\doibase 10.1103/PhysRevD.72.114512} {\bibfield  {journal}
  {\bibinfo  {journal} {Phys. Rev.}\ }\textbf {\bibinfo {volume} {D72}},\
  \bibinfo {pages} {114512} (\bibinfo {year} {2005})},\ \Eprint
  {http://arxiv.org/abs/hep-lat/0411030} {hep-lat/0411030} \BibitemShut
  {NoStop}%
\bibitem [{\citenamefont {Shamir}(2005)}]{Shamir:2004zc}%
  \BibitemOpen
  \bibfield  {author} {\bibinfo {author} {\bibfnamefont {Y.}~\bibnamefont
  {Shamir}},\ }\href {\doibase 10.1103/PhysRevD.71.034509} {\bibfield
  {journal} {\bibinfo  {journal} {Phys. Rev.}\ }\textbf {\bibinfo {volume}
  {D71}},\ \bibinfo {pages} {034509} (\bibinfo {year} {2005})},\ \Eprint
  {http://arxiv.org/abs/hep-lat/0412014} {hep-lat/0412014} \BibitemShut
  {NoStop}%
\bibitem [{\citenamefont {{D\"urr}}(2005)}]{Durr:2005ax}%
  \BibitemOpen
  \bibfield  {author} {\bibinfo {author} {\bibfnamefont {S.}~\bibnamefont
  {{D\"urr}}},\ }\href@noop {} {\bibfield  {journal} {\bibinfo  {journal}
  {PoS}\ }\textbf {\bibinfo {volume} {LAT2005}},\ \bibinfo {pages} {021}
  (\bibinfo {year} {2005})},\ \Eprint {http://arxiv.org/abs/hep-lat/0509026}
  {hep-lat/0509026} \BibitemShut {NoStop}%
\bibitem [{\citenamefont {Shamir}(2007)}]{Shamir:2006nj}%
  \BibitemOpen
  \bibfield  {author} {\bibinfo {author} {\bibfnamefont {Y.}~\bibnamefont
  {Shamir}},\ }\href {\doibase 10.1103/PhysRevD.75.054503} {\bibfield
  {journal} {\bibinfo  {journal} {Phys. Rev.}\ }\textbf {\bibinfo {volume}
  {D75}},\ \bibinfo {pages} {054503} (\bibinfo {year} {2007})},\ \Eprint
  {http://arxiv.org/abs/hep-lat/0607007} {hep-lat/0607007} \BibitemShut
  {NoStop}%
\bibitem [{\citenamefont {Sharpe}(2006)}]{Sharpe:2006re}%
  \BibitemOpen
  \bibfield  {author} {\bibinfo {author} {\bibfnamefont {S.~R.}\ \bibnamefont
  {Sharpe}},\ }\href@noop {} {\bibfield  {journal} {\bibinfo  {journal} {PoS}\
  }\textbf {\bibinfo {volume} {LAT2006}},\ \bibinfo {pages} {022} (\bibinfo
  {year} {2006})},\ \Eprint {http://arxiv.org/abs/hep-lat/0610094}
  {hep-lat/0610094} \BibitemShut {NoStop}%
\bibitem [{\citenamefont {Bernard}\ \emph {et~al.}(2008)\citenamefont
  {Bernard}, \citenamefont {Golterman}, \citenamefont {Shamir},\ and\
  \citenamefont {Sharpe}}]{Bernard:2007eh}%
  \BibitemOpen
  \bibfield  {author} {\bibinfo {author} {\bibfnamefont {C.}~\bibnamefont
  {Bernard}}, \bibinfo {author} {\bibfnamefont {M.}~\bibnamefont {Golterman}},
  \bibinfo {author} {\bibfnamefont {Y.}~\bibnamefont {Shamir}}, \ and\ \bibinfo
  {author} {\bibfnamefont {S.~R.}\ \bibnamefont {Sharpe}},\ }\href {\doibase
  10.1103/PhysRevD.77.114504} {\bibfield  {journal} {\bibinfo  {journal}
  {Phys.Rev.}\ }\textbf {\bibinfo {volume} {D77}},\ \bibinfo {pages} {114504}
  (\bibinfo {year} {2008})},\ \Eprint {http://arxiv.org/abs/0711.0696}
  {arXiv:0711.0696 [hep-lat]} \BibitemShut {NoStop}%
\bibitem [{\citenamefont {Kronfeld}(2007)}]{Kronfeld:2007ek}%
  \BibitemOpen
  \bibfield  {author} {\bibinfo {author} {\bibfnamefont {A.~S.}\ \bibnamefont
  {Kronfeld}},\ }\href@noop {} {\bibfield  {journal} {\bibinfo  {journal}
  {PoS}\ }\textbf {\bibinfo {volume} {LAT2007}},\ \bibinfo {pages} {016}
  (\bibinfo {year} {2007})},\ \Eprint {http://arxiv.org/abs/0711.0699}
  {arXiv:0711.0699 [hep-lat]} \BibitemShut {NoStop}%
\bibitem [{\citenamefont {Golterman}(2008)}]{Golterman:2008gt}%
  \BibitemOpen
  \bibfield  {author} {\bibinfo {author} {\bibfnamefont {M.}~\bibnamefont
  {Golterman}},\ }\href@noop {} {\bibfield  {journal} {\bibinfo  {journal}
  {PoS}\ }\textbf {\bibinfo {volume} {CONFINEMENT8}},\ \bibinfo {pages} {014}
  (\bibinfo {year} {2008})},\ \Eprint {http://arxiv.org/abs/0812.3110}
  {arXiv:0812.3110 [hep-ph]} \BibitemShut {NoStop}%
\bibitem [{\citenamefont {Donald}\ \emph {et~al.}(2011)\citenamefont {Donald},
  \citenamefont {Davies}, \citenamefont {Follana},\ and\ \citenamefont
  {Kronfeld}}]{Donald:2011if}%
  \BibitemOpen
  \bibfield  {author} {\bibinfo {author} {\bibfnamefont {G.~C.}\ \bibnamefont
  {Donald}}, \bibinfo {author} {\bibfnamefont {C.~T.~H.}\ \bibnamefont
  {Davies}}, \bibinfo {author} {\bibfnamefont {E.}~\bibnamefont {Follana}}, \
  and\ \bibinfo {author} {\bibfnamefont {A.~S.}\ \bibnamefont {Kronfeld}},\
  }\href {\doibase 10.1103/PhysRevD.84.054504} {\bibfield  {journal} {\bibinfo
  {journal} {Phys. Rev.}\ }\textbf {\bibinfo {volume} {D84}},\ \bibinfo {pages}
  {054504} (\bibinfo {year} {2011})},\ \Eprint {http://arxiv.org/abs/1106.2412}
  {arXiv:1106.2412 [hep-lat]} \BibitemShut {NoStop}%
\bibitem [{\citenamefont {Wilson}(1977)}]{Wilson:1977nj}%
  \BibitemOpen
  \bibfield  {author} {\bibinfo {author} {\bibfnamefont {K.~G.}\ \bibnamefont
  {Wilson}},\ }in\ \href@noop {} {\emph {\bibinfo {booktitle} {New Phenomena in
  Subnuclear Physics}}},\ \bibinfo {editor} {edited by\ \bibinfo {editor}
  {\bibfnamefont {A.}~\bibnamefont {Zichichi}}}\ (\bibinfo  {publisher}
  {Plenum},\ \bibinfo {address} {New York},\ \bibinfo {year}
  {1977})\BibitemShut {NoStop}%
\bibitem [{\citenamefont {Sheikholeslami}\ and\ \citenamefont
  {Wohlert}(1985)}]{Sheikholeslami:1985ij}%
  \BibitemOpen
  \bibfield  {author} {\bibinfo {author} {\bibfnamefont {B.}~\bibnamefont
  {Sheikholeslami}}\ and\ \bibinfo {author} {\bibfnamefont {R.}~\bibnamefont
  {Wohlert}},\ }\href {\doibase 10.1016/0550-3213(85)90002-1} {\bibfield
  {journal} {\bibinfo  {journal} {Nucl. Phys.}\ }\textbf {\bibinfo {volume}
  {B259}},\ \bibinfo {pages} {572} (\bibinfo {year} {1985})}\BibitemShut
  {NoStop}%
\bibitem [{\citenamefont {El-Khadra}\ \emph {et~al.}(1997)\citenamefont
  {El-Khadra}, \citenamefont {Kronfeld},\ and\ \citenamefont
  {Mackenzie}}]{ElKhadra:1996mp}%
  \BibitemOpen
  \bibfield  {author} {\bibinfo {author} {\bibfnamefont {A.~X.}\ \bibnamefont
  {El-Khadra}}, \bibinfo {author} {\bibfnamefont {A.~S.}\ \bibnamefont
  {Kronfeld}}, \ and\ \bibinfo {author} {\bibfnamefont {P.~B.}\ \bibnamefont
  {Mackenzie}},\ }\href {\doibase 10.1103/PhysRevD.55.3933} {\bibfield
  {journal} {\bibinfo  {journal} {Phys. Rev.}\ }\textbf {\bibinfo {volume}
  {D55}},\ \bibinfo {pages} {3933} (\bibinfo {year} {1997})},\ \Eprint
  {http://arxiv.org/abs/hep-lat/9604004} {hep-lat/9604004} \BibitemShut
  {NoStop}%
\bibitem [{\citenamefont {Bernard}\ \emph {et~al.}(2011)\citenamefont {Bernard}
  \emph {et~al.}}]{Bernard:2010fr}%
  \BibitemOpen
  \bibfield  {author} {\bibinfo {author} {\bibfnamefont {C.}~\bibnamefont
  {Bernard}} \emph {et~al.} (\bibinfo {collaboration} {Fermilab Lattice and
  MILC}),\ }\href {\doibase 10.1103/PhysRevD.83.034503} {\bibfield  {journal}
  {\bibinfo  {journal} {Phys. Rev.}\ }\textbf {\bibinfo {volume} {D83}},\
  \bibinfo {pages} {034503} (\bibinfo {year} {2011})},\ \Eprint
  {http://arxiv.org/abs/1003.1937} {arXiv:1003.1937 [hep-lat]} \BibitemShut
  {NoStop}%
\bibitem [{\citenamefont {Sommer}(1994)}]{Sommer:1993ce}%
  \BibitemOpen
  \bibfield  {author} {\bibinfo {author} {\bibfnamefont {R.}~\bibnamefont
  {Sommer}},\ }\href {\doibase 10.1016/0550-3213(94)90473-1} {\bibfield
  {journal} {\bibinfo  {journal} {Nucl.Phys.}\ }\textbf {\bibinfo {volume}
  {B411}},\ \bibinfo {pages} {839} (\bibinfo {year} {1994})},\ \Eprint
  {http://arxiv.org/abs/hep-lat/9310022} {arXiv:hep-lat/9310022 [hep-lat]}
  \BibitemShut {NoStop}%
\bibitem [{\citenamefont {Bernard}\ \emph {et~al.}(2000)\citenamefont {Bernard}
  \emph {et~al.}}]{Bernard:2000gd}%
  \BibitemOpen
  \bibfield  {author} {\bibinfo {author} {\bibfnamefont {C.~W.}\ \bibnamefont
  {Bernard}} \emph {et~al.} (\bibinfo {collaboration} {MILC}),\ }\href@noop {}
  {\bibfield  {journal} {\bibinfo  {journal} {Phys. Rev.}\ }\textbf {\bibinfo
  {volume} {D62}},\ \bibinfo {pages} {034503} (\bibinfo {year} {2000})},\
  \Eprint {http://arxiv.org/abs/hep-lat/0002028} {hep-lat/0002028} \BibitemShut
  {NoStop}%
\bibitem [{\citenamefont {Bazavov}\ \emph {et~al.}(2012)\citenamefont {Bazavov}
  \emph {et~al.}}]{Bazavov:2011aa}%
  \BibitemOpen
  \bibfield  {author} {\bibinfo {author} {\bibfnamefont {A.}~\bibnamefont
  {Bazavov}} \emph {et~al.} (\bibinfo {collaboration} {Fermilab Lattice and
  MILC}),\ }\href {\doibase 10.1103/PhysRevD.85.114506} {\bibfield  {journal}
  {\bibinfo  {journal} {Phys. Rev.}\ }\textbf {\bibinfo {volume} {D85}},\
  \bibinfo {pages} {114506} (\bibinfo {year} {2012})},\ \Eprint
  {http://arxiv.org/abs/1112.3051} {arXiv:1112.3051 [hep-lat]} \BibitemShut
  {NoStop}%
\bibitem [{\citenamefont {Wingate}\ \emph {et~al.}(2003)\citenamefont
  {Wingate}, \citenamefont {Shigemitsu}, \citenamefont {Davies}, \citenamefont
  {Lepage},\ and\ \citenamefont {Trottier}}]{Wingate:2003nn}%
  \BibitemOpen
  \bibfield  {author} {\bibinfo {author} {\bibfnamefont {M.}~\bibnamefont
  {Wingate}}, \bibinfo {author} {\bibfnamefont {J.}~\bibnamefont {Shigemitsu}},
  \bibinfo {author} {\bibfnamefont {C.~T.~H.}\ \bibnamefont {Davies}}, \bibinfo
  {author} {\bibfnamefont {G.~P.}\ \bibnamefont {Lepage}}, \ and\ \bibinfo
  {author} {\bibfnamefont {H.~D.}\ \bibnamefont {Trottier}},\ }\href {\doibase
  10.1103/PhysRevD.80.014503} {\bibfield  {journal} {\bibinfo  {journal} {Phys.
  Rev.}\ }\textbf {\bibinfo {volume} {D67}},\ \bibinfo {pages} {054505}
  (\bibinfo {year} {2003})},\ \Eprint {http://arxiv.org/abs/hep-lat/0211014}
  {hep-lat/0211014} \BibitemShut {NoStop}%
\bibitem [{\citenamefont {Kronfeld}(2000)}]{Kronfeld:2000ck}%
  \BibitemOpen
  \bibfield  {author} {\bibinfo {author} {\bibfnamefont {A.~S.}\ \bibnamefont
  {Kronfeld}},\ }\href {\doibase 10.1103/PhysRevD.62.014505} {\bibfield
  {journal} {\bibinfo  {journal} {Phys. Rev.}\ }\textbf {\bibinfo {volume}
  {D62}},\ \bibinfo {pages} {014505} (\bibinfo {year} {2000})},\ \Eprint
  {http://arxiv.org/abs/hep-lat/0002008} {hep-lat/0002008} \BibitemShut
  {NoStop}%
\bibitem [{\citenamefont {Harada}\ \emph {et~al.}(2002)\citenamefont {Harada},
  \citenamefont {Hashimoto}, \citenamefont {Kronfeld},\ and\ \citenamefont
  {Onogi}}]{Harada:2001fj}%
  \BibitemOpen
  \bibfield  {author} {\bibinfo {author} {\bibfnamefont {J.}~\bibnamefont
  {Harada}}, \bibinfo {author} {\bibfnamefont {S.}~\bibnamefont {Hashimoto}},
  \bibinfo {author} {\bibfnamefont {A.~S.}\ \bibnamefont {Kronfeld}}, \ and\
  \bibinfo {author} {\bibfnamefont {T.}~\bibnamefont {Onogi}},\ }\href
  {\doibase 10.1103/PhysRevD.65.094514} {\bibfield  {journal} {\bibinfo
  {journal} {Phys. Rev.}\ }\textbf {\bibinfo {volume} {D65}},\ \bibinfo {pages}
  {094514} (\bibinfo {year} {2002})},\ \Eprint
  {http://arxiv.org/abs/hep-lat/0112045} {hep-lat/0112045} \BibitemShut
  {NoStop}%
\bibitem [{\citenamefont {Nobes}\ and\ \citenamefont
  {Trottier}(2006)}]{Nobes:2005dz}%
  \BibitemOpen
  \bibfield  {author} {\bibinfo {author} {\bibfnamefont {M.}~\bibnamefont
  {Nobes}}\ and\ \bibinfo {author} {\bibfnamefont {H.}~\bibnamefont
  {Trottier}},\ }\href@noop {} {\bibfield  {journal} {\bibinfo  {journal}
  {PoS}\ }\textbf {\bibinfo {volume} {LAT2005}},\ \bibinfo {pages} {209}
  (\bibinfo {year} {2006})},\ \Eprint {http://arxiv.org/abs/hep-lat/0509128}
  {hep-lat/0509128} \BibitemShut {NoStop}%
\bibitem [{\citenamefont {Brodsky}\ \emph {et~al.}(1983)\citenamefont
  {Brodsky}, \citenamefont {Lepage},\ and\ \citenamefont
  {Mackenzie}}]{Brodsky:1982gc}%
  \BibitemOpen
  \bibfield  {author} {\bibinfo {author} {\bibfnamefont {S.~J.}\ \bibnamefont
  {Brodsky}}, \bibinfo {author} {\bibfnamefont {G.~P.}\ \bibnamefont {Lepage}},
  \ and\ \bibinfo {author} {\bibfnamefont {P.~B.}\ \bibnamefont {Mackenzie}},\
  }\href {\doibase 10.1103/PhysRevD.28.228} {\bibfield  {journal} {\bibinfo
  {journal} {Phys. Rev.}\ }\textbf {\bibinfo {volume} {D28}},\ \bibinfo {pages}
  {228} (\bibinfo {year} {1983})}\BibitemShut {NoStop}%
\bibitem [{\citenamefont {Lepage}\ and\ \citenamefont
  {Mackenzie}(1993)}]{Lepage:1992xa}%
  \BibitemOpen
  \bibfield  {author} {\bibinfo {author} {\bibfnamefont {G.~P.}\ \bibnamefont
  {Lepage}}\ and\ \bibinfo {author} {\bibfnamefont {P.~B.}\ \bibnamefont
  {Mackenzie}},\ }\href {\doibase 10.1103/PhysRevD.48.2250} {\bibfield
  {journal} {\bibinfo  {journal} {Phys. Rev.}\ }\textbf {\bibinfo {volume}
  {D48}},\ \bibinfo {pages} {2250} (\bibinfo {year} {1993})},\ \Eprint
  {http://arxiv.org/abs/hep-lat/9209022} {hep-lat/9209022} \BibitemShut
  {NoStop}%
\bibitem [{\citenamefont {Harada}\ \emph {et~al.}(2003)\citenamefont {Harada},
  \citenamefont {Hashimoto}, \citenamefont {Kronfeld},\ and\ \citenamefont
  {Onogi}}]{Harada:2002jh}%
  \BibitemOpen
  \bibfield  {author} {\bibinfo {author} {\bibfnamefont {J.}~\bibnamefont
  {Harada}}, \bibinfo {author} {\bibfnamefont {S.}~\bibnamefont {Hashimoto}},
  \bibinfo {author} {\bibfnamefont {A.~S.}\ \bibnamefont {Kronfeld}}, \ and\
  \bibinfo {author} {\bibfnamefont {T.}~\bibnamefont {Onogi}},\ }\href
  {\doibase 10.1103/PhysRevD.67.014503} {\bibfield  {journal} {\bibinfo
  {journal} {Phys. Rev.}\ }\textbf {\bibinfo {volume} {D67}},\ \bibinfo {pages}
  {014503} (\bibinfo {year} {2003})},\ \Eprint
  {http://arxiv.org/abs/hep-lat/0208004} {hep-lat/0208004} \BibitemShut
  {NoStop}%
\bibitem [{\citenamefont {Hornbostel}\ \emph {et~al.}(2003)\citenamefont
  {Hornbostel}, \citenamefont {Lepage},\ and\ \citenamefont
  {Morningstar}}]{Hornbostel:2002af}%
  \BibitemOpen
  \bibfield  {author} {\bibinfo {author} {\bibfnamefont {K.}~\bibnamefont
  {Hornbostel}}, \bibinfo {author} {\bibfnamefont {G.~P.}\ \bibnamefont
  {Lepage}}, \ and\ \bibinfo {author} {\bibfnamefont {C.}~\bibnamefont
  {Morningstar}},\ }\href {\doibase 10.1103/PhysRevD.67.034023} {\bibfield
  {journal} {\bibinfo  {journal} {Phys. Rev.}\ }\textbf {\bibinfo {volume}
  {D67}},\ \bibinfo {pages} {034023} (\bibinfo {year} {2003})},\ \Eprint
  {http://arxiv.org/abs/hep-ph/0208224} {hep-ph/0208224} \BibitemShut {NoStop}%
\bibitem [{\citenamefont {Falk}\ and\ \citenamefont
  {Neubert}(1993)}]{Falk:1992wt}%
  \BibitemOpen
  \bibfield  {author} {\bibinfo {author} {\bibfnamefont {A.~F.}\ \bibnamefont
  {Falk}}\ and\ \bibinfo {author} {\bibfnamefont {M.}~\bibnamefont {Neubert}},\
  }\href {\doibase 10.1103/PhysRevD.47.2965} {\bibfield  {journal} {\bibinfo
  {journal} {Phys. Rev.}\ }\textbf {\bibinfo {volume} {D47}},\ \bibinfo {pages}
  {2965} (\bibinfo {year} {1993})},\ \Eprint
  {http://arxiv.org/abs/hep-ph/9209268} {hep-ph/9209268} \BibitemShut {NoStop}%
\bibitem [{\citenamefont {Mannel}(1994)}]{Mannel:1994kv}%
  \BibitemOpen
  \bibfield  {author} {\bibinfo {author} {\bibfnamefont {T.}~\bibnamefont
  {Mannel}},\ }\href {\doibase 10.1103/PhysRevD.50.428} {\bibfield  {journal}
  {\bibinfo  {journal} {Phys. Rev.}\ }\textbf {\bibinfo {volume} {D50}},\
  \bibinfo {pages} {428} (\bibinfo {year} {1994})},\ \Eprint
  {http://arxiv.org/abs/hep-ph/9403249} {hep-ph/9403249} \BibitemShut {NoStop}%
\bibitem [{\citenamefont {Luke}(1990)}]{Luke:1990eg}%
  \BibitemOpen
  \bibfield  {author} {\bibinfo {author} {\bibfnamefont {M.~E.}\ \bibnamefont
  {Luke}},\ }\href {\doibase 10.1016/0370-2693(90)90568-Q} {\bibfield
  {journal} {\bibinfo  {journal} {Phys. Lett.}\ }\textbf {\bibinfo {volume}
  {B252}},\ \bibinfo {pages} {447} (\bibinfo {year} {1990})}\BibitemShut
  {NoStop}%
\bibitem [{\citenamefont {Aubin}\ and\ \citenamefont
  {Bernard}(2006)}]{Aubin:2005aq}%
  \BibitemOpen
  \bibfield  {author} {\bibinfo {author} {\bibfnamefont {C.}~\bibnamefont
  {Aubin}}\ and\ \bibinfo {author} {\bibfnamefont {C.}~\bibnamefont
  {Bernard}},\ }\href {\doibase 10.1103/PhysRevD.73.014515} {\bibfield
  {journal} {\bibinfo  {journal} {Phys. Rev.}\ }\textbf {\bibinfo {volume}
  {D73}},\ \bibinfo {pages} {014515} (\bibinfo {year} {2006})},\ \Eprint
  {http://arxiv.org/abs/hep-lat/0510088} {hep-lat/0510088} \BibitemShut
  {NoStop}%
\bibitem [{\citenamefont {Laiho}\ and\ \citenamefont {Van~de
  Water}(2006)}]{Laiho:2005ue}%
  \BibitemOpen
  \bibfield  {author} {\bibinfo {author} {\bibfnamefont {J.}~\bibnamefont
  {Laiho}}\ and\ \bibinfo {author} {\bibfnamefont {R.~S.}\ \bibnamefont {Van~de
  Water}},\ }\href {\doibase 10.1103/PhysRevD.73.054501} {\bibfield  {journal}
  {\bibinfo  {journal} {Phys. Rev.}\ }\textbf {\bibinfo {volume} {D73}},\
  \bibinfo {pages} {054501} (\bibinfo {year} {2006})},\ \Eprint
  {http://arxiv.org/abs/hep-lat/0512007} {hep-lat/0512007} \BibitemShut
  {NoStop}%
\bibitem [{\citenamefont {Colangelo}\ \emph {et~al.}(2011)\citenamefont
  {Colangelo} \emph {et~al.}}]{Colangelo:2010et}%
  \BibitemOpen
  \bibfield  {author} {\bibinfo {author} {\bibfnamefont {G.}~\bibnamefont
  {Colangelo}} \emph {et~al.},\ }\href {\doibase
  10.1140/epjc/s10052-011-1695-1} {\bibfield  {journal} {\bibinfo  {journal}
  {Eur. Phys. J.}\ }\textbf {\bibinfo {volume} {C71}},\ \bibinfo {pages} {1695}
  (\bibinfo {year} {2011})},\ \Eprint {http://arxiv.org/abs/1011.4408}
  {arXiv:1011.4408 [hep-lat]} \BibitemShut {NoStop}%
\bibitem [{\citenamefont {Bazavov}\ \emph
  {et~al.}(2010{\natexlab{b}})\citenamefont {Bazavov} \emph
  {et~al.}}]{Bazavov:2010hj}%
  \BibitemOpen
  \bibfield  {author} {\bibinfo {author} {\bibfnamefont {A.}~\bibnamefont
  {Bazavov}} \emph {et~al.} (\bibinfo {collaboration} {MILC}),\ }\href@noop {}
  {\bibfield  {journal} {\bibinfo  {journal} {PoS}\ }\textbf {\bibinfo {volume}
  {LATTICE2010}},\ \bibinfo {pages} {074} (\bibinfo {year}
  {2010}{\natexlab{b}})},\ \Eprint {http://arxiv.org/abs/1012.0868}
  {arXiv:1012.0868 [hep-lat]} \BibitemShut {NoStop}%
\bibitem [{\citenamefont {Be\v{c}irevi\'c}\ and\ \citenamefont
  {Sanfilippo}(2013)}]{Becirevic:2012pf}%
  \BibitemOpen
  \bibfield  {author} {\bibinfo {author} {\bibfnamefont {D.}~\bibnamefont
  {Be\v{c}irevi\'c}}\ and\ \bibinfo {author} {\bibfnamefont {F.}~\bibnamefont
  {Sanfilippo}},\ }\href {\doibase 10.1016/j.physletb.2013.03.004} {\bibfield
  {journal} {\bibinfo  {journal} {Phys. Lett.}\ }\textbf {\bibinfo {volume}
  {B721}},\ \bibinfo {pages} {94} (\bibinfo {year} {2013})},\ \Eprint
  {http://arxiv.org/abs/1210.5410} {arXiv:1210.5410 [hep-lat]} \BibitemShut
  {NoStop}%
\bibitem [{\citenamefont {Can}\ \emph {et~al.}(2013)\citenamefont {Can},
  \citenamefont {Erkol}, \citenamefont {Oka}, \citenamefont {Ozpineci},\ and\
  \citenamefont {Takahashi}}]{Can:2012tx}%
  \BibitemOpen
  \bibfield  {author} {\bibinfo {author} {\bibfnamefont {K.~U.}\ \bibnamefont
  {Can}}, \bibinfo {author} {\bibfnamefont {G.}~\bibnamefont {Erkol}}, \bibinfo
  {author} {\bibfnamefont {M.}~\bibnamefont {Oka}}, \bibinfo {author}
  {\bibfnamefont {A.}~\bibnamefont {Ozpineci}}, \ and\ \bibinfo {author}
  {\bibfnamefont {T.~T.}\ \bibnamefont {Takahashi}},\ }\href {\doibase
  10.1016/j.physletb.2012.12.050} {\bibfield  {journal} {\bibinfo  {journal}
  {Phys. Lett.}\ }\textbf {\bibinfo {volume} {B719}},\ \bibinfo {pages} {103}
  (\bibinfo {year} {2013})},\ \Eprint {http://arxiv.org/abs/1210.0869}
  {arXiv:1210.0869 [hep-lat]} \BibitemShut {NoStop}%
\bibitem [{\citenamefont {Anastassov}\ \emph {et~al.}(2002)\citenamefont
  {Anastassov} \emph {et~al.}}]{Anastassov:2001cw}%
  \BibitemOpen
  \bibfield  {author} {\bibinfo {author} {\bibfnamefont {A.}~\bibnamefont
  {Anastassov}} \emph {et~al.} (\bibinfo {collaboration} {CLEO}),\ }\href
  {\doibase 10.1103/PhysRevD.65.032003} {\bibfield  {journal} {\bibinfo
  {journal} {Phys. Rev.}\ }\textbf {\bibinfo {volume} {D65}},\ \bibinfo {pages}
  {032003} (\bibinfo {year} {2002})},\ \Eprint
  {http://arxiv.org/abs/hep-ex/0108043} {hep-ex/0108043} \BibitemShut {NoStop}%
\bibitem [{\citenamefont {Lees}\ \emph
  {et~al.}(2013{\natexlab{a}})\citenamefont {Lees} \emph
  {et~al.}}]{Lees:2013uxa}%
  \BibitemOpen
  \bibfield  {author} {\bibinfo {author} {\bibfnamefont {J.~P.}\ \bibnamefont
  {Lees}} \emph {et~al.} (\bibinfo {collaboration} {BaBar}),\ }\href {\doibase
  10.1103/PhysRevLett.111.111801} {\bibfield  {journal} {\bibinfo  {journal}
  {Phys. Rev. Lett.}\ }\textbf {\bibinfo {volume} {111}},\ \bibinfo {pages}
  {111801} (\bibinfo {year} {2013}{\natexlab{a}})},\ \Eprint
  {http://arxiv.org/abs/1304.5009} {arXiv:1304.5009 [hep-ex]} \BibitemShut
  {NoStop}%
\bibitem [{\citenamefont {Lees}\ \emph
  {et~al.}(2013{\natexlab{b}})\citenamefont {Lees} \emph
  {et~al.}}]{Lees:2013zna}%
  \BibitemOpen
  \bibfield  {author} {\bibinfo {author} {\bibfnamefont {J.~P.}\ \bibnamefont
  {Lees}} \emph {et~al.} (\bibinfo {collaboration} {BaBar}),\ }\href {\doibase
  10.1103/PhysRevD.88.052003} {\bibfield  {journal} {\bibinfo  {journal} {Phys.
  Rev.}\ }\textbf {\bibinfo {volume} {D88}},\ \bibinfo {pages} {052003}
  (\bibinfo {year} {2013}{\natexlab{b}})},\ \Eprint
  {http://arxiv.org/abs/1304.5657} {arXiv:1304.5657 [hep-ex]} \BibitemShut
  {NoStop}%
\bibitem [{\citenamefont {Flynn}\ \emph {et~al.}(2013)\citenamefont {Flynn}
  \emph {et~al.}}]{Flynn:2013kwa}%
  \BibitemOpen
  \bibfield  {author} {\bibinfo {author} {\bibfnamefont {J.~M.}\ \bibnamefont
  {Flynn}} \emph {et~al.},\ }\href@noop {} {\  (\bibinfo {year} {2013})},\
  \Eprint {http://arxiv.org/abs/1311.2251} {arXiv:1311.2251 [hep-lat]}
  \BibitemShut {NoStop}%
\bibitem [{\citenamefont {Detmold}\ \emph {et~al.}(2012)\citenamefont
  {Detmold}, \citenamefont {Lin},\ and\ \citenamefont
  {Meinel}}]{Detmold:2012ge}%
  \BibitemOpen
  \bibfield  {author} {\bibinfo {author} {\bibfnamefont {W.}~\bibnamefont
  {Detmold}}, \bibinfo {author} {\bibfnamefont {C.~J.~D.}\ \bibnamefont {Lin}},
  \ and\ \bibinfo {author} {\bibfnamefont {S.}~\bibnamefont {Meinel}},\ }\href
  {\doibase 10.1103/PhysRevD.85.114508} {\bibfield  {journal} {\bibinfo
  {journal} {Phys. Rev.}\ }\textbf {\bibinfo {volume} {D85}},\ \bibinfo {pages}
  {114508} (\bibinfo {year} {2012})},\ \Eprint {http://arxiv.org/abs/1203.3378}
  {arXiv:1203.3378 [hep-lat]} \BibitemShut {NoStop}%
\bibitem [{\citenamefont {Bazavov}\ \emph {et~al.}(2009)\citenamefont {Bazavov}
  \emph {et~al.}}]{Bazavov:2009ir}%
  \BibitemOpen
  \bibfield  {author} {\bibinfo {author} {\bibfnamefont {A.}~\bibnamefont
  {Bazavov}} \emph {et~al.} (\bibinfo {collaboration} {MILC}),\ }\href@noop {}
  {\bibfield  {journal} {\bibinfo  {journal} {PoS}\ }\textbf {\bibinfo {volume}
  {LAT2009}},\ \bibinfo {pages} {077} (\bibinfo {year} {2009})},\ \Eprint
  {http://arxiv.org/abs/0911.0472} {arXiv:0911.0472 [hep-lat]} \BibitemShut
  {NoStop}%
\bibitem [{\citenamefont {Arndt}\ and\ \citenamefont
  {Lin}(2004)}]{Arndt:2004bg}%
  \BibitemOpen
  \bibfield  {author} {\bibinfo {author} {\bibfnamefont {D.}~\bibnamefont
  {Arndt}}\ and\ \bibinfo {author} {\bibfnamefont {C.~J.~D.}\ \bibnamefont
  {Lin}},\ }\href {\doibase 10.1103/PhysRevD.70.014503} {\bibfield  {journal}
  {\bibinfo  {journal} {Phys. Rev.}\ }\textbf {\bibinfo {volume} {D70}},\
  \bibinfo {pages} {014503} (\bibinfo {year} {2004})},\ \Eprint
  {http://arxiv.org/abs/hep-lat/0403012} {hep-lat/0403012} \BibitemShut
  {NoStop}%
\bibitem [{\citenamefont {Amhis}\ \emph {et~al.}()\citenamefont {Amhis} \emph
  {et~al.}}]{Amhis:2012bh}%
  \BibitemOpen
  \bibfield  {author} {\bibinfo {author} {\bibfnamefont {Y.}~\bibnamefont
  {Amhis}} \emph {et~al.} (\bibinfo {collaboration} {Heavy Flavor Averaging
  Group}),\ }\href@noop {} {\enquote {\bibinfo {title} {{Averages of
  $b$-hadron, $c$-hadron, and $\tau$-lepton properties as of early 2012}},}\
  }\Eprint {http://arxiv.org/abs/1207.1158} {arXiv:1207.1158 [hep-ex]}
  \BibitemShut {NoStop}%
\bibitem [{\citenamefont {Barberio}\ and\ \citenamefont
  {W\c{a}s}(1994)}]{Barberio:1993qi}%
  \BibitemOpen
  \bibfield  {author} {\bibinfo {author} {\bibfnamefont {E.}~\bibnamefont
  {Barberio}}\ and\ \bibinfo {author} {\bibfnamefont {Z.}~\bibnamefont
  {W\c{a}s}},\ }\href {\doibase 10.1016/0010-4655(94)90074-4} {\bibfield
  {journal} {\bibinfo  {journal} {Comput. Phys. Commun.}\ }\textbf {\bibinfo
  {volume} {79}},\ \bibinfo {pages} {291} (\bibinfo {year} {1994})}\BibitemShut
  {NoStop}%
\bibitem [{\citenamefont {{Richter-W\c as}}(1993)}]{RichterWas:1992qb}%
  \BibitemOpen
  \bibfield  {author} {\bibinfo {author} {\bibfnamefont {E.}~\bibnamefont
  {{Richter-W\c as}}},\ }\href {\doibase 10.1016/0370-2693(93)90062-M}
  {\bibfield  {journal} {\bibinfo  {journal} {Phys. Lett.}\ }\textbf {\bibinfo
  {volume} {B303}},\ \bibinfo {pages} {163} (\bibinfo {year}
  {1993})}\BibitemShut {NoStop}%
\bibitem [{\citenamefont {Ginsberg}(1966)}]{Ginsberg:1966zz}%
  \BibitemOpen
  \bibfield  {author} {\bibinfo {author} {\bibfnamefont {E.~S.}\ \bibnamefont
  {Ginsberg}},\ }\href {\doibase 10.1103/PhysRev.142.1035} {\bibfield
  {journal} {\bibinfo  {journal} {Phys. Rev.}\ }\textbf {\bibinfo {volume}
  {142}},\ \bibinfo {pages} {1035} (\bibinfo {year} {1966})}\BibitemShut
  {NoStop}%
\bibitem [{\citenamefont {Cirigliano}\ \emph {et~al.}(2008)\citenamefont
  {Cirigliano}, \citenamefont {Giannotti},\ and\ \citenamefont
  {Neufeld}}]{Cirigliano:2008wn}%
  \BibitemOpen
  \bibfield  {author} {\bibinfo {author} {\bibfnamefont {V.}~\bibnamefont
  {Cirigliano}}, \bibinfo {author} {\bibfnamefont {M.}~\bibnamefont
  {Giannotti}}, \ and\ \bibinfo {author} {\bibfnamefont {H.}~\bibnamefont
  {Neufeld}},\ }\href {\doibase 10.1088/1126-6708/2008/11/006} {\bibfield
  {journal} {\bibinfo  {journal} {JHEP}\ }\textbf {\bibinfo {volume} {0811}},\
  \bibinfo {pages} {006} (\bibinfo {year} {2008})},\ \Eprint
  {http://arxiv.org/abs/0807.4507} {arXiv:0807.4507 [hep-ph]} \BibitemShut
  {NoStop}%
\bibitem [{\citenamefont {Schwanda}()}]{Schwanda:2013se}%
  \BibitemOpen
  \bibfield  {author} {\bibinfo {author} {\bibfnamefont {C.}~\bibnamefont
  {Schwanda}},\ }\href@noop {} {}\bibinfo {note} {{private communication
  (2013)}}\BibitemShut {NoStop}%
\bibitem [{\citenamefont {Aubert}\ \emph {et~al.}(2008)\citenamefont {Aubert}
  \emph {et~al.}}]{Aubert:2007qs}%
  \BibitemOpen
  \bibfield  {author} {\bibinfo {author} {\bibfnamefont {B.}~\bibnamefont
  {Aubert}} \emph {et~al.} (\bibinfo {collaboration} {BaBar}),\ }\href
  {\doibase 10.1103/PhysRevLett.100.231803} {\bibfield  {journal} {\bibinfo
  {journal} {Phys. Rev. Lett.}\ }\textbf {\bibinfo {volume} {100}},\ \bibinfo
  {pages} {231803} (\bibinfo {year} {2008})},\ \Eprint
  {http://arxiv.org/abs/0712.3493} {arXiv:0712.3493 [hep-ex]} \BibitemShut
  {NoStop}%
\bibitem [{\citenamefont {Adam}\ \emph {et~al.}(2003)\citenamefont {Adam} \emph
  {et~al.}}]{Adam:2002uw}%
  \BibitemOpen
  \bibfield  {author} {\bibinfo {author} {\bibfnamefont {N.}~\bibnamefont
  {Adam}} \emph {et~al.} (\bibinfo {collaboration} {CLEO}),\ }\href {\doibase
  10.1103/PhysRevD.67.032001} {\bibfield  {journal} {\bibinfo  {journal} {Phys.
  Rev.}\ }\textbf {\bibinfo {volume} {D67}},\ \bibinfo {pages} {032001}
  (\bibinfo {year} {2003})},\ \Eprint {http://arxiv.org/abs/hep-ex/0210040}
  {arXiv:hep-ex/0210040 [hep-ex]} \BibitemShut {NoStop}%
\bibitem [{\citenamefont {Aubert}\ \emph {et~al.}(2009)\citenamefont {Aubert}
  \emph {et~al.}}]{Aubert:2008yv}%
  \BibitemOpen
  \bibfield  {author} {\bibinfo {author} {\bibfnamefont {B.}~\bibnamefont
  {Aubert}} \emph {et~al.} (\bibinfo {collaboration} {BaBar}),\ }\href
  {\doibase 10.1103/PhysRevD.79.012002} {\bibfield  {journal} {\bibinfo
  {journal} {Phys. Rev.}\ }\textbf {\bibinfo {volume} {D79}},\ \bibinfo {pages}
  {012002} (\bibinfo {year} {2009})},\ \Eprint {http://arxiv.org/abs/0809.0828}
  {arXiv:0809.0828 [hep-ex]} \BibitemShut {NoStop}%
\bibitem [{\citenamefont {Gambino}\ and\ \citenamefont
  {Schwanda}(2013)}]{Gambino:2013rza}%
  \BibitemOpen
  \bibfield  {author} {\bibinfo {author} {\bibfnamefont {P.}~\bibnamefont
  {Gambino}}\ and\ \bibinfo {author} {\bibfnamefont {C.}~\bibnamefont
  {Schwanda}},\ }\href@noop {} {\  (\bibinfo {year} {2013})},\ \Eprint
  {http://arxiv.org/abs/1307.4551} {arXiv:1307.4551} \BibitemShut {NoStop}%
\bibitem [{\citenamefont {Oktay}\ and\ \citenamefont
  {Kronfeld}(2008)}]{Oktay:2008ex}%
  \BibitemOpen
  \bibfield  {author} {\bibinfo {author} {\bibfnamefont {M.~B.}\ \bibnamefont
  {Oktay}}\ and\ \bibinfo {author} {\bibfnamefont {A.~S.}\ \bibnamefont
  {Kronfeld}},\ }\href {\doibase 10.1103/PhysRevD.78.014504} {\bibfield
  {journal} {\bibinfo  {journal} {Phys. Rev.}\ }\textbf {\bibinfo {volume}
  {D78}},\ \bibinfo {pages} {014504} (\bibinfo {year} {2008})},\ \Eprint
  {http://arxiv.org/abs/0803.0523} {arXiv:0803.0523 [hep-lat]} \BibitemShut
  {NoStop}%
\bibitem [{\citenamefont {Caprini}\ \emph {et~al.}(1998)\citenamefont
  {Caprini}, \citenamefont {Lellouch},\ and\ \citenamefont
  {Neubert}}]{Caprini:1997mu}%
  \BibitemOpen
  \bibfield  {author} {\bibinfo {author} {\bibfnamefont {I.}~\bibnamefont
  {Caprini}}, \bibinfo {author} {\bibfnamefont {L.}~\bibnamefont {Lellouch}}, \
  and\ \bibinfo {author} {\bibfnamefont {M.}~\bibnamefont {Neubert}},\ }\href
  {\doibase 10.1016/S0550-3213(98)00350-2} {\bibfield  {journal} {\bibinfo
  {journal} {Nucl. Phys.}\ }\textbf {\bibinfo {volume} {B530}},\ \bibinfo
  {pages} {153} (\bibinfo {year} {1998})},\ \Eprint
  {http://arxiv.org/abs/hep-ph/9712417} {hep-ph/9712417} \BibitemShut {NoStop}%
\bibitem [{\citenamefont {Qiu}\ \emph {et~al.}(2013)\citenamefont {Qiu},
  \citenamefont {DeTar}, \citenamefont {El-Khadra}, \citenamefont {Kronfeld},
  \citenamefont {Laiho},\ and\ \citenamefont {{Van de Water}}}]{Qiu:2013ofa}%
  \BibitemOpen
  \bibfield  {author} {\bibinfo {author} {\bibfnamefont {S.-W.}\ \bibnamefont
  {Qiu}}, \bibinfo {author} {\bibfnamefont {C.}~\bibnamefont {DeTar}}, \bibinfo
  {author} {\bibfnamefont {A.~X.}\ \bibnamefont {El-Khadra}}, \bibinfo {author}
  {\bibfnamefont {A.~S.}\ \bibnamefont {Kronfeld}}, \bibinfo {author}
  {\bibfnamefont {J.}~\bibnamefont {Laiho}}, \ and\ \bibinfo {author}
  {\bibfnamefont {R.~S.}\ \bibnamefont {{Van de Water}}},\ }\href@noop {}
  {\bibfield  {journal} {\bibinfo  {journal} {PoS}\ }\textbf {\bibinfo {volume}
  {Lattice 2013}},\ \bibinfo {pages} {385} (\bibinfo {year} {2013})},\ \Eprint
  {http://arxiv.org/abs/1312.0155} {arXiv:1312.0155 [hep-lat]} \BibitemShut
  {NoStop}%
\bibitem [{\citenamefont {Aubin}\ and\ \citenamefont
  {Bernard}(2003{\natexlab{a}})}]{Aubin:2003mg}%
  \BibitemOpen
  \bibfield  {author} {\bibinfo {author} {\bibfnamefont {C.}~\bibnamefont
  {Aubin}}\ and\ \bibinfo {author} {\bibfnamefont {C.}~\bibnamefont
  {Bernard}},\ }\href {\doibase 10.1103/PhysRevD.68.034014} {\bibfield
  {journal} {\bibinfo  {journal} {Phys. Rev.}\ }\textbf {\bibinfo {volume}
  {D68}},\ \bibinfo {pages} {034014} (\bibinfo {year} {2003}{\natexlab{a}})},\
  \Eprint {http://arxiv.org/abs/hep-lat/0304014} {hep-lat/0304014} \BibitemShut
  {NoStop}%
\bibitem [{\citenamefont {Aubin}\ and\ \citenamefont
  {Bernard}(2003{\natexlab{b}})}]{Aubin:2003uc}%
  \BibitemOpen
  \bibfield  {author} {\bibinfo {author} {\bibfnamefont {C.}~\bibnamefont
  {Aubin}}\ and\ \bibinfo {author} {\bibfnamefont {C.}~\bibnamefont
  {Bernard}},\ }\href {\doibase 10.1103/PhysRevD.68.074011} {\bibfield
  {journal} {\bibinfo  {journal} {Phys. Rev.}\ }\textbf {\bibinfo {volume}
  {D68}},\ \bibinfo {pages} {074011} (\bibinfo {year} {2003}{\natexlab{b}})},\
  \Eprint {http://arxiv.org/abs/hep-lat/0306026} {hep-lat/0306026} \BibitemShut
  {NoStop}%
\bibitem [{\citenamefont {Antonelli}\ \emph {et~al.}(2010)\citenamefont
  {Antonelli} \emph {et~al.}}]{Antonelli:2009ws}%
  \BibitemOpen
  \bibfield  {author} {\bibinfo {author} {\bibfnamefont {M.}~\bibnamefont
  {Antonelli}} \emph {et~al.},\ }\href {\doibase 10.1016/j.physrep.2010.05.003}
  {\bibfield  {journal} {\bibinfo  {journal} {Phys. Rept.}\ }\textbf {\bibinfo
  {volume} {494}},\ \bibinfo {pages} {197} (\bibinfo {year} {2010})},\ \Eprint
  {http://arxiv.org/abs/0907.5386} {arXiv:0907.5386 [hep-ph]} \BibitemShut
  {NoStop}%
\end{thebibliography}%
